\newif\ifpagelimit

\pagelimittrue
\newif\ifdouble
\doubletrue
\ifdouble
\documentclass[conference,letterpaper]{IEEEtran}
\else
\documentclass[lettersize,onecolumn]{IEEEtran}
\fi

\usepackage[utf8]{inputenc} 
\usepackage[T1]{fontenc}
\usepackage{url}
\usepackage{ifthen}
\usepackage{cite}
\usepackage{comment}
\usepackage[cmex10]{amsmath} 

\allowdisplaybreaks 

\usepackage{graphicx}
\usepackage{subcaption} 
\usepackage{algorithm}
\usepackage{algorithmic}
\usepackage{setspace}
\usepackage[colorlinks=true, allcolors=blue]{hyperref}
\usepackage{amssymb}
\usepackage{amsfonts}
\usepackage{xcolor}
\usepackage{acronym}
\usepackage{soul}
\usepackage{enumitem}
\usepackage{tikz}
\usepackage{amsthm}
\usepackage{float}
\usepackage{dblfloatfix}
\usetikzlibrary{calc}
\usetikzlibrary{automata, positioning, arrows.meta}
\tikzset{
    ->,
    >=Stealth, 
    node distance=2cm,
    every state/.style={thick},  
    initial text =,
}

\usepackage{subcaption}
\acrodef{rtt}[RTT]{Round-Trip Time}
\acrodef{fec}[FEC]{Forward Error Correction}
\acrodef{acrlnc}[AC-RLNC]{Adaptive and Causal Network Coding}
\acrodef{bs-acrlnc}[BS-AC-RLNC]{Blank Space AC-RLNC}
\acrodef{net-acrlnc}[NET]{Network AC-RLNC}
\acrodef{bsp}[BSP]{Blank Space Period}

\newtheorem{theorem}{Theorem}
\newtheorem{lemma}{Lemma}
\newtheorem{corollary}{Corollary}
\newtheorem{definition}{Definition}
\newtheorem{remark}{Remark}

\newcommand{\ale}[1]{\textcolor{red}{ [ #1 -- Alejandro ] \normalsize}}
\newcommand{\off}[1]{}
\newcommand\rg[1]{{\color{black} #1}}
\newcommand\al[1]{{\color{red} #1}}

\newcommand\blfootnote[1]{%
    \begingroup
    \renewcommand\thefootnote{}%
    \footnote{#1}%
     \addtocounter{footnote}{-1}%
    \endgroup
}

\begin{document}
\title{Blank Space: Adaptive Causal Coding for Streaming Communications Over Multi-Hop Networks\vspace{-0.4cm}}


\ifdouble
\author{%
   \IEEEauthorblockN{Rivka Gitik\IEEEauthorrefmark{1},
   Adina Waxman\IEEEauthorrefmark{1}$^{\ddagger}$,
                     Shai Ginzach\IEEEauthorrefmark{2},
                     Aviel Glam\IEEEauthorrefmark{2},
                     and Alejandro Cohen\IEEEauthorrefmark{1}}
   \IEEEauthorblockA{\IEEEauthorrefmark{1}%
                      Faculty of ECE, Technion, Israel, \{rivkagitik, adina.waxman\}@campus.technion.ac.il, alecohen@technion.ac.il}  
    \IEEEauthorblockA{\IEEEauthorrefmark{2}%
                     Rafael, Israel, \{shaigi, avielg\}@rafael.co.il}
\vspace{-1.1cm}}
\fi

\maketitle


\begin{abstract}


In this work, we introduce Blank Space Adaptive Causal Random Linear Network Coding (BS-AC-RLNC), a novel coding scheme designed to mitigate the triplet trade-off between throughput-delay-efficiency in multi-hop networks. BS-AC-RLNC leverages the physical limitations of the network, considering the bottleneck from each node to the destination. 
%
In particular, this approach introduces a light-computational re-encoding algorithm, called \acf{net-acrlnc}, implemented independently at intermediate nodes. \ac{net-acrlnc} adaptively adjusts the \ac{fec} rates and schedules idle periods.
It incorporates two distinct suspension mechanisms: 1) Blank Space Period, accounting for the forward-channels bottleneck, and 2) No-New No-FEC approach, based on data availability. We present theoretical lower and upper bounds on in-order delivery delay, goodput, and throughput; in the case of in-order delay, we further derive a mean bound. These analytical results are extended to the multicast scenario, providing a broader understanding of the algorithm's performance under diverse network conditions.
The experimental results achieve significant improvements in resource efficiency, demonstrating a $20\%$ reduction in channel usage compared to baseline RLNC solutions\off{, with even greater reductions at higher erasure rates}. Notably, these efficiency gains are achieved while maintaining competitive throughput and delay performance, ensuring improved resource utilization does not compromise network performance.\blfootnote{$\ddagger$ This paper was presented in part at the IEEE International Symposium on Information Theory in 2025.}
\end{abstract}

\vspace{-0.2cm}
\section{Introduction}
\vspace{-0.2cm}
The exponential growth in streaming applications, demanding high data rates and low latency, has pushed wireless connectivity beyond traditional point-to-point schemes toward multi-hop networks, where intermediate nodes cooperate and share the common medium. However, as these networks evolve, ensuring efficient use of power consumption and spectrum utilization while maintaining high goodput and low delay is essential. The difficulty increases when the wireless channel is noisy, dynamic, and unknown. These conditions are all common to both modern and future terrestrial and non-terrestrial networks, such as Unmanned Aerial Vehicles (UAV) networks \cite{ yanmaz2018drone}, Mobile Ad-Hoc Networks (MANETs) \cite{kumar2012overview} and Wireless Sensor Networks (WSN) \cite{radi2012multipath}. 

Network Coding (NC) \cite{ahlswede2000network, li2003linear}, particularly Random Linear Network Coding (RLNC) \cite{ho2006random,dong2024throughput}, has demonstrated the ability to achieve the min-cut max-flow capacity in multi-hop networks. RLNC was combined with Automatic Repeat Request (ARQ) protocols under the packet erasure channel, using the concept of recoding and a sliding-window approach to network coding \cite{4595268, 5061931, 5688180}. 
However, these solutions struggle to meet the ultra-low delay requirements due to their large blocklength regime. 
Alternative approaches like rateless \cite{luby2002lt,shokrollahi2006raptor} and stemming codes for point-to-point communications \cite{joshi2012playback,cloud2015coded,gabriel2018multipath,9775949,7249034,8638958} have been proposed to address delay concerns. TCP-based congestion control was studied in\cite{cloud2013network, 6883489}. However, these solutions remain sensitive to channel variations as they lack adaptation mechanisms \cite{8767270,7117455,9834704,9174225,8835153,10064107}.

Adaptive and Causal RLNC (AC-RLNC), recently proposed in \cite{cohen2020adaptive, cohen2020adaptiveMH}, employs an adaptive-size sliding window on coded packets based on channel state estimation. In particular, it facilitates communication with zero error probability, a property that has been the focus of extensive investigation in the classical literature \cite{shannon2003zero,gallager2003simple,lovasz1979shannon} and more recently under the finite block length regime \cite{sahai2008block,polyanskiy2011feedback}. AC-RLNC utilizes two key mechanisms: an a-priori Forward Error Correction (FEC) mechanism that compensates for expected erasures, and an a-posteriori feedback-based FEC mechanism that handles lost packets \cite{cohen2020adaptive,cohen2021bringing,cohen2022broadcast,dias2023sliding}. By leveraging per-packet acknowledgments, AC-RLNC is more suitable for Ultra-Reliable Low Latency Communications (URLLC) \cite{ali2021urllc}.

\ifdouble
\begin{figure}
    \centering
    \includegraphics[width=1\linewidth]{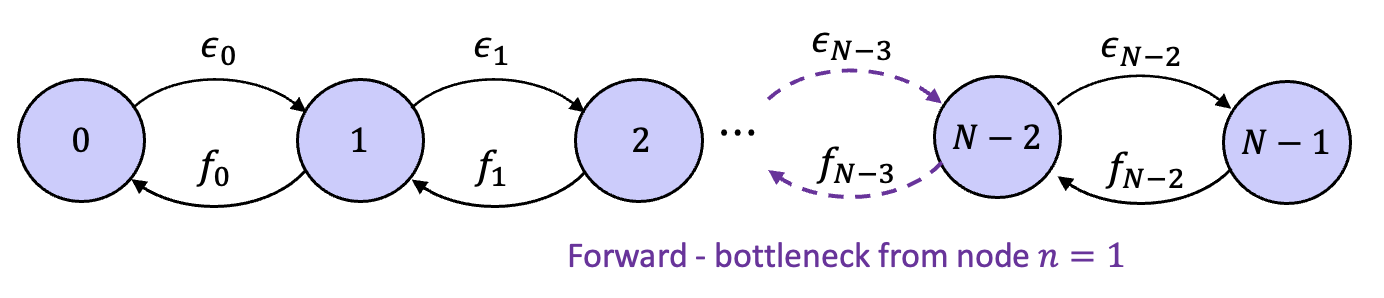}
    \vspace{-0.7cm}
    \caption{\small 
    Illustration of a multi-hop communication system with $N$ nodes. The dashed line denotes the \textit{forward bottleneck from node 1} to the destination, as estimated by the proposed \ac{bs-acrlnc}. In this example, for node $n = 1$, the estimated forward bottleneck $\epsilon_{N-3}$ is the largest among the set $\{\epsilon_1, \ldots, \epsilon_{N-2}\}$, though it is not necessarily larger than $\epsilon_0$.}
    \label{fig:network}
    \vspace{-0.6cm}
\end{figure}
\else\fi

In this work, we propose a coding-based solution that mitigates the trade-off between throughput-delay-efficiency in multi-hop networks. Optimizing this trade-off remains an open question, and to the best of our knowledge, this is the first work addressing this problem using RLNC.
We propose \acf{bs-acrlnc}, a novel adaptive and causal solution that generalizes AC-RLNC to achieve the desired triplet trade-off in multi-hop networks.

In \ac{bs-acrlnc}, intermediate nodes may employ \acf{net-acrlnc}, a light-computational re-encoding algorithm.
\ac{net-acrlnc} enables nodes to adaptively adjust their re-transmission rates based on channel conditions (similarly to the source in \ac{acrlnc}) and schedule idle periods to prevent ineffective transmissions.  
The nodes manage their sliding coding windows based on decoding assessment rather than actual packet decoding, eliminating the decoding computational overhead.
\ac{net-acrlnc} introduces two transmission suspension mechanisms: 1) Blank Space Period, which considers the forward channels bottleneck - see Fig.~\ref{fig:network}, and 2) No-New No-FEC approach, which stems from the \ac{bs-acrlnc} algorithm structure of independently operating nodes, based on the current buffer state and content.
These two mechanisms effectively reduce the channel usage rate and increase the goodput, while maintaining high delivery rates and low in-order delivery delay.

\rg{Related work has also considered chain-topology networks with batch RLNC. In particular, Dong et al.~[7] provide analytical expressions for maximum throughput and end-to-end latency in a line network under link outages. While their analysis focuses on outage events, our work instead tracks channel conditions in real time, taking into account node bottlenecks and dynamically adjusting transmissions to mitigate the throughput–delay–efficiency tradeoff.}

We provide a theoretical analysis of the proposed coding algorithm by deriving lower and upper bounds on key performance metrics, including in-order delivery delay, goodput, and throughput. In the case of in-order delay, we further establish a bound on the expected (mean) delay, offering more detailed insight into latency behavior. These analytical results are extended to the multicast setting over multi-hop erasure channels, where receivers experience independent losses across shared paths. This extension demonstrates the algorithm's robustness in multi-hop scenarios and enhances our understanding of its performance under varying conditions within this constrained yet practically relevant class of network topologies.

Our experimental results demonstrate a substantial improvement in resource efficiency while reducing channel usage. For example, for a setting with a network bottleneck of $40\%$ erasure rate, the channel usage drops by $20\%$ for the end-to-end network, and up to $40\%$ for individual nodes, compared to a baseline RLNC implementation. This usage rate further decreases as the global bottleneck erasure rate increases. Notably, the increase in efficiency is achieved while maintaining competitive performance in the rate-delay trade-off for URLLC, ensuring that improved resource utilization does not come at the cost of degraded network performance.

The structure of the paper is as follows. Section~\ref{sec:SysModel} introduces the system model and defines the problem. Section~\ref{sec:Algorithm} details the proposed \ac{bs-acrlnc} for multi-hop networks. In Section~\ref{section:analyticalResults}, we present theoretical lower and upper bounds for in-order delivery delay, goodput, and throughput; for delay, we also derive the mean bound. These results are extended to the multi-cast scenario in Section~\ref{section:multicast}. Section~\ref{section:evaluation} presents simulation results for the \ac{bs-acrlnc} protocol, where the theoretical bounds are demonstrated and evaluated. Finally, Section~\ref{section:conclusions} concludes the paper and outlines directions for future work. The notation used throughout the paper is summarized in Tables~\ref{fig:table1} and~\ref{fig:table_analytical_results} \ifpagelimit in the supplementary material of this work\else\fi.
\vspace{-0.2cm}
\section{System Model and Preliminaries}
\label{sec:SysModel}
This section covers our system model (Subsection~\ref{subsec:SysModel}), problem formulation (Subsection~\ref{subsec:Problem}), and adaptive and causal network coding preliminaries (Subsection~\ref{subsec:Preliminaries}).
\vspace{-0.15cm}
\subsection{System Model}
\label{subsec:SysModel}
Consider \rg{the network layer, or more generally any layer operating at the packet level,} of a multi-hop real-time slotted communication system with $N$ nodes, ranging from node $0$ (source) to node $N-1$ (destination). Each pair of adjacent nodes $(n, n+1)$ communicates over a directional noisy forward channel, denoted as $e_n$, for $n \in \{ 0,1,...,N-2 \}$. A noiseless\footnote{\rg{The noiseless-feedback assumption may be relaxed, e.g., via cumulative feedback~\cite{malak2019tiny} over noisy channels. Since the feedback is binary and can be transmitted at a low rate, the resulting increase in slot duration is minimal and does not significantly affect system performance.}} feedback channel, $f_n$, used to send acknowledgments, is available in the reverse direction. 
This multi-hop network is illustrated in Fig.~\ref{fig:network}.
\rg{Time is slotted and indexed by $t \in \{0,1,\dots,T-1\}$, where $T \in \mathbb{N}$ denotes the finite time horizon. At each time slot $t$, node $n$ can transmit a packet over its forward channel $e_n$.
}
These forward channels are modeled as i.i.d. packet-level Binary Erasure Channels (BECs)\footnote{Following~\cite{cohen2020adaptive,cohen2020adaptiveMH}, we adopt the BEC terminology. Related literature also refers to similar packet-level abstractions as erasure or packet erasure channels~\cite{karzand2015fec,shi2013whether}.} - For each channel $e_n$,
packets are either perfectly received with probability $1-\epsilon_n$, or completely erased with probability $\epsilon_n$ at any time slot $t < T$.
The corresponding channel rates in the asymptotic regime where $T \rightarrow \infty$ are $r_n = 1- \epsilon_n$, the capacity of the BEC channel \cite{PacketBEC}. 

For any node $n$, we define its  forward-channels bottleneck ${BN}_n$, as the channel index with the highest erasure rate along the remaining forward path, that is,
\vspace{-0.1cm}
\begin{equation}
\textstyle    {BN}_n = \underset{{n \leq i \leq N-2} } {\text{argmax}}\left\{  \epsilon_i\right\}.
    \label{eqn: BN_n}
\end{equation}
\vspace{-0.1cm}
The global bottleneck in the multi-hop network, denoted by ${\rm BN}$, is therefore ${BN}_0$. 

We do not assume prior knowledge of the forward channel's erasure rate. However, each node does have access to delayed local feedback and may use it to estimate the channel's statistics.
Specifically, each pair of nodes has its own round-trip time denoted by ${\rm RTT}_n$. 
For simplicity, we assume symmetric propagation delays of ${{\rm RTT}_n}/{2}$ in both forward and feedback channels. The global propagation delay from the source to the destination is therefore, ${\rm RTT}/{2}$ where ${\rm RTT} = \sum_{n=0}^{N-2}{\rm RTT}_n$.
Upon receiving a packet from node $n$, node $n+1$ transmits either an acknowledgment (ACK) or a negative-acknowledgment (NACK) message back to node $n$ over the noiseless feedback channel, $f_n$. 
For a packet transmitted at time $t$, we denote its feedback message from node $n+1$ as $F_{t}^{n+1}$, where $F_t^{n+1}=1$ indicates an ACK and $F_{t}^{n+1}=0$ indicates a NACK.
For a batch of \rg{the $M$ last} received feedback messages, the estimated erasure rate\footnote{\rg{This estimation is initialized by a default value for $t < {\rm RTT}_n-M$.}} of the forward channel $e_n$, is given by
\begin{equation}
\textstyle \hat{\epsilon}_n = 1-\frac{1}{M} \sum_{i={\rg{t-{\rm RTT}_n-M}}}^\rg{t-{\rm RTT}_n} {F_i^{\rg{n+1}}}.
\label{eqn:eps_hat}
\end{equation}
\off{\rg{Let \(M(t)\) denote the number of feedback messages received up to the current time \(t\). Accordingly,} the estimated erasure rate of the forward channel $e_n$, is given by 
\begin{equation}
\textstyle \hat{\epsilon}_n = 1-\frac{1}{M\rg{(t)}} \sum_{i=0}^{M\rg{(t)}} {F_i^n}.
\label{eqn:eps_hat}
\end{equation}}

Each node has a buffer to store incoming packets, which may arrive one per time slot (including the source). 
The source operates as an encoder, which encodes available raw data packets and transmits a coded packet to the next node. Each intermediate node operates as a re-encoder, capable of encoding incoming coded packets into new coded packets. We distinguish between \rg{\textit{raw data packet}}, $p_i$ (the $i$th data packet arrived at the source) and \textit{coded packets} (any coded packet transmitted over any of the forward channels). The packets are denoted by $c_t^n$, representing a coded packet transmitted from node $n$ at time $t$. For any raw data packet $p_i$, we define $T_1(p_i)$ as the arrival time at the source node and $T_d(p_i)$ as its decoding time at the destination. $d(t)$ represents the cumulative number of raw data packets decoded at the destination by time $t$.

Any transmitting node $n$ begins operation after an initial delay of $\sum_{i=0}^{n-1}\frac{{\rm RTT}_i}{2}$, since this is the first time it may receive a packet. During operation, a node may pause transmission at any time slot, in which case no packet is sent. The number of these idle slots for node $n$ is denoted by $O_n\rg{(t)}$ (excluding the initial delay period).
\rg{In this notation, \(O_n(T)\) represents the number of idle slots over the time horizon \(T\) caused by the node’s own transmission decisions, and does not include idle time due to the initial propagation delay.}

\vspace{-0.3cm}
\subsection{Problem Formulation}
\label{subsec:Problem}
\vspace{-0.09cm}
Our goal herein is threefold: 1) maximize data delivery, 2) minimize in-order delivery delay, and 3) minimize channel usage. 
We will use the following metrics.

\noindent 1) {\bf Normalized Goodput, $\eta$:}  The total amount of raw data delivered to the destination\rg{, measured in number of raw data packets}, divided by the total amount of \rg{packets} transmitted by the source in this period. Since we consider a slotted transmission with a fixed number of $T$ time slots where one packet may be transmitted at each slot and a constant packet size, the normalized goodput can be calculated as,
\begin{equation}
\textstyle    \eta \triangleq \frac{d(T)}{T - O_0\rg{(T)} - \frac{\rm RTT}{2}},
    \label{eqn:eta_n}
\end{equation}
\rg{where \(T\) is reduced by the number of idle slots at the first node and by the global propagation delay.}\off{\ale{Fix here}{\color{blue}{Thank you! done}}}
recalling $d(T)$ is the total number of raw data packets decoded by the end of the network operation.

\rg{Note that in this metric, the traditional denominator is normalized to exclude both deliberately introduced idle periods and unavoidable initial propagation delays, as neither contributes to active data transmission despite extending the total elapsed time. 
This normalization is essential in our framework, where congestion is proactively mitigated through the intentional insertion of idle slots, effectively suspending transmission to prevent persistent congestion cycles and associated packet losses. As this mechanism deliberately increases the transmission duration, the conventional goodput metric would underestimate system performance. The normalized goodput, therefore, provides a more faithful efficiency measure by accounting only for effective transmission opportunities.\off{A complementary metric, the Channel Usage Rate, will be introduced later to provide an additional perspective on the efficiency aspect of the system.}}


\noindent 2) {\bf Delivery Rate, $R_{\rm{del}}$:} The average raw data packets delivered to the destination in units of \rg{raw data packets per slot,}
\rg{for a chosen \( T' \) such that \( {T}/{T'} \in \mathbb{N} \).}
\begin{equation}\label{eqn:rDel}
\textstyle    R_{\rm{del}} = \frac{1}{T/T'} \sum_{i=1}^{T/T'} (d(i\cdot T')-d((i-1)\cdot T')).
\end{equation} 
\rg{This metric quantifies the average number of raw data packets successfully delivered to the destination over $T'$. To compute it, the time horizon $T$ is divided into $T/T'$ consecutive intervals, and for each interval, the number of received raw data packets is counted. The delivery rate is then defined as the mean over all intervals.
}

\noindent 3) {\bf Channel Usage Rate, $U$, $U_n$:} The ratio of transmitting time slots to overall transmission opportunities in the entire network and per forward channel, respectively:
\begin{equation}
\textstyle    U = 1 - \frac{\sum_{i=0}^{N-2}O_i\rg{(T)}}{\sum_{i=0}^{N-2}\left(T-(N-2-i)\frac{{\rm RTT}_i}{2} \right)} \text{, and   } U_n = 1 - \frac{O_n\rg{(T)}}{T-\sum_{i=0}^{n-1}\frac{{\rm RTT}_i}{2}}.
\label{eqn:channel_use}
\end{equation}
\rg{where $O_i(T)$ denotes the number of idle slots at node $i$, excluding the initial propagation delay.}
We note that the transmission opportunity (i.e., the denominator) of each node considers the operation interval $T$ and the initial propagation delay until each node in the network. 
That is, the transmission opportunity in the entire network with $N-1$ nodes is 
\ifdouble
\vspace{-0.2cm}
\begin{align*}
\textstyle &\underbrace{T}_{\text{node } 0}+\underbrace{\left(T-\frac{{\rm RTT}_0}{2}\right)}_{\text{node }  1}+\cdots
&\quad+\underbrace{\left(T-\sum_{k=0}^{N-3}\frac{{\rm RTT}_k}{2}\right)}_{\text{node }  N-2}\\
&\Rightarrow \sum_{i=0}^{N-2} T-(N-2-i)\frac{{\rm RTT}_i}{2}
\end{align*}
\fi
and for each node $n$ in the network by
\begin{equation*}
\textstyle    T-\underbrace{\frac{{\rm RTT}_0}{2}}_{\text{hop } 0}-\cdots-\underbrace{\frac{{\rm RTT}_{n-1}}{2}}_{\text{hop } n-1} \Rightarrow T-\sum_{i=0}^{n-1}\frac{{\rm RTT}_i}{2}.
\end{equation*}
\vspace{-0.3cm}


\noindent 4) {\bf In-Order Delivery Delay, $D$:}
The difference between the time slot in which a raw data packet arrives at the source node and the time slot in which it is decoded in order at the destination. \rg{As will be shown in Subsection \ref{subsec:Preliminaries}, the proposed solution \off{\ale{solution?}{\color{blue}{Done}}} inherently enforces in-order delivery.} That is, the in-order delivery delay per raw data packet of index $i$ is $D_i \triangleq T_d(p_i) - T_1(p_i)$.
To evaluate network performance, we analyze delay throughout the network. Let $d$ denote the total number of decoded raw data packets in the operation interval $T$. We propose two metrics for delay performance.

4.1) {\bf Mean Delay, $D^{\text{mean}}$:} The {\em average} delay of all raw data packets sent in the network
\vspace{-0.1cm}
\begin{subequations}
\label{eqn:D_mean_max}
    \begin{equation}
\textstyle    D^{\text{mean}} \triangleq  \frac{1}{d}\sum_{i=1}^{d}D_i.
    \label{eqn:D_mean}
    \end{equation}
This measure represents the network’s ability to send the entirety of the data.

4.2) {\bf Maximum Delay, $D^{\text{max}}$:} The {\em maximum} delay of all raw data packets sent in the network
\vspace{-0.1cm}
\begin{equation}
\textstyle    D^{\text{max}} \triangleq \max_{i\in\{1,\ldots, d\}} D_i
   . \label{eqn:D_max}
    \end{equation}
\end{subequations}
This quantity represents the worst-case guaranteed quality of service in a real-time streaming scenario.

While both normalized goodput and delivery rate assess data delivery performance, each one focuses on a different aspect. The goodput measures successful data delivery, accounting for the propagation delay and idle periods, reflecting protocol performance relative to its own limitations. The delivery rate quantifies the amount of successfully delivered data regardless of resource usage, offering insight into how effectively the protocol manages overall transmission. We use this metric to evaluate scheduling efficiency and the impact of idle periods. More specifically, if suspending transmissions does not affect the delivery rate, it indicates that these transmissions were unnecessary in the first place.

\rg{To conclude, our goal is to develop an adaptive coding scheme, $\mathcal{C}$, that constructs encoded packets from raw data packets while leveraging past feedback information, with the aim of maximizing high goodput and high delivery rate while decreasing the channel usage rate without increasing the in-order delivery delay. Accordingly, the design problem can be expressed as
\vspace{-0.1cm}
\begin{equation*}
    \begin{aligned}
        \arg &\max_{\mathcal{C}} \quad  \eta(\mathcal{C}) \\
        &\text{s.t.} \quad  R_{del}(\mathcal{C}) \geq R_{target}, \text{ } U(\mathcal{C}) \leq U_{target}\\
        & \qquad D^{mean}(\mathcal{C}) \leq D^{mean}_{target}, \text{ and } D^{\max}(\mathcal{C}) \leq D^{\max}_{target}.
    \end{aligned}
\end{equation*}
Here, $R_{target}$, $U_{target}$, $D^{mean}_{target}$, and $D^{\max}_{target}$ denote target performance levels determined by system requirements.
It is important to note that the above optimization formulation is primarily intended as a systematic framework for selecting the design parameters. The coding scheme $\mathcal{C}$ introduced in Section~\ref{sec:Algorithm} is parameterized by $\hat{\epsilon}_n$, ${\rm RTT}_n$, and additional quantities specified therein. These parameters are chosen based on the analytical delay bounds established in Section~\ref{section:analyticalResults}, which guide their tuning to meet the desired performance objectives.}
\vspace{-0.1cm}
\subsection{Preliminaries}
\label{subsec:Preliminaries}
Our solution extends the single-path AC-RLNC originally presented in \cite{cohen2020adaptive} and makes comparisons with the multipath-multihop work in \cite{cohen2020adaptiveMH}. We will briefly review both schemes. The parameters in this Subsection are defined in the same manner as in Subsection~\ref{subsec:SysModel}, omitting the unnecessary node index $n$.

\subsubsection{Single Path (SP) AC-RLNC}
\label{subsec:AC}
SP-AC-RLNC uses sliding window RLNC \cite{swrlnc} to code raw data packets into linear combinations. 
Let $p_i \in \mathbb{F}_q^{\ell}$ represent raw data packet of length $\ell$ over finite field $\mathbb{F}_q$, with data index $0 \leq i \leq T$.
The linear combination transmitted at time $t$ is
\begin{equation}
\label{eqn:dof}
\textstyle  c_{t}=\sum_{i=w_t^{\min}}^{w_t^{\max}}\mu_{i}p_{i},
\end{equation}
where $\mu_i \in \mathbb{F}_q$ are random coefficients drawn uniformly, and $w_t^{\min}$, $w_t^{\max}$ define the \rg{effective} sliding window bounds\footnote{\rg{Storage requirements under a queuing theoretic framework for BS-AC-RLNC, are beyond the scope of this paper; see~\cite{shrader2007queueing,domanovitz2022information,enenche2023network} for RLNC under queuing theory. This remains an interesting avenue for future work.}}.
{\color{black}{While the maximum sliding window is limited by $w \geq w_t^{\max}-w_t^{\min}$, we define its current effective size, as $\overline{w} = w_t^{\max}-w_t^{\min}$}.} \rg{Usually is assumed that $w$ is selected to be equal to or larger than $\rm{RTT}$ to obtain the required throughput-delay tradeoff \cite{swrlnc,dias2023sliding,cohen2022broadcast}.}\off{\ale{Updated. Please check.}} Any linearly independent combination adds a Degree Of Freedom (DoF) to the linear system formed by the coded packets. Using a sufficiently large field, these combinations are linearly independent with high probability \cite{ho2006random,geil2008field}. A Gaussian elimination is used for decoding at the destination.\footnote{\rg{For random linear (network) coding over a sufficiently large finite field $\mathbb{F}_q$ of size $q$, the receiver can decode a generation of $w$ raw data packets with high probability by performing Gaussian elimination on the linear system formed by any set of $w$ coded packets~\cite{ho2006random}.}} 
\off{\ale{I think the following paragraph needs to move to Subsection~\ref{subsec:Preliminaries}. This is part of the solution. Not the problem}{\color{blue}{Thank you! Done}}}
\rg{Each coded packet carries a coefficient vector\footnote{\rg{The coding coefficients (or their seed) can be incorporated into the TCP/IP packet headers with
moderate overhead in practice~\cite{cohen2020bringing}.}} along with a linear combination, as define in \eqref{eqn:dof}, of \(\overline{w}\) raw data packets, which introduce additional overhead. 
In our framework, this overhead is assumed to incur polynomial time complexity\footnote{\rg{This assumption is aligned with our modeling framework, where the number of raw data packets participating in the RLNC process is bounded by the parameter \(w\), which allows for polynomial time decoding complexity. This is in contrast to solutions suggested in the literature, where large blocks may be required to achieve a desired reliability performance. Practically \(w\) is selected to be small~\cite[Chapter~4]{medard2025network}; otherwise, the quantitative impact of finite packet-length overhead needs to be considered. We leave this extension for future work.}} \cite{cohen2020adaptive,cohen2020adaptiveMH, patterson2014and,katti2008xors} and is therefore not included in the performance metrics defined in Subsection~\ref{subsec:Problem}.}

\rg{Upon receiving an ACK from node $n+1$ for coded packet $c_k^n$, node $n$ infers whether node $n+1$ has accumulated a sufficient number of coded packets such that successful decoding of the raw data packet $p_{w_t^{\min}}, p_{w_t^{\min}+1}, \dots$ has been possible.
When this holds, node $n$ advances the lower window bound $w_t^{\min}$ to the smallest index $j$ such that the raw data packet $p_j$ has not yet been decodable at node $n+1$. Node $n$ then evicts from its buffer all packets that do not involve raw data packet $p_j$ or any subsequent raw data packet.

Upon receiving a coded packet $c_{k+1}^{n-1}$ from node $n-1$, node $n$ inspects its payload to identify newly observed raw data packets, possibly more than one. If $m$ new raw data packets are observed, the node $n$ stores $c_{k+1}^{n-1}$ in its buffer. If $c_{k+1}^{n-1}$ contains no new raw data packets, node $n$ checks whether the stored packets suffice for decoding; if so, the packet is discarded, otherwise it is stored in the buffer. In all cases, an ACK is sent to node $n-1$.  

The re-encoding at intermediate nodes combines a subset of buffered packets whose raw data packets lie within the current efficient window $[w_t^{\min}, w_t^{\max}]$, as $c_t^n = \alpha c_l^{n-1} + \beta c_m^{n-1} + \dots$, where $\alpha, \beta \in \mathbb{F}_q$ are random coefficients and $l, m < t$ denote earlier time slots.}

\rg{We consider each coded packet as a DoF, which may refer to either a \emph{new} or the \emph{same} DoF; the following definitions distinguish between these two cases.
\begin{definition}\label{def:new}
A coded packet is referred to as a \emph{new} DoF if it is a transmission that contains new raw data packets. When a new DoF is transmitted, $w_{t}^{\max}$ is updated according to the new raw data packets included in the new DoF. 
\end{definition}
\begin{definition}\label{def:same}
A coded packet is referred to as the \emph{same} DoF if it is a retransmission of the same raw data packets (i.e., raw data packets that were already transmitted in a coded packet) with a newly generated random linear combination.
\end{definition}}

The source uses different coefficients each time~\cite{geil2008field,chen2021exact,chen2020decoding} and transmits either a new DoF or re-transmits the same DoF. The re-transmissions, also referred to as Forward Error Correction (FEC) transmissions, compensate for potential erasures by providing redundant information.
To guide these FEC transmissions, the sender estimates channel rate $r$ and erasure rate $\epsilon$ using feedback up to time $t$, with $\hat{r}$ and $\hat{\epsilon}$ calculated by \eqref{eqn:eps_hat} for $M=t-{\rm RTT}$. These estimates inform the following FEC mechanisms:

\begin{enumerate}[wide, labelwidth=0.3cm, labelindent=1pt, label=\textcolor{blue}{M\arabic*}, itemsep=1em]

\vspace{-0.05cm}

\item \label{itm:a-FEC}
\emph{\underline{A-Priori FEC}:} 
Similar to \cite{michel2022flec,cohen2022broadcast}, to compensate for expected erasures without feedback information, the sender transmits $\lceil \hat{\epsilon} \cdot c_{t}^{\text{new}} \rfloor$ FEC packets every \(\rm RTT\) time slots, where $c_{t}^{\text{new}}$ denotes the number of new DoFs (belonging to the current transmission window) transmitted in the last \( \rm RTT\) period.

\vspace{-0.25cm}

\item \label{itm:p-FEC}
\emph{\underline{Posterior FEC}:} 
At each time $t$, the sender compares the ratio between missing DoFs and received DoFs (the DoF rate). When the channel rate falls below this DoF rate, the destination lacks sufficient DoFs to decode the received coded packets, and a re-transmission is suggested.
\rg{We calculate the DoF rate following \cite{9834410}, considering two transmission regions:
The "known area" of acknowledged packets sent more than one $\rm RTT$ ago, and the "unknown area" of unacknowledged packets sent within the last $\rm RTT$ period.
We estimate missing DoFs by combining the confirmed missing DoFs from the known area, denoted by $md_t^{\text{nack}}$, with the estimated missing DoFs in the unknown area, given by $\hat{\epsilon} \cdot c_t^{\text{new}}$. Similarly, the number of received DoFs is estimated by combining the acknowledged transmissions in the known area, $ad_t^{\text{ack}}$, with the estimated number of successfully received retransmissions, $\hat{r} \cdot c_t^{\text{same}}$, yielding}
\begin{equation}
\label{eqn:delta_new}
\textstyle    \Delta_t \triangleq 
    \frac{md_t^{\text{nack}}+\hat{\epsilon} \cdot c_{t}^{\text{new}}} { ad_t^{\text{ack}}+\hat{r} \cdot c_{t}^{\text{same}}} - 1.
\end{equation}
\rg{This formulation provides a practical interpretation of the DoF rate as an estimate of the effective decoding progress under delayed feedback \cite{cohen2020adaptiveMH,9834410}.} A re-transmission is triggered when $\Delta_t - th > 0$, where $th$ is a tunable threshold parameter. \rg{In this case, the node re-transmits \emph{same} DoF (See Definition~\ref{def:same}), i.e., a newly generated random linear combination of the same raw data packets.}

\vspace{-0.25cm}
\item \label{itm:EOW}
\emph{\underline{End-Of-Window (EOW)}:} 
Limits the delay using a maximum sliding window of size $w$. When $w_t^{\max}-w_t^{\min}\geq w$, the sender repeats the same DoF until decoding is acknowledged.
\rg{By construction, the sliding window mechanism together with the acknowledgment process enforces in-order decoding and delivery, so that raw data packet $p_i$ is always delivered before or simultaneously with raw data packet $p_{i+1}$.}
\end{enumerate}
\vspace{-0.08cm}
\smallskip
\subsubsection{Multipath-Multi-hop (MP-MH) AC-RLNC}
\label{subsec:MPMH}
In \cite{cohen2020adaptiveMH}, a solution addressing the throughput-delay tradeoff in MP-MH networks is presented, without considering efficiency in terms of channel usage. For a single-path-multi-hop scenario, the algorithm implements the AC-RLNC scheme at the source, using the global $\rm RTT$ for destination acknowledgments to manage the DoF window and schedule FEC transmissions. Intermediate nodes employ a re-encoding approach, where all incoming packets are mixed without additional window management or FEC scheduling. 
The algorithm achieves over $90\%$ of channel capacity in the non-asymptotic regime while maintaining strict delay constraints required for URLLC.
We will use this approach as a baseline comparison.

\begin{figure}
    \vspace{-0.2cm}
    \centering
    \includegraphics[width=0.5\linewidth]{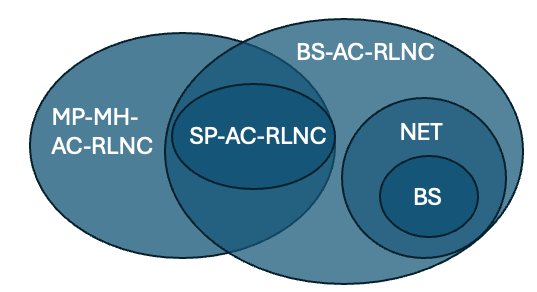}
    \caption{\small \rg{BS-AC-RLNC, derived from SP-AC-RLNC~\cite{swrlnc}, integrates the lightweight sliding-window module NET with the BS component at intermediate nodes. It addresses multi-hop networks, whereas MP-MH-AC-RLNC~\cite{cohen2020adaptive} additionally supports multipath networks. When NET is unavailable, BS-AC-RLNC can rely on traditional communication methods, making it more broadly deployable \cite{cohen2021bringing}.}}
    \vspace{-0.6cm}
    \label{fig:hierarchy}
\end{figure}
\vspace{-0.1cm}
\section{Blank Space AC-RLNC}
\label{sec:Algorithm}
\vspace{-0.08cm}
In this section, we detail our proposed \acf{bs-acrlnc} protocol for multi-hop networks. \rg{A detailed description of the operations performed by each node is provided by an example in Appendix~\ref{appendix:example}.} The solution leverages the network's physical limitations, considering the estimated global and forward-channels bottleneck information, as defined in \eqref{eqn: BN_n}, to improve the efficiency (i.e., reduce channel usage) while maintaining competitive performance in the rate-delay trade-off.
\ac{bs-acrlnc} is designed based on the AC-RLNC protocol (see Section~\ref{subsec:AC} \rg{and Fig.~\ref{fig:hierarchy}}). In particular, \rg{it incorporates} a new light-computational sliding window module called \acf{net-acrlnc} that is applied at intermediate nodes\footnote{\label{comm:NoCode}
In case \ac{net-acrlnc} can't be applied at some nodes\off{ in the proposed \ac{bs-acrlnc} protocol (or at the source with encoding)}, those nodes can deploy traditional communication methods to transmit and forward information in the multi-hop network. This provides flexibility in deployment.}. The \ac{net-acrlnc} consists three key features:
1) Each node can re-encode and send \rg{new or same} DoFs over the forward channels. Based on per-packet feedback, the node schedules re-transmissions using the three FEC mechanisms~\ref{itm:a-FEC}-\ref{itm:EOW}.
2) To manage the sliding window, nodes only track the DoFs received by the next node, instead of fully decoding\footnote{In the case of single-source multicast, which is relevant to many emerging applications, e.g., \cite{ yanmaz2018drone,kumar2012overview,radi2012multipath}, each node can decode the data transmitted.}. This approach, termed \textit{semi-decoding}, eliminates the decoding computational overhead.\label{term:semi_dec}
3) Within the \ac{net-acrlnc} module, we introduce two mechanisms to pause transmissions:
3.1) The first, called \textit{\acf{bsp}}, identifies optional idle periods that emerge from re-transmissions generated at subsequent nodes. These re-transmissions create intervals where sending data would result in unnecessary buffering. This behavior is described in detail in Fig.~\ref{fig:bl_sp}.  Thus, considering the capacity limit of the remaining path,\off{(see \eqref{eqn: BN_n})} nodes can safely suspend transmissions, creating 'blank spaces' in the transmission timelines. 
3.2) The second stems from \ac{bs-acrlnc} independent nodes operation. When neither FEC transmissions are required (\ref{itm:a-FEC}-\ref{itm:EOW}) nor new data is available in the buffer,\footnote{
New data is with respect to its own sliding window, regardless of whether it arrived as FEC or new from the previous node. \rg{For more details see Subsection~\ref{subsec:Preliminaries}.} } \rg{and the next node has acknowledged a sufficient number of DoF to ensure successful decoding of all raw data packets, including the most recent raw data packet referenced by \(w_t^{\max}\)}, nodes can safely pause transmission - a mechanism we call \textit{No-New No-FEC}.
The full NET operation given in Alg~\ref{alg:full_alg} is detailed next, with a representative example provided in Appendix~\ref{appendix:example}.
\begin{figure}
    \vspace{-0.2cm}
    \centering
    \resizebox!{4.3cm}{\begin{tikzpicture}
\foreach \x [count=\i, evaluate=\i as \label using int(\i-1)] in {0,2,4,6} {
   \node[draw, circle, fill=blue!20, minimum size=1cm] (c\i) at (\x,0) {\label};
}
\foreach \i in {1,2,3} {
   \pgfmathtruncatemacro{\next}{\i + 1}
   \draw[-{Stealth[length=3mm]}] (c\i) -- (c\next);
}
\draw ($(c1)+(0,-0.7)$) -- node[right, midway] {new(0)} ++(0,-0.7);
\draw [brown] ($(c1)+(0,-1.4)$) -- node[right, midway, brown] {FEC(0)} ++(0,-0.7);
\draw[purple!60, dash pattern=on 4pt off 2pt] ($(c1)+(0,-2.1)$) -- node[right, midway, purple!60] {BS(1)} ++(0,-0.5);
\draw[blue, dash pattern=on 4pt off 2pt] ($(c1)+(0,-2.6)$) -- node[right, midway, blue] {BS(2)} ++(0,-0.7);
\draw[green!50!black] ($(c1)+(0,-3.2)$) -- node[right, midway, green!50!black] {new(0)} ++(0,-0.7);

\draw ($(c2)+(0,-1.2)$) -- node[right, midway] {new(0)} ++(0,-0.7);


\draw[purple!60] ($(c2)+(0,-1.9)$) -- node[right, midway, purple!60] {FEC(1)} ++(0,-0.5);
\draw[blue, dash pattern=on 4pt off 2pt] ($(c2)+(0,-2.4)$) -- node[right, midway, blue] {BS(2)} ++(0,-0.7);
\draw [brown] ($(c2)+(0,-3.1)$) -- node[right, midway, brown] {FEC(0)} ++(0,-0.7);

\draw[green!50!black] ($(c2)+(0,-3.8)$) -- node[right, midway, green!50!black] {new(0)} ++(0,-0.7);
\draw ($(c3)+(0,-1.7)$) -- node[right, midway] {new(0)} ++(0,-0.7);
\draw[blue] ($(c3)+(0,-2.4)$) -- node[right, midway, blue] {FEC(2)} ++(0,-0.7);
\draw[purple!60] ($(c3)+(0,-3.1)$) -- node[right, midway, purple!60] {FEC(1)} ++(0,-0.5);
\draw[brown]($(c3)+(0,-3.6)$) -- node[right, midway, brown] {FEC(0)} ++(0,-0.7);
\draw[green!50!black] ($(c3)+(0,-4.3)$) -- node[right, midway, green!50!black] {new(0)} ++(0,-0.7);
\end{tikzpicture}} 
    \caption{\small
    The diagram illustrates how FEC periods at intermediate nodes create blank spaces at preceding nodes.
    Vertical lines represent each node transmission timeline, with packets labeled by type and generator node
    (The labels indicate the information added to the output DoF at each node).
    The colors indicate period lengths corresponding to each channel erasure rate. Dashed lines mark blank-space regions where packet transmissions may be ineffective.
    For example, FEC(0) represents a FEC period generated at node $0$ with length $ \lceil \hat{\epsilon}_0 \cdot {\rm RTT}_0 \rfloor$ and sent to node 1. Then, node $1$ first generates its own FEC period, FEC(1), before processing any DoF from FEC(0). This creates the time window BS(1) in node 0's timeline, where any transmitted packets will not be immediately processed.
    Similarly, BS(2) propagates from node $2$'s FEC period. Since node $2$ cannot process packets during this time, node $1$ suspends transmission, which in turn allows node $0$ to pause as well. 
    \vspace{-0.1cm}
    }
    \label{fig:bl_sp}
    \vspace{-0.6cm}
\end{figure}

\noindent {\bf Network AC-RLNC (NET):}
At time slot $t$, node $n$ may receive feedback (line~\ref{line:inputs}) and 1) update its erasure rate estimation, 2) identify the forward bottleneck (line~\ref{line:estimation}), and 3) eliminate semi-decoded packets from the buffer (line~\ref{line:wmin}). The node then determines a transmission according to one of three states:
1) The a-priori FEC period (line~\ref{line:apriori}) is activated every ${\rm RTT}_n$ time slots (within the node's initial delay) and schedules FEC transmissions according to~\ref{itm:a-FEC}.
2) Once all a-priori FEC packets are sent, the blank-space period (BSP) is activated (lines~\ref{line:fec_check}-\ref{line:blank_end}). Its full operation is described in the next paragraph.
3) When any other period is inactive, the node operates in a No-New No-FEC mode. It evaluates FEC necessity by the end-of-window and posterior criteria (line~\ref{line:posterior}). If FEC is unnecessary and a new packet exists in the buffer, it transmits a new DoF; otherwise, it pauses transmission (line~\ref{line:empty_tran}).

%

\setlength{\textfloatsep}{1pt}
\begin{algorithm}
\caption{Network AC-RLNC (NET)}
\label{alg:full_alg}
\begin{footnotesize}
\begin{algorithmic}[1]
    
    \STATE \textbf{Optional Inputs:} 
    From Previous Node: Input packet $c^{n-1}_{t-{\rm RTT}_{n-1}}$ (or $p_t$ for the first encoding node$^{\ref{comm:NoCode}}$, i.e., a source who encodes using \eqref{eqn:dof}); 
    From Next Node: Aggregated feedback $AF^n_t = \{F^{i}_{t'} |{t' \leq t - \sum_{k=i}^{N-1}{\rm RTT}_{k}}, \space i \in \{n+1, \ldots, N-1\}\}$. \label{line:inputs}
    
    \STATE \textbf{Optional Outputs:} To Previous Node: Output aggregated feedback $\{F^n_t,AF^n_t\}$; 
    \smallskip
    To Next Node: Output packet $c^{n}_t$. \label{line:outputs}
    
    \STATE \textbf{Estimation Update:} According to the feedback, update $\{\hat{\epsilon}_i \}_{n\leq i \leq N-2}$ using \eqref{eqn:eps_hat}, and the ${BN_n}$ \eqref{eqn: BN_n}. \label{line:estimation}
    
    \STATE \textbf{Eliminate Packets:} Using the feedback, assess and eliminate "semi-decoded" (\ref{term:semi_dec}) packets from the buffer. Update $w_{min}$ to the oldest packet.
    \label{line:wmin}

    \colorbox{gray!10}{\parbox{\dimexpr\linewidth-2\fboxsep\relax}{%
    \STATE \textbf{A-Priori FEC period -~\ref{itm:a-FEC}.} \label{line:apriori}
    }}
    
    \smallskip
    \colorbox{blue!10}{\parbox{\dimexpr\linewidth-2\fboxsep\relax}{%
        \textbf{BLANK SPACE:} \label{line:blank_start}
        \IF{A-Priori FEC period is complete} \label{line:fec_check}
            \STATE Determine BS Basic Duration \ref{itm:BS-Period}:
            \smallskip
            \\$BS \gets \alpha \sum_{i=n+1}^{BN_n}{ {\rm RTT}_i \cdot\hat{\epsilon}_i}$ \label{line:bs_num}
            \smallskip
            \IF{$BS > 0$}
                \IF{$\Delta_t^{BS} > 1-\hat{\epsilon}_{BN_n}$ (BS Criterion \ref{itm:BS-Criterion})} \label{line:bs_crit}
                    \STATE Terminate BS period \label{line:bs_termin}
                
                \ELSE
                    \STATE Do not transmit \label{line:bs_empty}
                    \STATE $BS \gets BS-1$
                    \RETURN
                \ENDIF
            \ENDIF
        \ENDIF
    }} \label{line:blank_end}
    
    \smallskip
    \colorbox{gray!10}{\parbox{\dimexpr\linewidth-2\fboxsep\relax}{%
    \textbf{No-New No-FEC:}
    
    \IF{ $w_{max}-w_{min}>w$ (EOW -~\ref{itm:EOW}) or $\Delta_t \geq 0$ (Posterior FEC -~\ref{itm:p-FEC}) \label{line:eow}} \label{line:posterior}
        \STATE Send a re-transmission
    \ELSIF{New packet available in Buffer} \label{line:buffer_check}
        \STATE Increment $w_{\text{max}}$ by the new raw data packets; Send new DoF \label{line:send_new}
    \ELSE
        \STATE Do not transmit \label{line:empty_tran}
    \ENDIF }}
    \RETURN \label{line:final_return}
\end{algorithmic}
\end{footnotesize}
\end{algorithm}

\vspace{0.3cm}
\noindent {\bf BS Period (BSP):}
During this period, the node suspends transmissions. Once initiated, the period's duration is set as described in~\ref{itm:BS-Period} (line~\ref{line:bs_num}). Each time slot, the node evaluates if early termination is needed by the \ac{bsp}-criterion detailed in~\ref{itm:BS-Criterion} (lines~\ref{line:bs_crit}-\ref{line:bs_termin}).

\vspace{-0.05cm}

\begin{enumerate}[wide, labelwidth=0.3cm, labelindent=1pt, label=\textcolor{blue}{M-BS-\arabic*}, itemsep=1em]

\item \label{itm:BS-Period}
\textit{\underline{\ac{bsp} Duration}:} %
The \ac{bsp} duration is determined based on the a priori FEC periods of subsequent nodes in the network and considering the following:
1) While nodes transmit FEC relative to their newly transmitted DoFs ($c_{t}^{\text{new}}$,~\ref{itm:a-FEC}), this information isn't directly available through feedback links. However, estimation of the subsequent links' erasure rates is available by backward feedback aggregation $AF^n_t$ (lines~\ref{line:inputs}-\ref{line:outputs}).
That is, the destination node sends back an ACK/NACK message to node N-1, which in turn sends back two things: its own feedback message, along with the feedback message it received from the destination. Then node N-1 sends both messages to node N-2, which adds its own feedback and passes all three messages back. This process continues backwards through the chain, where each node sends back its own feedback message $F^n_t$ and the collection of messages it received from the next node, denoted by $AF^n_t$. Specifically, $AF^n_t$ contains acknowledgments $F_{t'}^i$ from nodes $n+1$ to $N-1$ at times $t'$ not exceeding $t - \sum_{k=i}^{N-1}{\rm RTT}_{k}$. The erasure rate of each preceding channel $\hat{\epsilon}_n$ is then computed as the mean of all its corresponding acknowledgments in $AF^n_t$ by \eqref{eqn:eps_hat}.
%
2) The forward-channels bottleneck link, which requires the highest FEC transmission, constrains the local theoretical capacity and contributes the most \ac{bsp} slots. Any packet at this link should be forwarded immediately to maintain high transmission rates, regardless of subsequent channel conditions.
Therefore, the \ac{bsp} duration considers two factors: the basic window of new DoFs for each $i$th hop - ${\rm RTT}_i$, and the \rg{actual} estimated erasure rates \rg{which track the links' conditions by the feedback acknowledgment (see Eq.~\eqref{eqn:eps_hat})} up to the forward bottleneck, resulting, for node $n$ at time $t$ as
\begin{equation}
 \textstyle   BS_n(t) = \begin{cases}
\textstyle         \alpha \sum_{i=n+1}^{BN_n} {\rm RTT}_i \cdot\hat{\epsilon}_i & \hspace{-0.2cm}\text{if } n<BN_n, \text{ using Eq.\hspace{-0.1cm}~\eqref{eqn: BN_n}} \\
\textstyle        0 & \hspace{-0.2cm}\text{if } n=BN_n
    \end{cases}
    \label{eqn:BS_num}
\end{equation}
where $\alpha$ is a tunable parameter used to relax the evaluation error. \rg{In particular, if $\hat{\epsilon}_i$ is underestimated, selecting $\alpha>1$ provides compensation for this bias. Conversely, if $\hat{\epsilon}_i$ is overestimated, choosing $\alpha<1$ mitigates this effect.}
Note that if channel $e_n$ is the forward bottleneck itself, we set $BS_n(t)=0$, eliminating any transmission suspension.

\vspace{-0.2cm}
\item \label{itm:BS-Criterion}
\textit{\underline{BS Termination Criterion}:}
Transmission suspension can be considered a packet erasure, allowing us to analyze this through the DoF rate framework given in \eqref{eqn:delta_new}. We want to identify the critical point where further suspension would degrade the data rate performance.
To model this, we set $md_{t}^{\text{nack}}=1$, representing guaranteed packet erasure, with the remaining \ac{bsp} window available for re-transmissions compensation. Since only FEC are considered, $c_{t}^{\text{new}}=0$, and with uncertain delivery, $ad_{t}^{\text{ack}}=0$. 
Finally, $c_{t}^{\text{same}}=BS_n(t)$, representing the number of re-transmissions with unknown delivery status.
While a node's transmission rate is constrained by its forward channels' bottleneck, channels with similar erasure rates can have a significant impact as well. To quantify this, we consider both the erasure rate differential between the current channel and its bottleneck, and their distance in hops, denoted by $h = BN_n - n$.
\rg{This sets the \ac{bsp}-DoF rate by
\begin{equation} \label{eqn:delta_bs}
\textstyle    \Delta_t^{BS} = \left[ \left( 1-\hat{\epsilon}_n \right) BS(t)+\kappa\cdot h \cdot  \ln(\hat{\epsilon}_{BN_n}-\hat{\epsilon}_n) \right] ^{-1},
\end{equation}
where $\hat{\epsilon}_{BN_n}\geq \hat{\epsilon}_n$ and $\kappa >0$ is a tunable weight\footnote{\rg{The logarithmic correction in \eqref{eqn:delta_bs} captures the nonlinear sensitivity to the erasure-rate disparity from the bottleneck. As $\hat{\epsilon}_{BN_n}-\hat{\epsilon}_n \to 0$, it diverges to $0^-$, thereby preventing BSP termination; its scaling is absorbed into $\kappa$.}}.} 
To ensure suspending transmission doesn't reduce the rate below the capacity, we compare the \ac{bsp} DoF rate to the forward bottleneck rate. 
Thus, the \ac{bsp} is terminated when,
\begin{equation}
\textstyle    \Delta_t^{BS} > 1-\hat{\epsilon}_{BN_n}.
    \label{BS_crit}
\end{equation}
This criterion identifies the critical point where the remaining transmission slots, $BS_n(t)$, are the slots needed to compensate for the "missing" DoF. Any further pause would likely result in a rate reduction.\off{Since $\Delta_t^{BS}$ is defined as the inverse of the effective DoF compensation cost, larger values correspond to lower compensation burden. Hence, suspension is allowed only while the effective DoF rate remains below the forward bottleneck rate.} \rg{Note that this condition also acts as a safeguard against overly large choices of $\alpha$ in Eq.~\eqref{eqn:BS_num}.}
\end{enumerate}

\section{Analytical Results}\label{section:analyticalResults}
The algorithm is based on the principle that each successive pair of nodes, $n$ and $n+1$, is responsible for transmitting all relevant information so that node $n+1$ can fully reconstruct the raw data relayed through node $n$. This design impacts system performance metrics, such as delay, goodput, and throughput.  In this section, we present the theoretical lower and upper bounds for these metrics. 
\rg{For delay, we derive the mean bound in addition to the worst-case upper bound and the best-case lower bound, as it provides a tighter and more representative characterization of system performance on average.}
This section is divided into subsections, each focused on different aspects of the analysis. \rg{A comprehensive sensitivity analysis of $w$ and $\alpha$ with respect to delay, throughput, and normalized goodput bounds is provided in Subsection~\ref{appendix:sensitivityAnalysis}.} The notation used throughout this section is summarized in Table~\ref{fig:table_analytical_results} \ifpagelimit in the supplementary material of this work\else\fi. The analytical results are evaluated and illustrated using the example presented in Section~\ref{section:evaluation}. \rg{For further clarification, the toy example in Appendix~\ref{appendix:example} illustrates the analysis presented in this section.}

\subsection{Delay Lower Bound}\label{subsection:DelayLowerBound}

This subsection introduces a lower bound on the end-to-end delay, followed by a tighter bound on the algorithm implementation.

\begin{theorem}\label{theorem:delay}
The \rg{expected} end-to-end delay $D_{\min}$ is lower bounded by
\vspace{-0.2cm}
\begin{align}
    D_{\min} \geq \sum_{i=0}^{N-2}\left(\frac{{\rm RTT}_i}{2}+ \frac{1}{1-\epsilon_i}\right).
    \label{LowerBound}
\end{align}
\end{theorem}
\begin{proof}
\rg{We note that all the delays are measured in time slots.} The minimum achievable delay is determined by two principal components, the propagation delay and the packet erasure rate across each link. \rg{The term ${1}/{1-\epsilon_i}$ is the expected number of transmission slots required for one successful packet delivery over link $i$ under i.i.d. erasures, while ${\rm RTT}_i/{2}$ accounts for one-way propagation.} The impact of the erasures is mitigated through the use of a-priori FEC, as described in~(\ref{itm:a-FEC}). Specifically, the effective transmission rate becomes $1 - \epsilon_i$ with a corresponding a-priori FEC of $\epsilon_i$.
\end{proof}

According to the algorithm implementation, any node $n$ initially transmits the input packet $c_t$, followed by its own FEC packet. It then suspends its transmission during the BSP. Subsequently, node $n$ forwards the FEC packets received from node $n-1$. In particular, the data received by node $n$ may not, in itself, be sufficient to reconstruct the original input, since the node begins by transmitting its received data and its own FEC before utilizing the incoming FEC to complete the reconstruction of its own input. In the following, a tighter analytical bound is provided to enhance the precision of the performance estimate. We start by establishing a supporting lemma, which serves as the basis for deriving the lower bound introduced by the algorithm.

\begin{lemma}\label{lemma:delta}
As $\hat{\epsilon}_n \to \hat{\epsilon}_{BN_n}$, it follows that $\Delta_t^{BS} \to 0^\rg{-}$, and in particular, \ifpagelimit $ \Delta_t^{BS} - 1 + \hat{\epsilon}_{BN_n} < 0, $ \fi
\rg{where $0^{-}$ denotes the left-sided limit approaching zero.}
\end{lemma}
\begin{proof}
\rg{As $\hat{\epsilon}_n \to \hat{\epsilon}_{BN_n}$, we have \ifpagelimit
$ h \ln(\hat{\epsilon}_{BN_n} - \hat{\epsilon}_n) \to -\infty. $ \fi
Since $(1 - \hat{\epsilon}_n) BS(t)$ remains bounded, for any fixed $\kappa$ it follows that \ifpagelimit $(1 - \hat{\epsilon}_n) BS(t) + \kappa \cdot h \ln(\hat{\epsilon}_{BN_n} - \hat{\epsilon}_n) \to -\infty, $ \fi
and therefore, \ifpagelimit
$ \left[(1 - \hat{\epsilon}_n) BS(t) +  \kappa \cdot h \ln(\hat{\epsilon}_{BN_n} - \hat{\epsilon}_n)\right]^{-1} \to 0^\rg{-}. $ \fi
This establishes the claim.}
\end{proof}
The expression in Lemma~\ref{lemma:delta} is written in terms of the estimated erasure rate to match the notation in~\eqref{eqn:delta_bs}, while the bound is stated in terms of the actual erasure probability, i.e., $\epsilon_n$ and $\epsilon_{BN_n} $.

\begin{corollary}\label{theorem:minDelayAlgo}
    The \rg{expected} end-to-end delay $D_{min}$ introduced by the implemented algorithm is lower bounded by
\begin{align}
    D_{\min} \geq \sum_{i=0}^{N-2}\left(\frac{{\rm RTT}_i}{2}+ \frac{1}{1-\epsilon_i}+ \overline{BS}\right), \label{TightLowerBound}
\end{align}
where $\overline{BS}=\frac{1}{N-1}\sum_{n=0}^{N-2}BS_n(t)$.
\end{corollary}
\begin{proof}
The correctness of the first two terms follows directly from Theorem~\ref{theorem:delay}. For the third term, namely $\overline{BS}$, observe that during the BSP, two conditions must be satisfied: (a) $BS_n(t) > 0$ (see~\eqref{eqn:BS_num}), and (b) $\Delta_t^{BS} - 1 + \hat{\epsilon}_{BN_n} \leq 0$ (see~\eqref{eqn:delta_bs}).  

Assuming that $\hat{\epsilon}_n$ is sufficiently close to $\hat{\epsilon}_{BN_n}$, condition (b) is satisfied by Lemma~\ref{lemma:delta}. Consequently, the term $\overline{BS}$ provides an accurate approximation for the BSP.
\end{proof}

\rg{Relaxing the assumption that $\hat{\epsilon}_n$ is sufficiently close to $\hat{\epsilon}_{BN_n}$ implies that the BSP terminates earlier (see line~\ref{line:bs_termin} in Algorithm~\ref{alg:full_alg}). In this case, the bound derived in~\eqref{LowerBound} is applied.}

\ifdouble
\begin{figure}[t]
    \centering
    \begin{subfigure}[b]{0.49\linewidth}
        \centering
        \includegraphics[width=\linewidth]{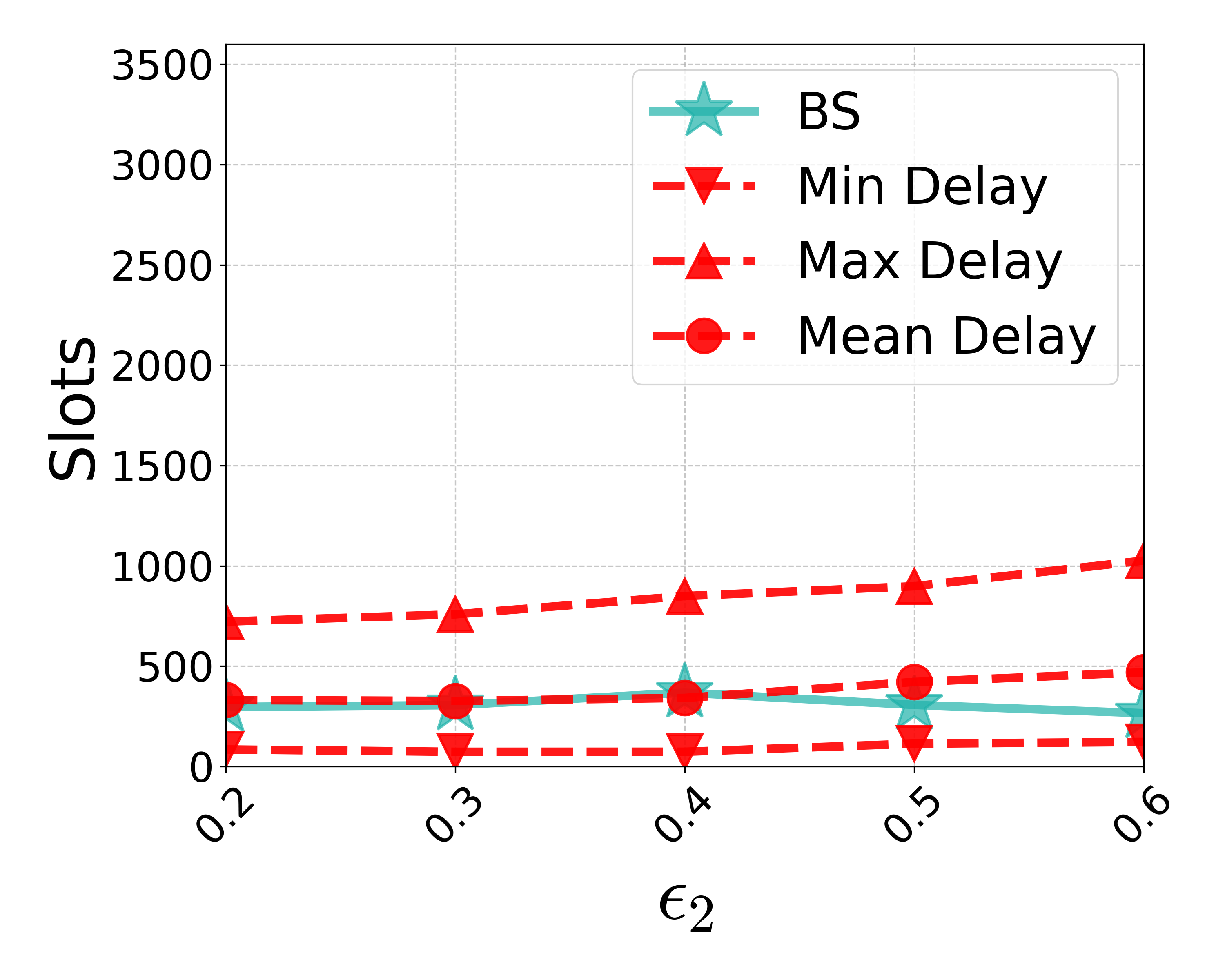}
        \caption{Mean Delay}
        \label{fig:meanDelayBound}
    \end{subfigure}
    \hfill
    \begin{subfigure}[b]{0.49\linewidth}
        \centering
        \includegraphics[width=\linewidth]{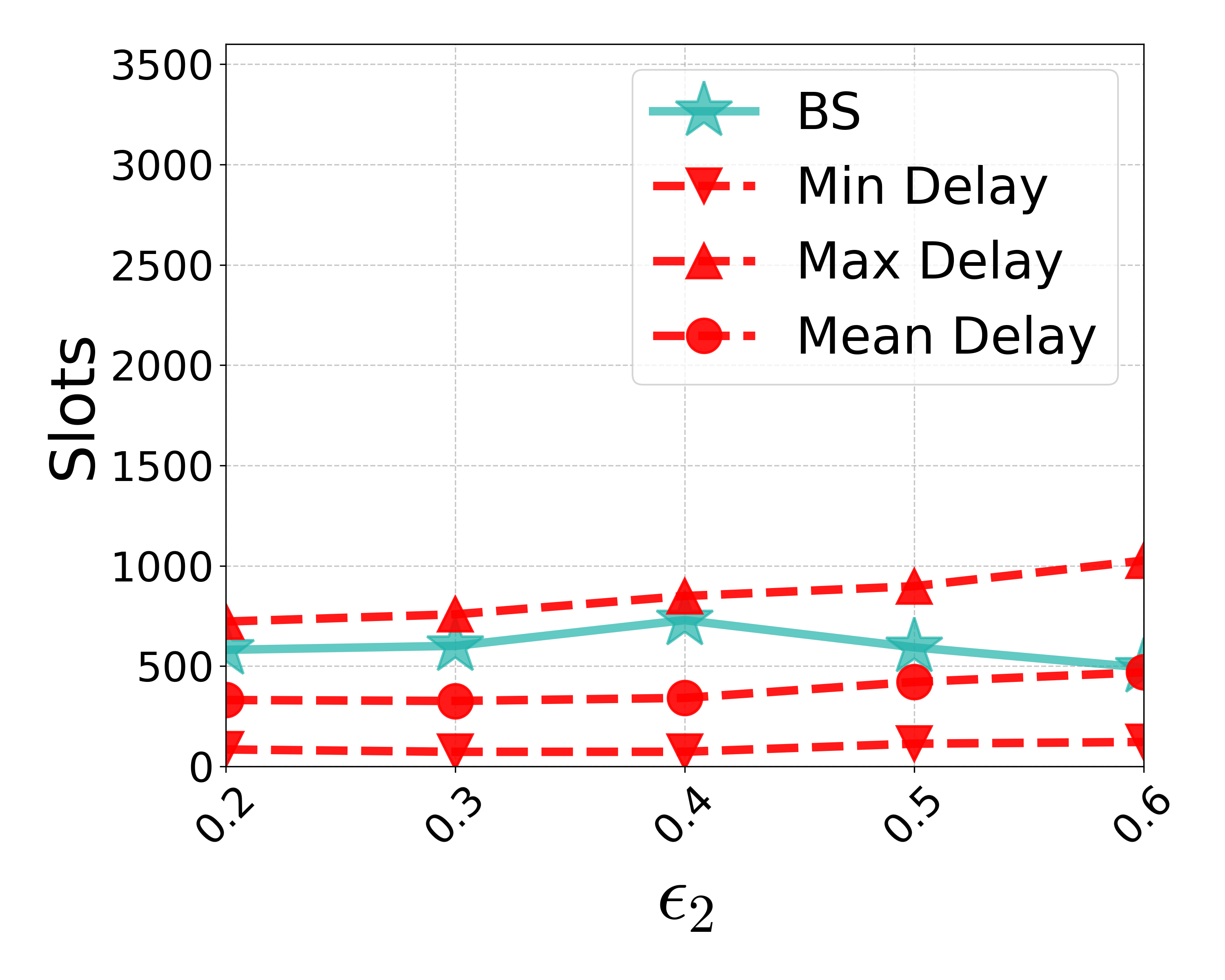}
        \caption{Max Delay}
        \label{fig:maxDelayBound}
    \end{subfigure}
    \hfill
    \vspace{-0.4cm}
    \caption{\small
    System performance of a 6-node network as a function of $\epsilon_2 \in [0.2, 0.6]$, with fixed erasure probabilities $\epsilon_0 = \epsilon_4 = 0.1$, $\epsilon_1 = 0.4$, and $\epsilon_3 = 0.3$. The left plot shows the mean delay, and the right plot shows the maximum delay, each depicted with solid lines. In both plots, the same set of analytical bounds, corresponding to the minimum, mean, and maximum delays, is shown with dashed lines for comparison.
    }
    \label{fig:three-subfigures}
\end{figure}
\fi

Figs.~\ref{fig:meanDelayBound} and~\ref{fig:maxDelayBound} show the lower bound of $D$ given in~\eqref{TightLowerBound} for the example discussed in Section~\ref{section:evaluation}. In accordance with Lemma~\ref{lemma:delta}, we neglect the condition $\Delta_t^{\text{BS}} - 1 + \hat{\epsilon}_{\text{BN}_n} > 0$ in the BSP analysis. The impact of this simplification is evident in the plots for $\epsilon_2 = 0.5$ and $\epsilon_2 = 0.6$, where the bounds increase while the algorithm-measured delay decreases. This behavior arises because, at these values, the rate disparity between the bottleneck node and the rest of the multi-hop path increases, causing the BSP to be halted earlier by the threshold $\Delta_t^{\text{BS}}$.

\subsection{Delay Mean Bound}

Here, we derive an upper bound on the expected delay.

\begin{theorem}\label{theorem:meanDelay}
The expected end-to-end delay~$D_{\text{mean}}$ satisfies
\begin{align*}
    D_{\mathrm{mean}} \leq\sum_{i=0}^{N-2}\left(\frac{1}{1-\epsilon_i}\left({\rm RTT}_i(1+\epsilon_i) + \overline{BS} +{\rm RTT}_i\right)\right).
\end{align*}
\end{theorem}
\begin{proof}
Consider a transmission from node~$n$ to node~$n{+}1$. The algorithm begins with an initial ${\rm RTT}_n$ transmission, followed by an a-priori FEC phase of duration ${\rm RTT}_n \cdot \epsilon_n$. The total propagation delay, including both forward transmission and feedback, is ${\rm RTT}_n$. If the link $e_n$ is not a BN link, a BSP is introduced. This cycle repeats until the first successful transmission. Due to the independent erasure process, the number of required attempts follows a geometric distribution. Hence, the expected number of attempts is ${1}/{(1 - \epsilon_n)}$.
\end{proof}
\vspace{-0.08cm}
Figs.~\ref{fig:meanDelayBound} and \ref{fig:maxDelayBound} depict the upper bound of the expected delay $D_{\text{mean}}$ for the example presented in Section~\ref{section:evaluation}.
\vspace{-0.08cm}
\subsection{Delay Upper Bound}

As described in \ref{itm:EOW}, the BS algorithm utilizes a maximum sliding window of size~$w$. The source begins by transmitting a single DoF, and in each subsequent time slot, it adds one new DoF based on the feedback received. A key advantage of this mechanism is that the receiver can begin decoding as soon as the first DoF arrives, avoiding the delay associated with waiting for all $w$ packets. Once the sliding window reaches its full size~$w$, the source continues transmitting $w$ DoF packets per time slot until the receiver acknowledges successful decoding. In the worst-case scenario, where erasures affect early transmissions, the receiver may need to wait until it has successfully received all $w$ DoFs before decoding the first message.

Accordingly, in the next theorem, we derive an upper bound assuming that all $w$ DoFs must be received before decoding can occur. To establish this result, we first present two supporting lemmas, which we then use to formally derive the bound.
\begin{lemma}\label{lemma:endToEndBSP}
    The end-to-end BSP delay is upper-bounded as follows
    \vspace{-0.4cm}
    \begin{align*}
        \alpha\sum_{i=0}^{N-2} (i+1) \cdot {\rm RTT}_i \cdot \hat{\epsilon}_i
    \end{align*}
\end{lemma}
\begin{proof}
For a given node~$n$, the BSP is upper bounded by
$$
\alpha\sum_{i = n+1}^{BN_n} {\rm RTT}_i \cdot \hat{\epsilon}_i \leq \alpha\sum_{i = n+1}^{N-2} {\rm RTT}_i \cdot \hat{\epsilon}_i,
$$
Summing this bound over all nodes yields the total end-to-end BSP delay.
\end{proof}
\begin{lemma}\label{lemma:deltaT}
An upper bound on the retransmission threshold $\Delta_t$ under continuous transmission is given by
    \begin{align*}
        \Delta_t <  {(w \cdot \epsilon_i)}/{(1 - \epsilon_i)}. 
    \end{align*}
\end{lemma}
\begin{proof}
By definition \eqref{eqn:delta_new}, the retransmission threshold satisfies
$$
\Delta_t < \frac{md^{\text{nack}}_t + \hat{\epsilon} \cdot c^{\text{new}}_t}{ad^{\text{ack}}_t + \hat{r} \cdot c^{\text{same}}_t}.
$$
Assuming that the denominator is strictly greater than one,
it follows that
$$
\frac{md^{\text{nack}}_t + \hat{\epsilon} \cdot c^{\text{new}}_t}{ad^{\text{ack}}_t + \hat{r} \cdot c^{\text{same}}_t} < md^{\text{nack}}_t + \hat{\epsilon} \cdot c^{\text{new}}_t.
$$
The term $ md^{\text{nack}}_t + \hat{\epsilon} \cdot c^{\text{new}}_t $
represents the total erasure incurred during the transmission of $w$ messages. This total loss is approximated by
$$
md^{\text{nack}}_t + \hat{\epsilon} \cdot c^{\text{new}}_t \approx \epsilon_i \cdot \left(w \cdot \frac{1}{1 - \epsilon_i}\right),
$$
which accounts for the effective loss rate after incorporating the a-priori FEC used during transmission.
\end{proof}

\begin{theorem}\label{theorem:maxDelay}
    The \rg{expected} end-to-end delay~$D_{\max}$ is upper bounded as follows 
    \begin{align*}
        D_{\max} \leq \sum_{i=0}^{N-2}\left({\rm RTT}_i + \frac{w}{1-\epsilon_i} + \frac{w\cdot\epsilon_i}{1-\epsilon_i} + \frac{\alpha \cdot w \cdot (i+1) \cdot \epsilon_{i}}{1 - \epsilon_i}\right).
    \end{align*}
\end{theorem}
\begin{proof}
We analyze each term in the delay expression individually.
\ifpagelimit
The first term, ${\rm RTT}_i$, captures the round-trip propagation delay, including feedback. The second term accounts for the transmission of $w$ messages and incorporates the overhead introduced by the a-priori FEC. The third term represents the retransmission factor $\Delta_t$, which is upper bounded as established in Lemma~\ref{lemma:deltaT}. Finally, the fourth term quantifies the delay due to BSPs. The total BSP delay is upper-bounded by the duration of a single BSP multiplied by the number of such periods.

Since a BSP is triggered every ${\rm RTT}_i(1 + \epsilon_i)$ time slots, the number of BSPs can be approximated by dividing the total number of message transmissions (as captured by the second and third terms) by this interval. This yields:
$$
\frac{w(1 + \epsilon_i)}{{\rm RTT}_i(1 + \epsilon_i)(1 - \epsilon_i)} = \frac{w}{{\rm RTT}_i(1 - \epsilon_i)}.
$$
Multiplying this quantity by the BSP duration at node~$i$, given by Lemma~\ref{lemma:endToEndBSP} as $\alpha(i+1){\rm RTT}_i \cdot \epsilon_i$, yields the total BSP delay contribution:
$$
\frac{w}{{\rm RTT}_i(1 - \epsilon_i)} \cdot \alpha(i+1){\rm RTT}_i \cdot \epsilon_i = \frac{\alpha \cdot w(i+1)\epsilon_i}{1 - \epsilon_i}.
$$
\fi
This completes the proof.
\end{proof}
\vspace{-0.2cm}
The upper bounds associated with the example detailed in Section~\ref{section:evaluation} are shown in Figs.~\ref{fig:meanDelayBound} and~\ref{fig:maxDelayBound}.
\vspace{-0.2cm}
\subsection{Normalized Goodput Bounds}

In this subsection, we derive upper and lower bounds for the goodput, as defined in~\eqref{eqn:eta_n}. To do so, we first establish bounds on the transmission rate and the idle period, as the goodput is expressed in terms of the transmission activity relative to the complement of the idle duration. The analysis is carried out on a per-link basis, since only the ratio between the transmission time and the non-idle (i.e., active) period is required. We begin by deriving the transmission bounds, then proceed to the idle period bounds. These are subsequently used to formulate and prove the corresponding goodput bounds.

The achievable per-link transmission rate is defined as the ratio of successfully transmitted packets to the maximum number of time slots required for their transmission between a pair of adjacent nodes. A lower bound on this rate is derived below.

\begin{lemma}\label{lemma:transmitionLowerBound}
    The achievable per-link transmission rate $\tau_{\min}$ is lower bounded as follows,
\begin{align}\label{tauLowerBound}
    \tau_{\min} \geq \frac{\min\{{\rm RTT}_i\}}{\min \{{\rm RTT}_i\}\cdot3/2 + \alpha\sum_{i=0}^{N-2} {\rm RTT}_i\cdot\epsilon_i}.
\end{align}
\end{lemma}
\begin{proof}
Consider a node $n$ that transmits to its immediate neighbor, node $n+1$. The total transmission duration includes: 
\ifpagelimit
1) ${\rm RTT}_n$ time slots for data transmission, 2) ${\rm RTT}_n \cdot \epsilon_n$ time slots for a-priori FEC, 3) ${\rm RTT}_n / 2$ time slots for propagation delay, and 4) $\alpha \sum_{i = n+1}^{BN_n} {\rm RTT}_i \cdot \epsilon_i$ time slots for cumulative BSP where the term $\alpha \sum_{i = n+1}^{N-2} {\rm RTT}_i \cdot \epsilon_i$ upper bound it.
\fi    

Thus, the total time required at node $n$ is:
$$
{\rm RTT}_n (1 + \epsilon_n) + \frac{{\rm RTT}_n}{2} + \alpha \sum_{i = n+1}^{N-2} {\rm RTT}_i \cdot \epsilon_i.
$$

For $\alpha \geq 1$, the combined FEC and BSP cost satisfies:
$$
{\rm RTT}_n \cdot \epsilon_n + \alpha \sum_{i = n+1}^{N-2} {\rm RTT}_i \cdot \epsilon_i
\leq \alpha \sum_{i = 0}^{N-2} {\rm RTT}_i \cdot \epsilon_i.
$$

The achievable transmission rate at node $n$ is therefore lower bounded by
\vspace{-0.2cm}
\ifpagelimit
$$
{{\rm RTT}_n}/
     {\left({\rm RTT}_n \cdot \frac{3}{2} + \alpha \sum_{i = 0}^{N-2} {\rm RTT}_i \cdot \epsilon_i\right)}.
$$
\fi
Since ${\rm RTT}_n \geq 1$ by definition, the minimum rate is achieved by selecting the smallest $\rm RTT$, i.e., $\min\{ {\rm RTT}_i \}$. This concludes the proof.
\end{proof}

The upper bound of transmission characterizes by the ratio of successfully transmitted packets to the minimum number of time slots required for their transmission. The following result provides a formal upper bound on the achievable per-link transmission rate.

\begin{lemma}
    The achievable per-link transmission rate $\tau_{\max}$ is upper bounded as follows,
\ifpagelimit
\begin{align}\label{tauUpperBound}
    \tau_{\max} \leq {(1 - \min\{\epsilon_i\})}/{(1 + \min\{\epsilon_i\})}.
\end{align}    
 \fi
\end{lemma}
\begin{proof}
    The highest achievable transmission rate is determined by the link with the smallest erasure probability, that is, $1 - \min\{\epsilon_i\}$. Consequently, the minimum number of time slots required to transmit one unit of information and it a-priori FEC is $1 + \min\{\epsilon_i\}$.
\end{proof}

Having bounded the per-link transmission rate, we now derive the bounds on the idle phase, which complements the active transmission period and directly impacts the achievable goodput.

\begin{lemma}
The per-link idle phase $\theta_{\max}$ satisfies
\begin{align} \label{thetaUpperBound}
    \theta_{\max} \leq \frac{\alpha\sum_{i=0}^{N-2}{\rm RTT}_i \cdot \epsilon_i}{\min \{{\rm RTT}_i\}\cdot3/2+ \alpha\sum_{i=0}^{N-2}{\rm RTT}_i \cdot \epsilon_i}.
\end{align}
\end{lemma} 
\begin{proof}
The correctness of the denominator follows Lemma~\ref{lemma:transmitionLowerBound}. The numerator represents an upper bound on the per-link BSP, as established in the preceding discussion. 
\end{proof}
By definition, the lower bound of the idle phase satisfies 
\begin{align}\label{thetaLowerBound}
    \theta_{\min} \geq 0.
\end{align}

With the transmission and idle phase rates bounded, we now proceed to derive the corresponding bounds on the goodput.

\begin{figure}[t]
    \centering
    \ifdouble
    \includegraphics[width=0.65\linewidth]{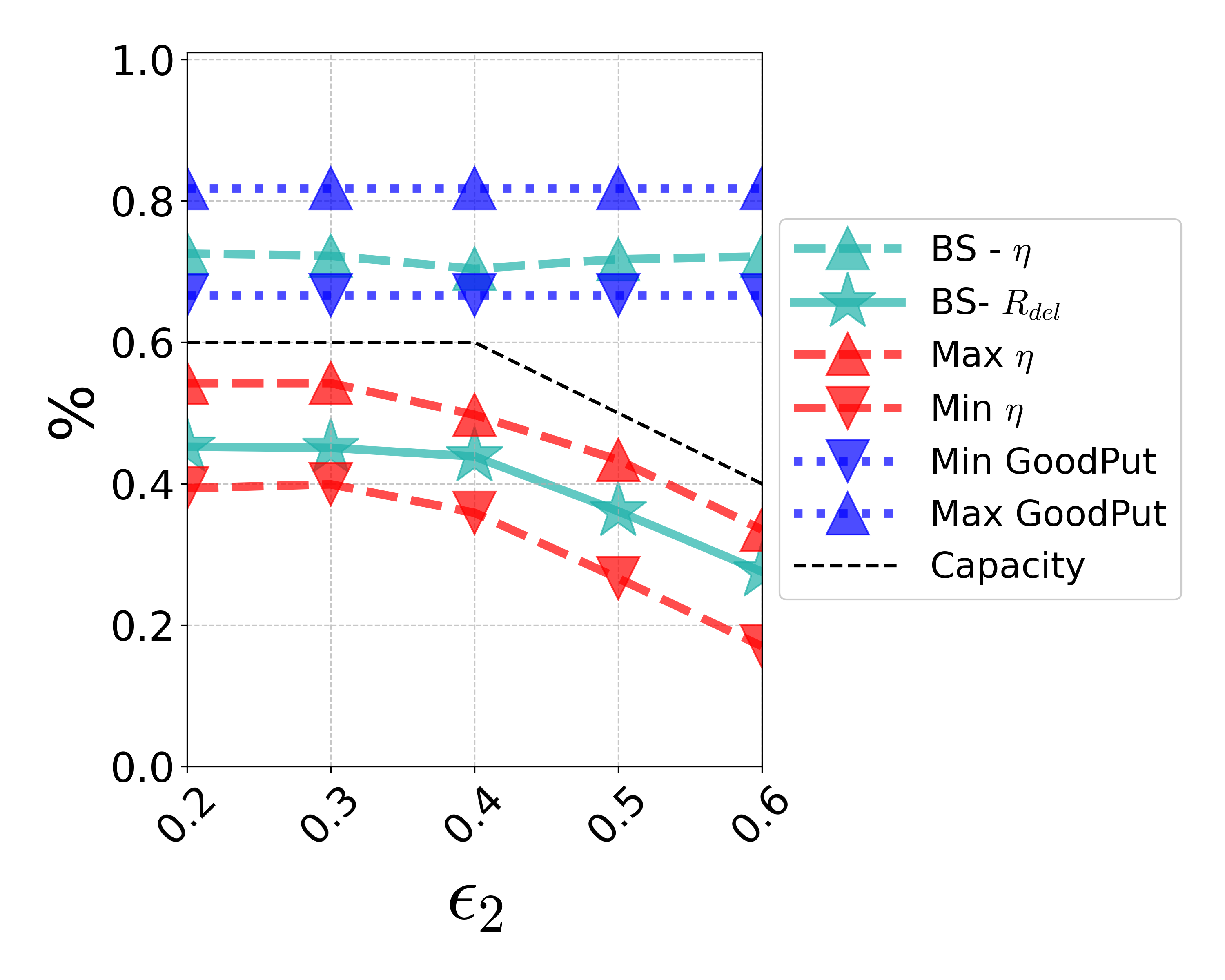}
    \fi
    \vspace{-0.4cm}
    \caption{\small System performance of a 6-node network as a function of $\epsilon_2 \in [0.2, 0.6]$, with fixed erasure probabilities $\epsilon_0 = \epsilon_4 = 0.1$, $\epsilon_1 = 0.4$, and $\epsilon_3 = 0.3$. The turquoise dashed line represents the goodput, as defined in \eqref{eqn:eta_n}, with its upper and lower bounds shown in blue dashed lines. The turquoise solid line depicts the throughput, computed according to \eqref{eqn:rDel}, with red dashed lines indicating its upper and lower bounds.}
    \label{fig:throughputGoodput}
\end{figure}

\begin{theorem}\label{theorem:goodputBounds}
The achievable normalized goodput, denoted by $\Gamma \triangleq \tau / (1-\theta)$, satisfies the following bounds
\ifdouble
\begin{align*}
\Gamma_{\min} &\geq
\frac{ \min \{ {\rm RTT}_i \} }{
    \frac{3}{2} \cdot \min \{ {\rm RTT}_i \}
    + \alpha \sum_{i=0}^{N-2} {\rm RTT}_i \cdot\epsilon_i
} /
\\
&\quad
\Bigg( 1 -
\frac{
    \alpha \sum_{i=0}^{N-2} {\rm RTT}_i \cdot \epsilon_i
}{
    \frac{3}{2} \cdot \min \{ {\rm RTT}_i \}
    + \alpha \sum_{i=0}^{N-2} {\rm RTT}_i \cdot\epsilon_i
} \Bigg),
\\[1ex]
\Gamma_{\max} &\leq
\frac{1 - \min \{ \epsilon_i \}}{1 + \min \{ \epsilon_i\}}.
\end{align*}
\fi
\end{theorem}
\begin{proof}
To establish the lower bound of $\Gamma$, we consider the case where $\tau = \tau_{\min}$, as defined in~\eqref{tauLowerBound}, corresponding to the scenario in which the BSP condition holds. Under this assumption, the upper bound on $\theta$, denoted $\theta_{\max}$, as given in~\eqref{thetaUpperBound}, is applicable. Substituting $\tau_{\min}$ and $\theta_{\max}$ into the expression for $\Gamma$ yields the desired lower bound.

To derive the upper bound of $\Gamma$, we consider $\tau_{\max}$ as defined in~\eqref{tauUpperBound}, along with the minimal value of $\theta$, which is zero, as given in~\eqref{thetaLowerBound}. Substituting these values into the expression for $\Gamma$ yields its upper bound.\end{proof}
\vspace{-0.08cm}
Fig.~\ref{fig:throughputGoodput} shows the goodput bounds for the example in Section~\ref{section:evaluation}.

\vspace{-0.1cm}
\subsection{Throughput Upper Bound}

In this subsection, we derive the upper bound on throughput. We begin by quantifying rate fluctuations, for which we introduce the Hellinger distance. Using this framework, we then derive both the per-link and end-to-end throughput upper bounds.

A node $n$ estimates its erasure rate at time slot $t$ based on feedback received from node $n+1$. Due to the round-trip delay of ${\rm RTT}_n$ time slots, this feedback reflects packets transmitted up to time $t{-}{\rm RTT}_n$. As a result, the node operates with outdated information and the actual erasure rate at time $t$ may differ from the estimated one. This challenge arising from estimating the erasure rate by feedback and its implications has been extensively discussed in~\cite{cohen2020adaptive,cohen2020adaptiveMH}.

Let $t^-$ denote the time slot ${\rm RTT}_n$ units before the current time $t$, that is, $t^- = t - {\rm RTT}_n$. We define $r_n(t^-)$ as the known erasure rate at node~$n$, measured up to time $t^-$. In contrast, $r_n(t)$ denotes the estimated rate at node~$n$ in the current time slot, which may differ from $r_n(t^-)$ due to variations in channel conditions. 
\vspace{-0.2cm}
\begin{lemma}\label{lemma:rateConsecutiveNodes}
    Consider two consecutive nodes $n$ and $n+1$ in the multi-hop communication chain. If the link $n$ connecting them constitutes the forward system's bottleneck, then the channel rate is upper bounded by,
    \vspace{-0.2cm}
    \begin{align*}
        r_n(t) \leq r_n(t^-) + \frac{\sqrt{V_n(t)}}{{\rm RTT}_n},
    \end{align*}
    where $V_n(t) = {\rm RTT}_n\cdot\epsilon_n(1-\epsilon_n)$.
\end{lemma}
\begin{proof}
    The result follows directly from the analysis presented in~\cite[Eq. (16)]{cohen2020adaptive} and \cite[Eq. (9)]{cohen2020adaptiveMH} which characterizes the behavior of the BEC in the point-to-point communication channel.
    \rg{For completeness, the full derivation is included in Appendix~\ref{appendix:lemma7} in the context of the proposed BS-AC-RLNC algorithm.}
    \off{ \al{where the channel variability in the finite regime $\left|r_n(t)-r_n(t^-)\right|$ is bounded by ${\sqrt{V_n(t)}}/{\text{RTT}_n}$ with high probability}} 
\end{proof}

The next corollary extends the upper bound to arbitrary (possibly non-consecutive) node pairs $j$ and $i$, with $j<i$, where node $j$ observes the erasure rate of node $i$.

\begin{corollary}
    An upper bound on the channel rate of node $i$ measured by node $j$, with $j<i$, can be expressed as    \begin{align}\label{equation:upperRateFromH}
        r_i(t) \leq r_i(t^-) + h\cdot\frac{\sqrt{V_i(t)}}{{\rm RTT}_i},
    \end{align}
    where $h$ denotes the number of hops between links $i$ and $j$.
\end{corollary}
\begin{proof}
    The result follows from Lemma~\ref{lemma:rateConsecutiveNodes}. In the case of non-consecutive nodes, the uncertainty in the observed throughput rate accumulates over the $h$ hops, since node $j$ receives feedback from node $i$ only after $h$ hops.
    \rg{Under the BEC model, the maximum attainable rate fluctuation of node $i$ over the uncertainty interval preceding the reception of its acknowledgment at node $j$ is upper bounded by $ h \cdot \sqrt{V_i(t)} / {\rm RTT}_i$.}
\end{proof}

\begin{remark}
A more accurate evaluation of the parameter $h$ is given by
\ifpagelimit
    $h = \left\lceil{\left(\sum_{k=j}^{i}{\rm RTT}_k\right)}/{{\rm RTT}_i}\right\rceil,$
\fi
offering a characterization that reflects cumulative round-trip times instead of merely counting the number of hops between nodes.
\end{remark}
To quantify potential deviations in the erasure rate over the $\rm RTT$ interval, we employ the Hellinger distance \cite{basu19972} as a statistical measure of distributional divergence. In contrast to \cite{cohen2020adaptiveMH,cohen2020adaptive}, which utilize the Bhattacharyya distance, we adopt the Hellinger distance due to its advantageous properties. Although both metrics assess the similarity between probability distributions, the Hellinger distance is a true metric bounded in the interval $[0,1]$, offering improved interpretability and numerical stability in our setting.

\begin{definition}\label{def:HD}
The \emph{Hellinger distance} between two rate distributions of node~$i$, at time slots $t^-$ and $t$, is given by,
\vspace{-0.2cm}
\ifdouble
\begin{multline*}
l(r_i(t^-), r_i(t)) = \frac{1}{\sqrt{2}} \cdot \\ { \sqrt{\left( \sqrt{r_i(t^-)} - \sqrt{r_i(t)} \right)^2 +
\left( \sqrt{1 - r_i(t^-)} - \sqrt{1 - r_i(t)} \right)^2 }}.
\end{multline*}
\fi
This corresponds to the divergence between two Bernoulli distributions parameterized by $r_i(t^-)$ and $r_i(t)$. Depending on the context, we distinguish two cases:

\ifpagelimit
\noindent 1) \textbf{Self-measured case:} When the distance is computed locally by node~$i$ using its own rate estimates, it is denoted by $l(r_i(t^-), r_i(t))$.\\
\noindent 2) \textbf{Remotely measured case:} When the distance is computed by another node~$j$ (with $j < i$), using an upper bound of $r_i(t)$, derived from~\eqref{equation:upperRateFromH}, it is denoted by $l_j(r_i(t^-), r_i(t))$.
\fi
\end{definition}

The next theorem extends the upper rate bound to the general case.

\begin{lemma}\label{lemma:nodeRateUpperBound}
    The channel rate of node $n$ is upper-bounded by
    \vspace{-0.2cm}
    \begin{align}\label{equation:upperRateBound}
         r_n(t)  \leq  r_n(t^-) + \frac{\sqrt{V_n(t)}}{{\rm RTT}_n} + \sum_{k=n+1}^{BN_n}l_n(r_k(t^-),r_k(t)).
    \end{align}
\end{lemma}
\begin{proof}
The channel rate at node~$n$ is constrained by both the local variability in $r_n(t)$ and the aggregate fluctuation over the forward channels that define the BSP. The local variability is upper-bounded by the channel rate $r_n(t^-)$ and the maximum fluctuation under the BEC model during ${\rm RTT}_n$ time slots, which is given by $\sqrt{V_n(t)} / {\rm RTT}_n$.

For the BSP bound, it is determined by the rates of the forward channels $\{e_k\}_{k = n+1}^{BN_n}$, as specified in~\eqref{eqn:BS_num}. Lower erasure rates along these channels result in a shorter BSP and thus a higher achievable channel rate. The uncertainty in these rates is quantified by the cumulative Hellinger distances, $\sum_{k = n+1}^{BN_n} l_n(r_k(t^-), r_k(t))$, which capture the maximum fluctuation. Combining these bounds yields the stated upper bound on $r_n(t)$.
\end{proof}
\vspace{-0.2cm}
Based on the rate bound in Lemma~\ref{lemma:nodeRateUpperBound}, we obtain the following result for the per-link throughput.

\begin{lemma}\label{lemma:linkUpperThroughput}
    The per-link upper throughput bound of channel~$n$, denoted by $\eta_n^{\max}$, satisfies
    \begin{align*}
        \eta_n^{\max} \leq 1 - \epsilon_n - l\left(r_n(t^-), r_n(t)\right).
    \end{align*}
\end{lemma}
\begin{proof}
    From Lemma~\ref{lemma:nodeRateUpperBound}, the maximum achievable rate at time slot $t$ is given by Eq.~\eqref{equation:upperRateBound}. Accordingly, the maximum possible rate loss is characterized by the discrepancy between $r_n(t^-)$ and $r_n(t)$, which is quantified using the Hellinger distance.
\end{proof}
\vspace{-0.2cm}
While Lemma~\ref{lemma:linkUpperThroughput} quantifies the maximum achievable rate at the link level, the next result characterizes the overall throughput limitation of the end-to-end system.

\begin{theorem}\label{theorem:throughputUpper}
    The end-to-end throughput $\eta_{\max}$ is upper bounded by $\eta_{\max} \leq \min\{\eta_n^{\max}\}$.
\end{theorem}
\begin{proof}
    The result follows directly from Lemma~\ref{lemma:linkUpperThroughput} and the max-flow min-cut theorem, as presented in \cite[Chapter~18]{yeung2008chapter18}.
\end{proof}

Fig.~\ref{fig:throughputGoodput} illustrates the derived throughput bounds for the example discussed in Section~\ref{section:evaluation}.

\subsection{Throughput Lower Bound}

To establish a lower throughput bound, we begin by analyzing the inefficiencies introduced by redundant retransmissions. These arise when a node exhausts its transmission window and
continues to transmit linear combinations of the current window's data until an acknowledgment is received. We then analyze the impact of extended BSPs, during which idle periods exceed the necessary duration. This prolonged inactivity leads to a reduction in effective throughput below the network bottleneck rate, thereby impairing overall system performance.

Let $n_i^w$ denote the total number of transmissions over channel $i$ during a given transmission window, and let $n_i^{\mathrm{EW}}$ represent the number of redundant transmissions that occur in channel $i$ at the end of the window, while the node continues to transmit until positive feedback arrives. In the following lemmas, we derive a lower bound on $n_i^w$ and an upper bound on $n_i^{\mathrm{EW}}$.

\begin{lemma}
A lower bound on the number of transmissions over channel $i$ during a transmission window, $n_i^w$, is given by
\begin{align*}
    n_i^w \geq (1+\epsilon_i)\cdot w + 1.
\end{align*}
\end{lemma}
\begin{proof}
    According to the transmission algorithm, during a window of size $w$, the node transmits $w$ source packets. The first packet is a linear combination of the first raw data packet; the second is a linear combination of the first and second raw data packets; and, in general, the $k$-th source packet is a linear combination of the first $k$ raw data packets, for $1 \leq k \leq w$. This is followed by $w \cdot \epsilon_i$ a-priori FEC packets. The final term, $1$, accounts for an additional erasure not covered by the a-priori FEC, ensuring that the window limit is reached.
\end{proof}
\vspace{-0.1cm}
Let $\operatorname{erf}(x)$ denote the error function, defined for nonnegative real values of $x$ as \ifpagelimit  $\operatorname{erf}(x) = \frac{2}{\sqrt{\pi}} \int_0^x e^{-t^2} \, dt$.\fi
This function corresponds to the probability that a real-valued random variable $Y$, which is normally distributed with mean $0$ and standard deviation $\frac{1}{\sqrt{2}}$, lies within the interval $[-x, x]$. Using this definition, we now proceed to derive the next lemma.

\begin{lemma}\label{lemma:RedundantTransmissions}
An upper bound on the number of redundant transmissions at the end of the window, $n_i^{\mathrm{EW}}$, over channel $i$, is given by
\begin{align*}
    n_i^{\mathrm{EW}} \leq \left(1 - \operatorname{erf}\left(\frac{1}{\sqrt{2}}\right)\right)(1 - \epsilon_i)\,{\rm RTT}_i + O\left(\frac{1}{\sqrt{t}}\right),
\end{align*}
where $t$ denotes the time at which the packet enabling successful decoding was transmitted.
\end{lemma}
\begin{proof}
    This result has previously been established in \cite[Theorem~2]{cohen2020adaptiveMH} \rg{and is included in Appendix~\ref{appendix:lemma11} for completeness in the context of the proposed BS-AC-RLNC algorithm herein.}
\end{proof}
\vspace{-0.2cm}
We now turn to a second source of inefficiency, prolonged BSP. Let $I_n$ denote the total BSP, measured in time slots, for channel $n$, and let $I^u_n$ denote the unnecessary part of this BSP. In the following lemmas, we derive upper and lower bounds on $I_n$ and $I^u_n$, respectively.

\begin{lemma}
    An upper bound on BSP over channel $n$, $I_n$, is given by
    \vspace{-0.2cm}
    \[
        I_n < \alpha \sum_{k=\rg{0}}^{N-2}{\rm RTT}_k \cdot \epsilon_k.
    \] 
\end{lemma}
\begin{proof}
    By definition \eqref{eqn:BS_num}, for channel $n$,
    \[
        \alpha \sum_{k=n+1}^{BN}{\rm RTT}_k \cdot \epsilon_k < \alpha \sum_{k=\rg{0}}^{N-2}{\rm RTT}_k \cdot \epsilon_k.
    \]
    \vspace{-0.2cm}
\end{proof}
\vspace{-0.3cm}
\begin{lemma}\label{lemma:unecessaryBSP}
    The unnecessary portion of the BS over channel~$n$, $I^u_n$, is lower bounded by
    \vspace{-0.2cm}
    \begin{align*}
        I^u_n \geq \frac{\left(1-\operatorname{erf}\left(\frac{1}{\sqrt{2}} \right)\right)}{2} \cdot(N-2-n) + O\left(\frac{1}{\sqrt{t}}\right),
    \end{align*}
    where $t$ denotes the starting time slot of the BSP.
\end{lemma}
\begin{proof}
A reduction in effective throughput below the network bottleneck rate requires a significant deviation of the erasure rate from its mean. This deviation is quantified by the term $1 - \operatorname{erf}(1/\sqrt{2})$, which corresponds to the probability that the erasure rate deviates by more than one standard deviation. The expression is divided by two to capture the case where the erasure rate is lower than its mean.

The Berry–Esseen theorem, applicable to sums of independent erasure events, implies that the distribution of the erasure rates converges to the normal distribution and the error in this approximation is upper bounded by $O(1/\sqrt{t})$ \cite[Theorem~13]{polyanskiy2010channel}. The factor $(N - 2 - n)$ represents an upper bound on the number of channels that define the BSP for channel~$n$.
\end{proof}

The preceding lemmas quantify the redundant transmissions resulting from the window limit being reached and the impact of extended BSP intervals. We now leverage these results to establish a lower bound on the achievable throughput.

\begin{lemma}\label{lemma:perlinkThroughputLower}
    The per-link throughput lower bound of channel $n$, denoted by $\eta_n^{\min}$, satisfies \ifpagelimit $\eta_n^{\min} \geq \eta_n^{max}\ - {n_n^{EW}}/{n_n^{w}} - {I^u_n}/{I_n}$. \fi
\end{lemma}
\begin{proof}
The claim follows as a direct consequence of Lemmas~\ref{lemma:linkUpperThroughput}--\ref{lemma:unecessaryBSP}. \rg{Under the assumption that $w \geq {\rm RTT}$, it holds that ${n_n^{EW}} / {n_n^{w}} < 1$.}
\end{proof}

\begin{theorem}\label{theorem:throughputLower}
    The end-to-end throughput $\eta_{\min}$ is lower bounded as follows, \ifpagelimit $\eta_{\min} \geq \min\{\eta_n^{\min}\}$.\fi
\end{theorem}
\begin{proof}
    The result follows directly from Lemma \ref{lemma:perlinkThroughputLower} and the max-flow min-cut theorem, as presented in \cite[Chapter~18]{yeung2008chapter18}.
\end{proof}

\begin{figure*}[!t]
    \centering
    \begin{subfigure}{0.19\linewidth}
        \centering
        \includegraphics[width=\linewidth]{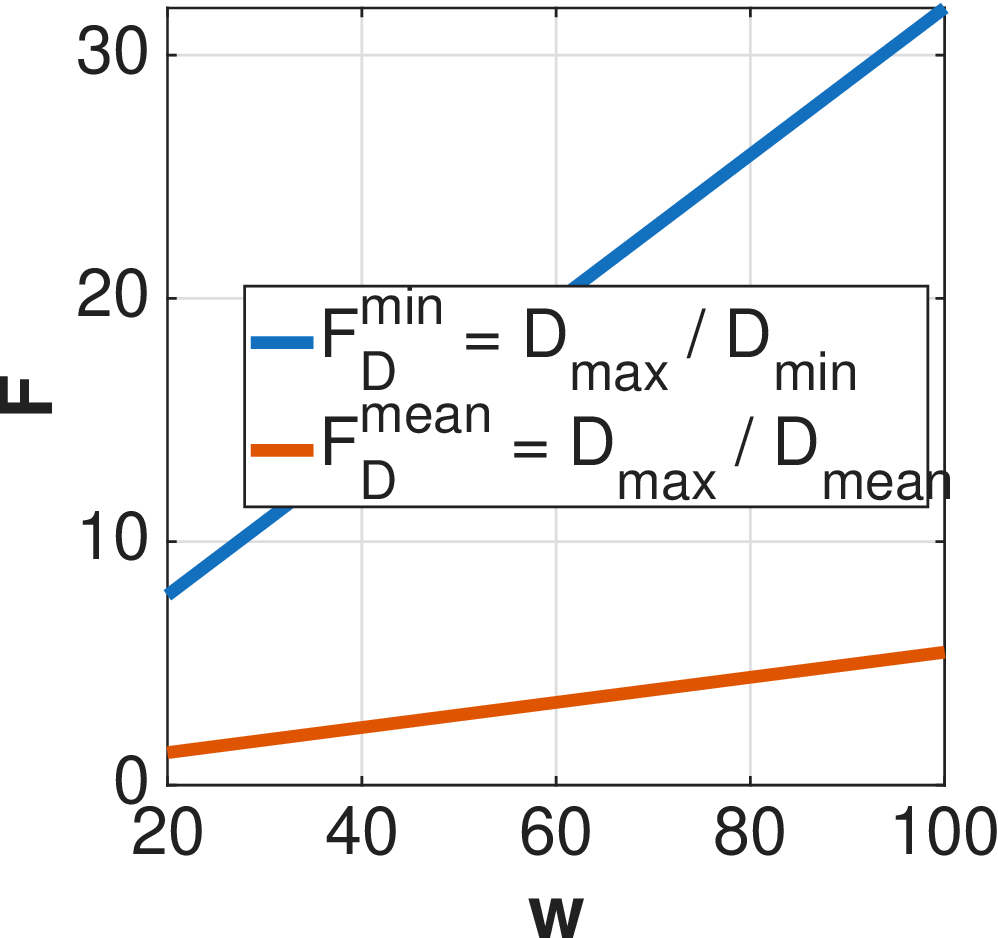}
        \caption{\small \rg{$w$ vs delay.}}
        \label{fig:delay_w}
    \end{subfigure}
    \begin{subfigure}{0.19\linewidth}
        \centering
        \includegraphics[width=\linewidth]{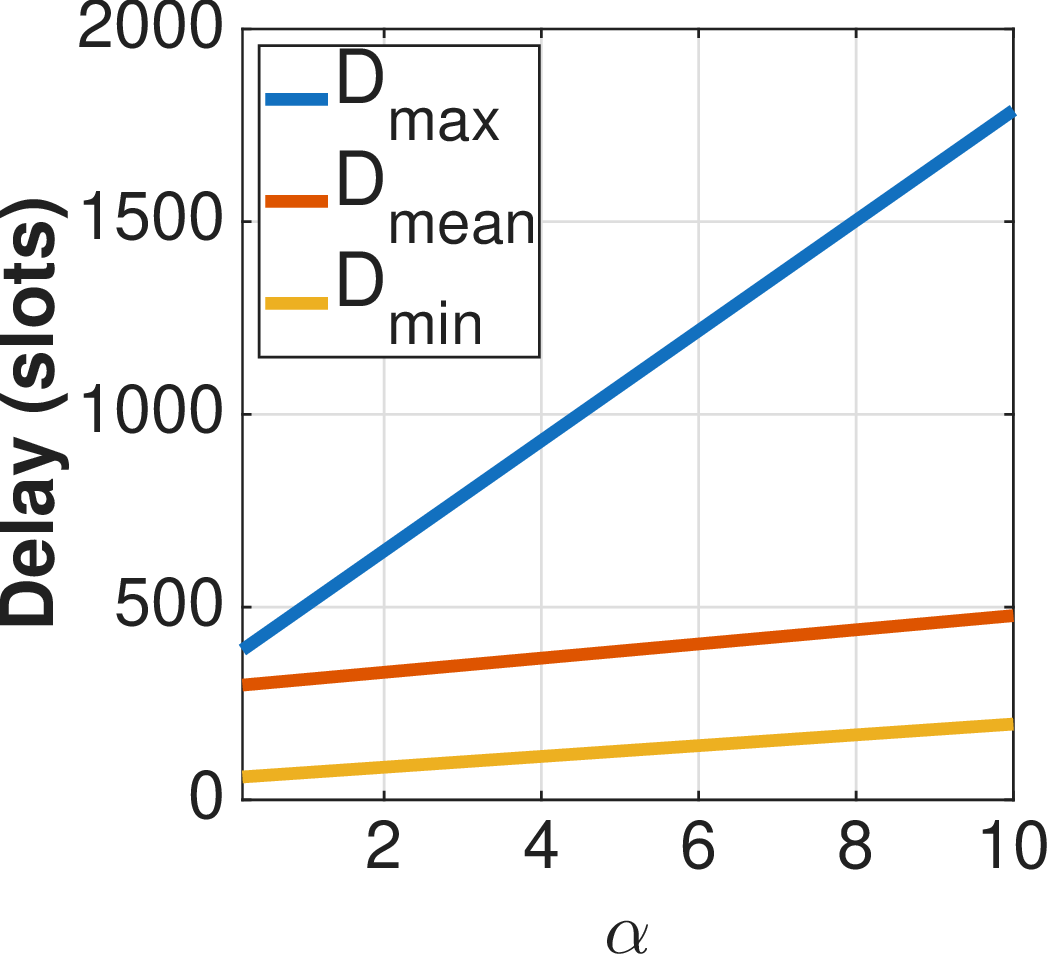}
        \caption{\small \rg{$\alpha$ vs delay.}}
        \label{fig:delay_alpha}
    \end{subfigure}
    \begin{subfigure}{0.19\linewidth}
        \centering
        \includegraphics[width=\linewidth]{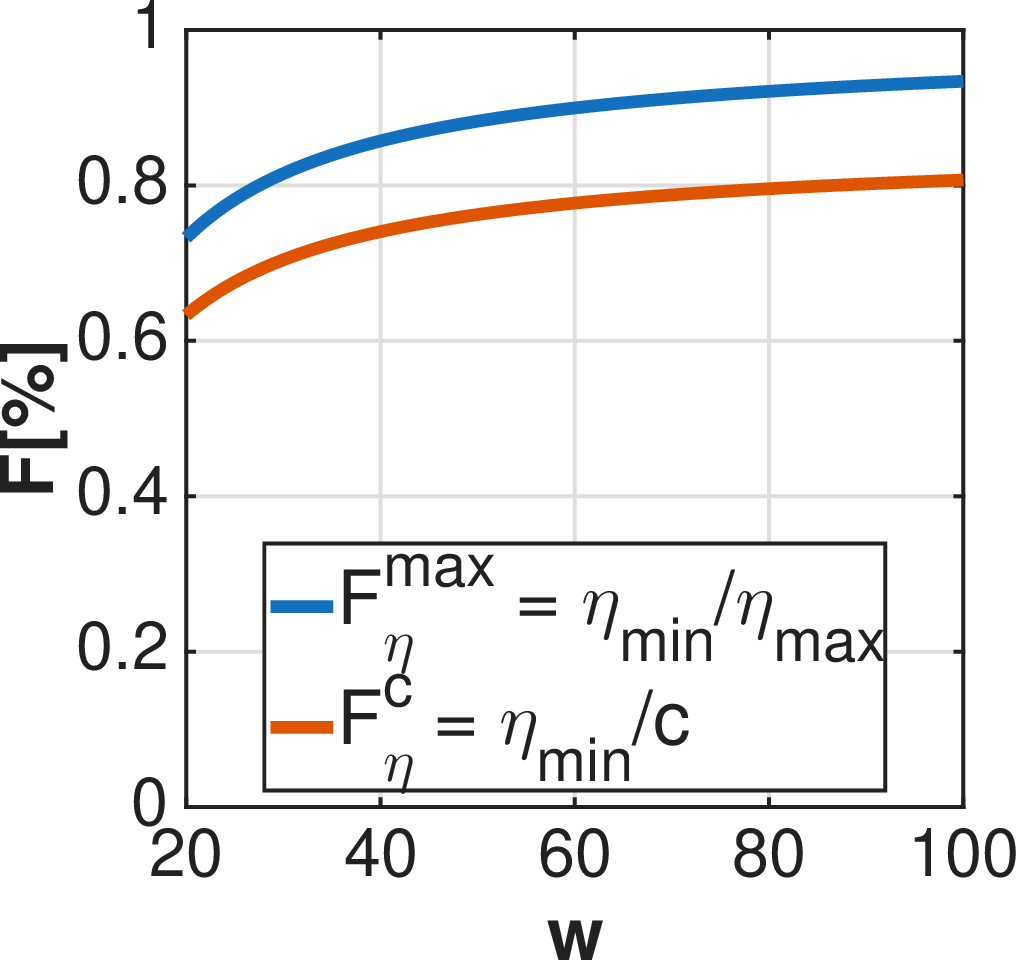}
        \caption{\small \rg{$w$ vs throughput.}}
        \label{fig:throughput_w}
    \end{subfigure}
    \begin{subfigure}{0.19\linewidth}
        \centering
        \includegraphics[width=\linewidth]{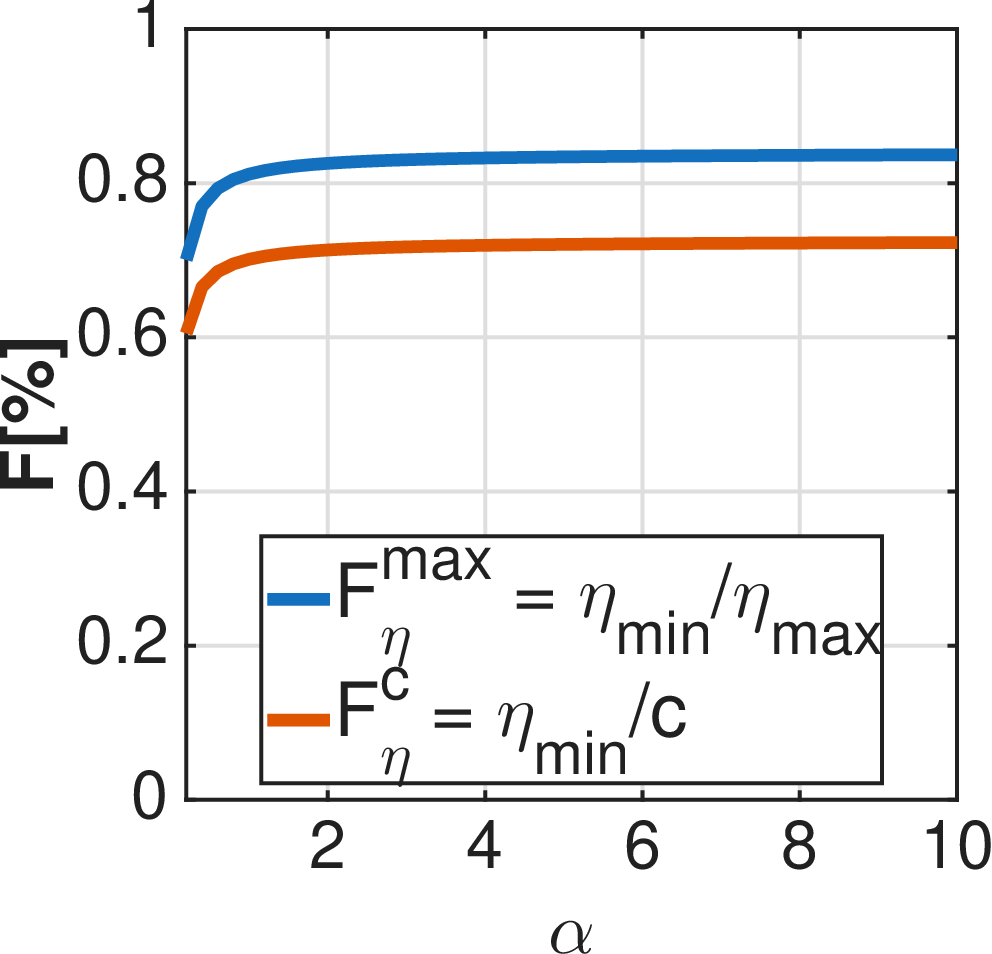}
        \caption{\small \rg{$\alpha$ vs throughput.}}
        \label{fig:throughput_alpha}
    \end{subfigure}
    \begin{subfigure}{0.19\linewidth}
        \centering
        \includegraphics[width=\linewidth]{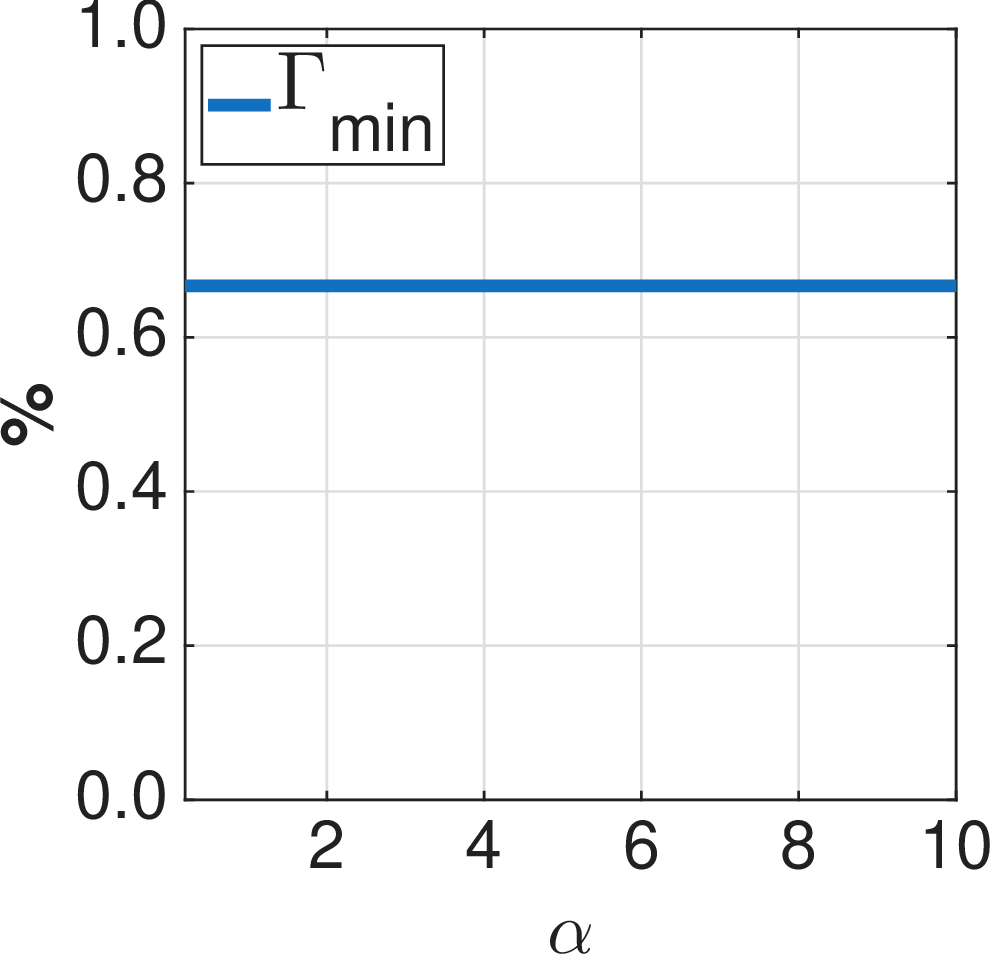}
        \caption{\small \rg{$\alpha$ vs goodput.}}
        \label{fig:goodput_alpha}
    \end{subfigure}
    \caption{\small \rg{Sensitivity analysis of delay, throughput, and normalized goodput bounds with respect to the maximum window size $w$ and the BS parameter $\alpha$.}}
    \vspace{-0.6cm}
    \label{fig:sensitivity}
\end{figure*}

\vspace{-0.7cm}
\subsection{\rg{Sensitivity Analysis of $w$ and $\alpha$}}\label{appendix:sensitivityAnalysis}

\rg{Here, we present a sensitivity analysis of $w$, the maximum window size, and $\alpha$, a key parameter of the BSP\footnote{\label{note:rtt} \rg{${\rm RTT} = 20$ reflects typical round-trip times reported in wireless communication systems~\cite{cohen2020adaptive,cohen2020adaptiveMH,dias2023sliding}.} \rg{For a sensitivity analysis with respect to $\rm RTT$, we refer the reader to~\cite{cohen2020adaptiveMH} and~\cite{cohen2022broadcast} due to the space limitations.}}.

Fig.~\ref{fig:sensitivity} illustrates the impact of $w$ and $\alpha$ on a six-hop network ($N=6$) with link erasure probabilities $\epsilon_0 = 0.1$, $\epsilon_1 = 0.4$, $\epsilon_2 = 0.2$, $\epsilon_3 = 0.7$, and $\epsilon_4 = 0.1$, and an $\rm RTT$ of $20$. Figs.~\ref{fig:delay_w} and~\ref{fig:throughput_w} illustrate how the delay and throughput bounds vary with $w$ from $20$ to $100$ at $\alpha = 2$. Figs.~\ref{fig:delay_alpha},~\ref{fig:throughput_alpha}, and~\ref{fig:goodput_alpha} show the effects of varying $\alpha$ from $0.2$ to $10$ at $w = 32$ on delay, throughput, and goodput bounds.

\off{The delay bounds $D_{\text{mean}}$ and $D_{\min}$ are independent of the maximum window size $w$.} Following~\cite{cohen2020adaptiveMH}, we define the scaling factors $F_D^{\min} = D_{\max}/D_{\min}$ and $F_D^{\text{mean}} = D_{\max}/D_{\text{mean}}$, which quantify the relative growth of the bounds as $w$ increases. Fig.~\ref{fig:delay_w} shows that increasing the maximum window size $w$ leads to longer delays. The dependence of the delay bounds on $\alpha$ is illustrated in Fig.~\ref{fig:delay_alpha}, which shows their growth as $\alpha$ increases.

\off{The throughput lower bound $\eta_{\min}$ is independent of $w$ and $\alpha$.} Moreover, we define the scaling factors $F_{\eta}^{\max} = \eta_{\min}/\eta_{\max}$ and $F_{\eta}^{c} = \eta_{\min}/c$, where $c$ denotes the network capacity, i.e. $c = 1 - \displaystyle\max_{i} \epsilon_i$. Figs.~\ref{fig:throughput_alpha} and \ref{fig:throughput_w} show that throughput increases with both $w$ and $\alpha$, exhibiting faster growth at smaller values and saturating at larger values.

\off{The upper bound of the normalized goodput is independent of both $\alpha$ and $w$.} For the lower bound, Fig.~\ref{fig:goodput_alpha} shows that $\Gamma_{\min}$ is insensitive to $\alpha$. The values were rounded to three decimal places to mitigate floating-point precision errors. This observation is consistent with the formulation of normalized goodput in Section~\ref{subsec:Problem}, which excludes idle slots.}
\vspace{-0.2cm}
\section{Multi-cast Analytical Results}\label{section:multicast}
\vspace{-0.1cm}
We consider a multi-cast setting in which the source node transmits common raw data to a designated destination as well as to all other nodes in the network. The system follows a multi-hop chain topology, as described in Subsection~\ref{subsec:SysModel}, where each node relays the message toward subsequent nodes along the chain. In this section, we derive analytical bounds on the delay, goodput, and throughput performance metrics.
\begin{corollary}
    For any node $n$ in the multi-cast chain, the delay is bounded as follows,
\ifdouble    
\begin{align*}
    D_{\min} &\geq \sum_{i=0}^{n-1}\left(\frac{{\rm RTT}_i}{2}+ \frac{1}{1-\epsilon_i} + \alpha\sum_{k=i}^{N-2} {\rm RTT}_i \cdot\epsilon_i\right)\\
    D_{\mathrm{mean}} &\leq\sum_{i=0}^{n-1}\frac{1}{1-\epsilon_i} \cdot \\ &\quad \left({\rm RTT}_i(1+\epsilon_i) +{\rm RTT}_i + \alpha\sum_{k=i}^{N-2}{\rm RTT}_i \cdot\epsilon_i\right) \\  D_{\max} &\leq \sum_{i=0}^{n-1}\left({\rm RTT}_i + \frac{w}{1-\epsilon_i} + \frac{w\cdot\epsilon_i}{1-\epsilon_i}\right) + \\ &\quad \sum_{k=0}^{N-2}\left(\frac{\alpha \cdot w \cdot (k+1) \cdot \epsilon_{k}}{1 - \epsilon_k}\right).
\end{align*}
\fi
\end{corollary}
\begin{proof}
This result follows directly from the structure of the chain network, Corollary~\ref{theorem:minDelayAlgo} and Theorems~\ref{theorem:meanDelay}--\ref{theorem:maxDelay}, with the summation truncated at node $n$, i.e. at channel $n-1$. The final term in each bound corresponds to the maximum BS, as each node in the chain computes its BSP relative to the final destination, rather than with respect to node~$n$.
\end{proof}

The minimum and mean delay bounds are depicted in the middle subplots of Figs.~\ref{fig:mc_2}--\ref{fig:mc_5}, while the maximum delay bounds appear in the rightmost subplots of the same figures. All results correspond to the example described in Section~\ref{section:evaluation}. As discussed in Subsection~\ref{subsection:DelayLowerBound}, the condition $\Delta_t^{\text{BS}}$ is omitted, following Lemma~\ref{lemma:delta}. The effect of this simplification is most apparent for $\epsilon_2 = 0.5$ and $\epsilon_2 = 0.6$, where the measured BSP decreases due to the threshold imposed by $\Delta_t^{\text{BS}}$.

Delay bounds are not shown for node~1, as the source transmits packets at a rate slightly below the global bottleneck. Since the derived bounds assume continuous transmission, they do not accurately capture the transient behavior that occurs at the initial hop.

The following two corollaries establish the multi-cast bounds for goodput and throughput.

\ifdouble
\begin{figure*}[tbp]
    \centering
    \begin{subfigure}{0.06\columnwidth}
        \raisebox{15ex}{
        \includegraphics[width=\linewidth]{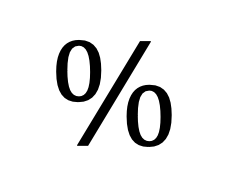}}
    \end{subfigure}
    \begin{subfigure}{0.32\columnwidth}
        \includegraphics[width=\linewidth]{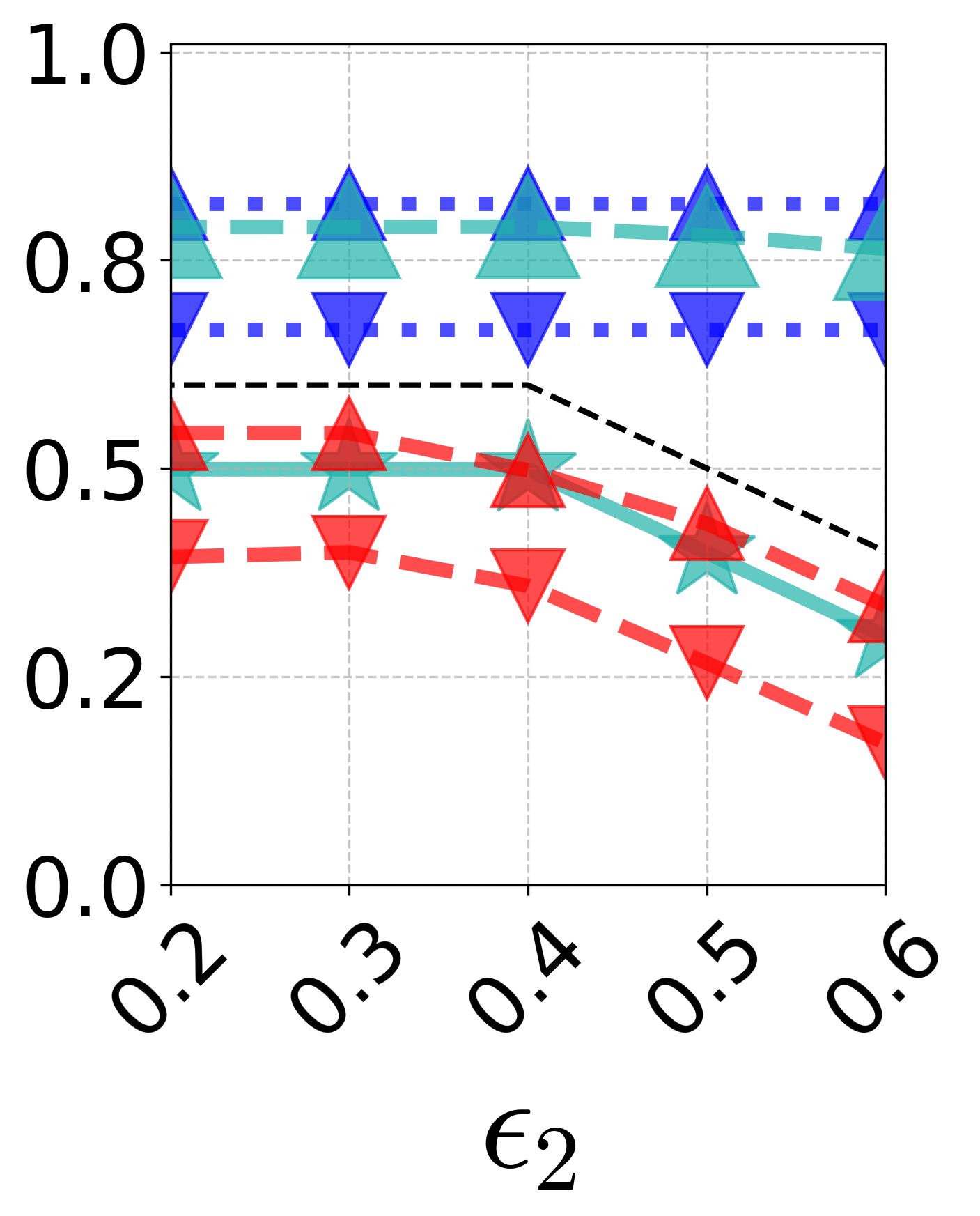}
        \caption{\small Node 1}
        \label{fig:1a}
    \end{subfigure}
    \hfill
    \begin{subfigure}{0.32\columnwidth}
        \includegraphics[width=\linewidth]{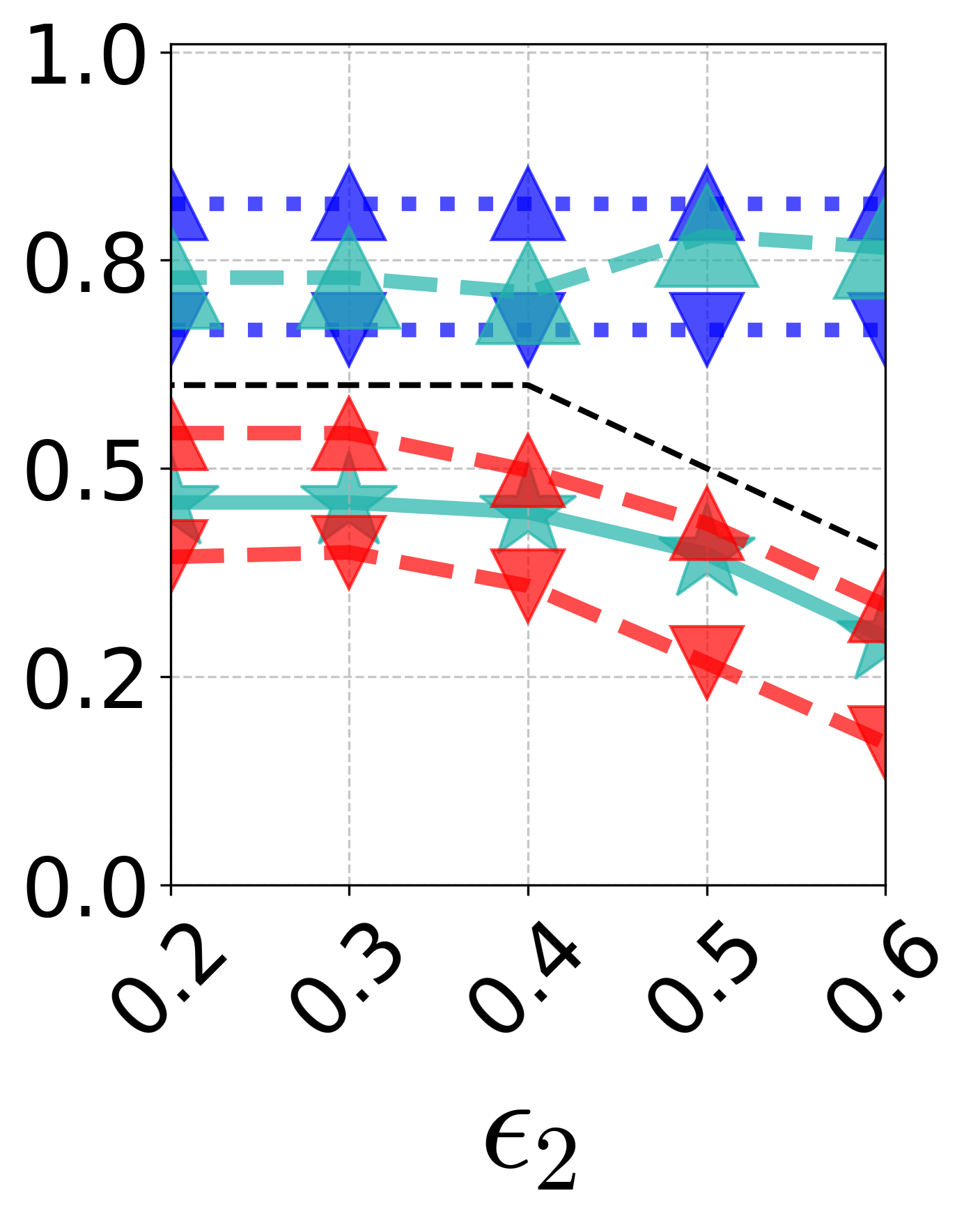}
        \caption{\small Node 2}
        \label{fig:1b}
    \end{subfigure}
    \hfill
    \begin{subfigure}{0.32\columnwidth}
        \includegraphics[width=\linewidth]{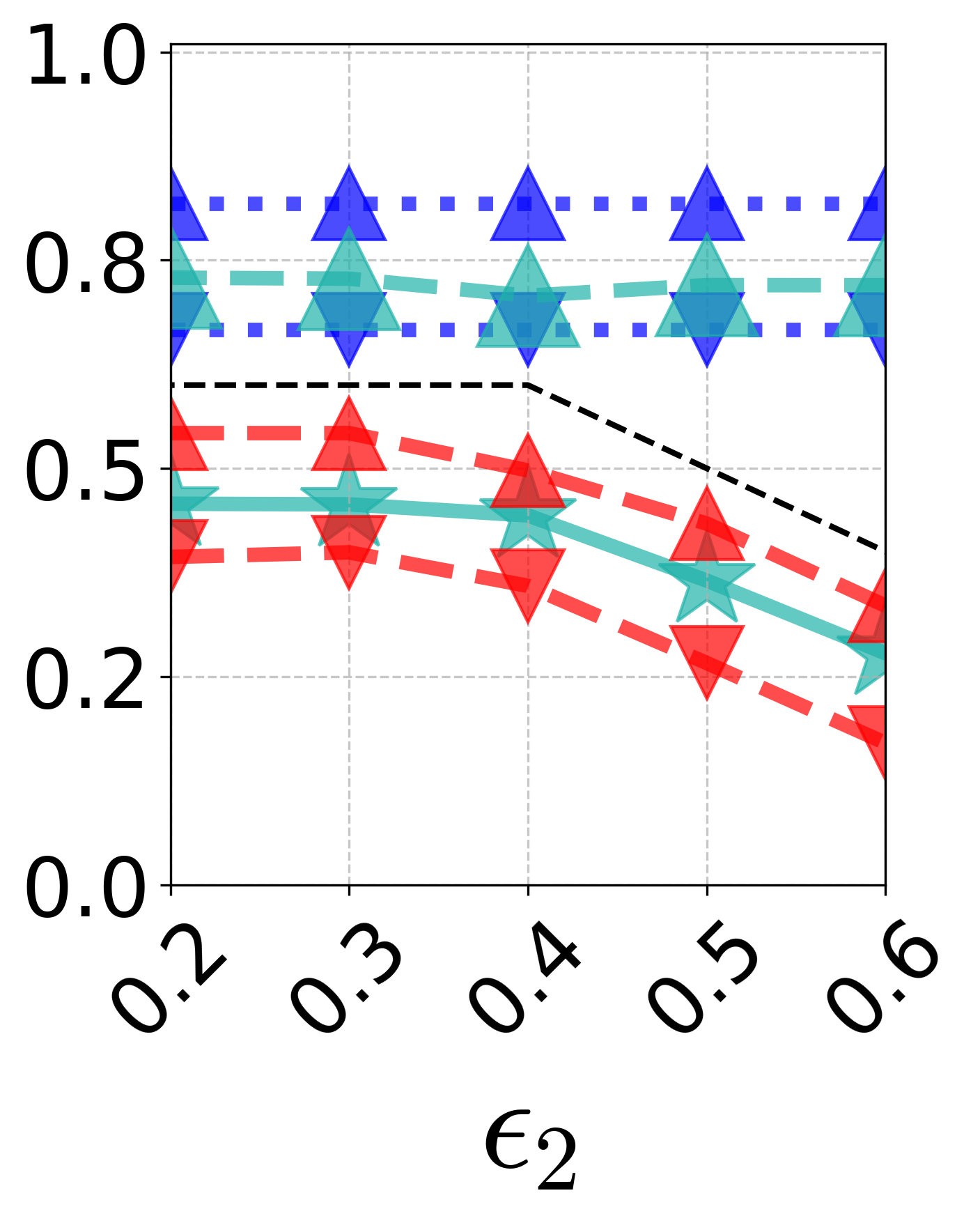}
        \caption{\small Node 3}
        \label{fig:1c}
    \end{subfigure}
    \hfill
    \begin{subfigure}{0.32\columnwidth}
        \includegraphics[width=\linewidth]{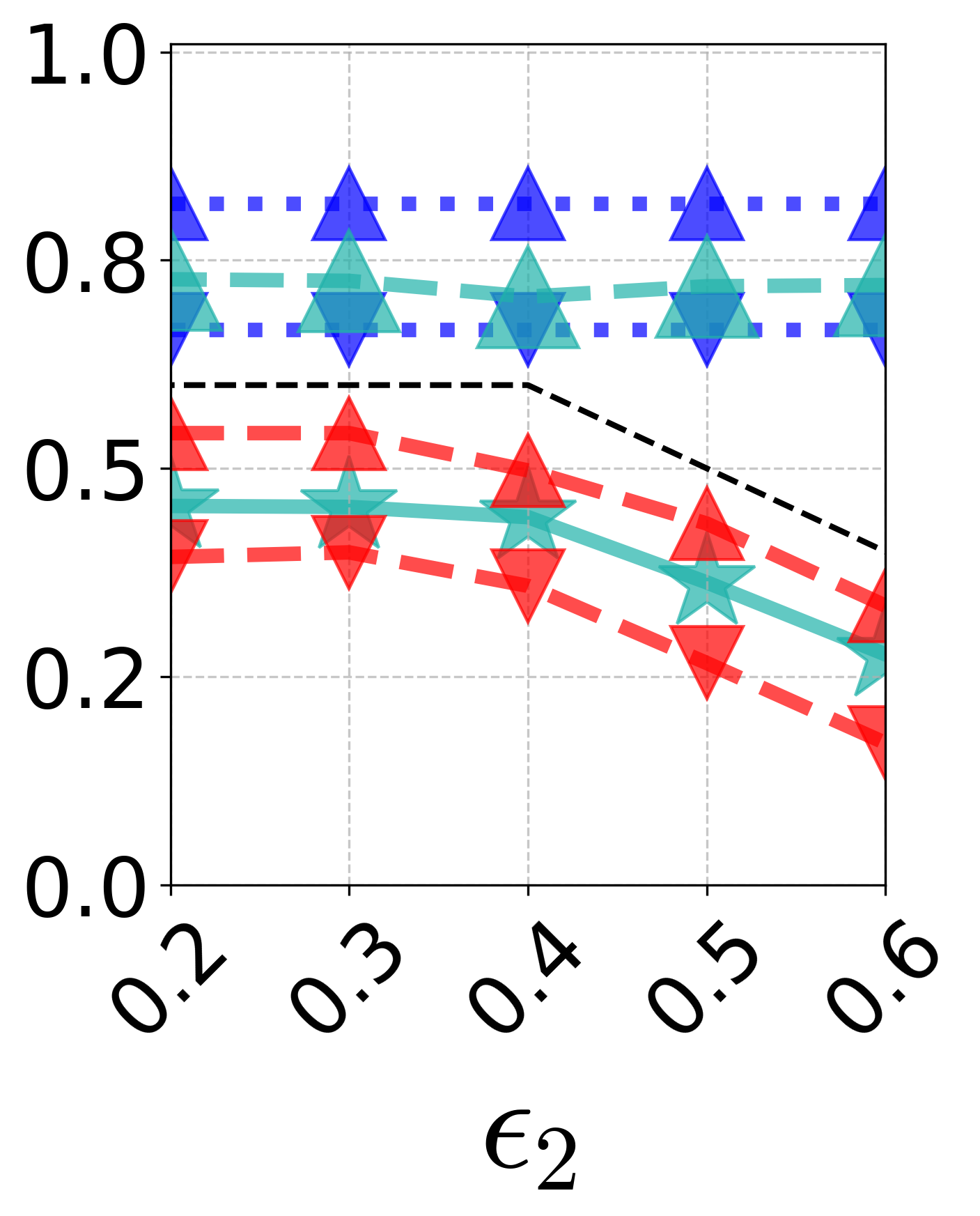}
        \caption{\small Node 4}
        \label{fig:1d}
    \end{subfigure}
    \hfill
    \begin{subfigure}{0.32\columnwidth}
        \includegraphics[width=\linewidth]{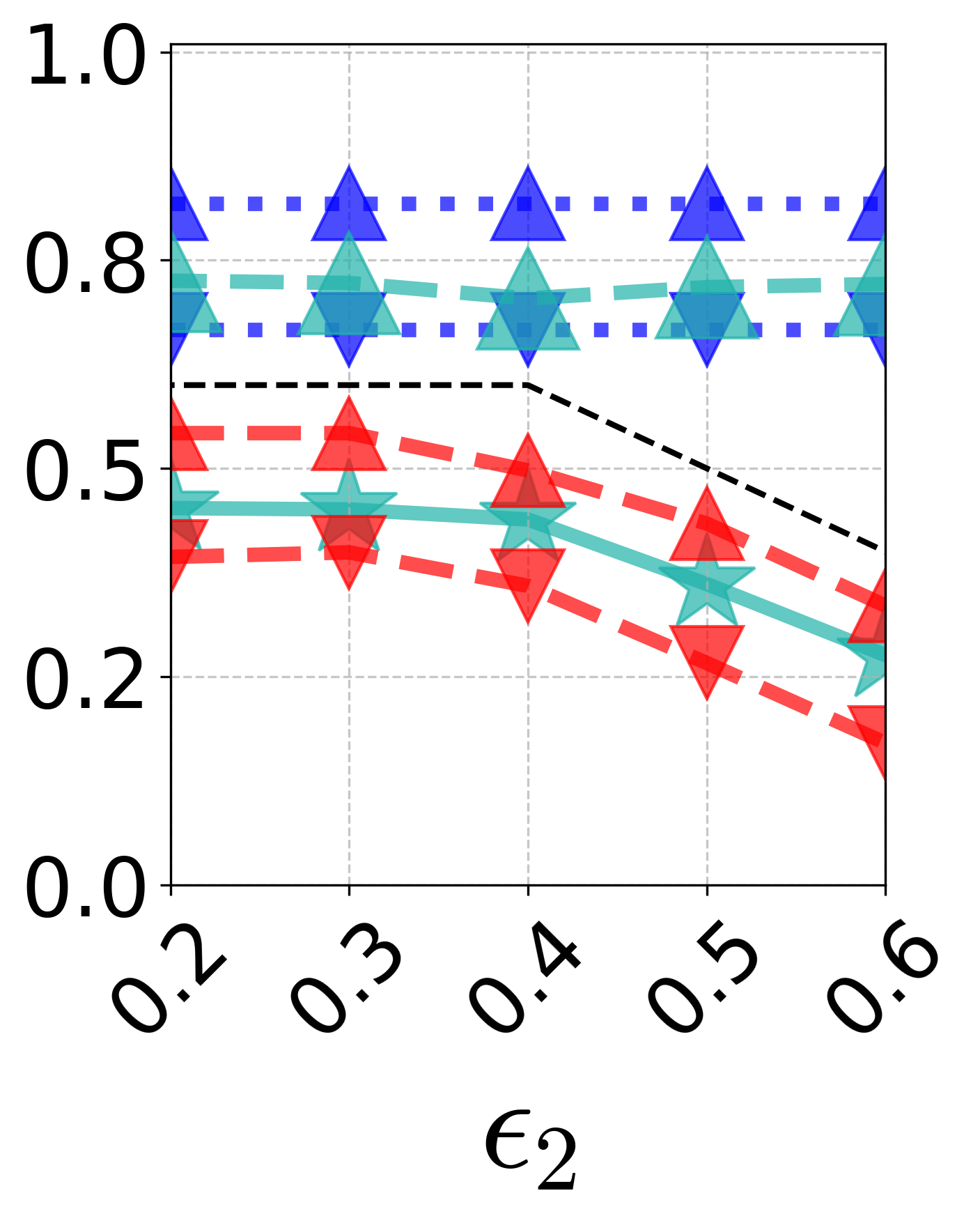}
        \caption{\small Node 5}
        \label{fig:1e}
    \end{subfigure}
    \hfill
    \begin{subfigure}{0.28\columnwidth}
        \vspace{-0.5cm}
        \raisebox{3ex}{\includegraphics[width=0.9\linewidth]{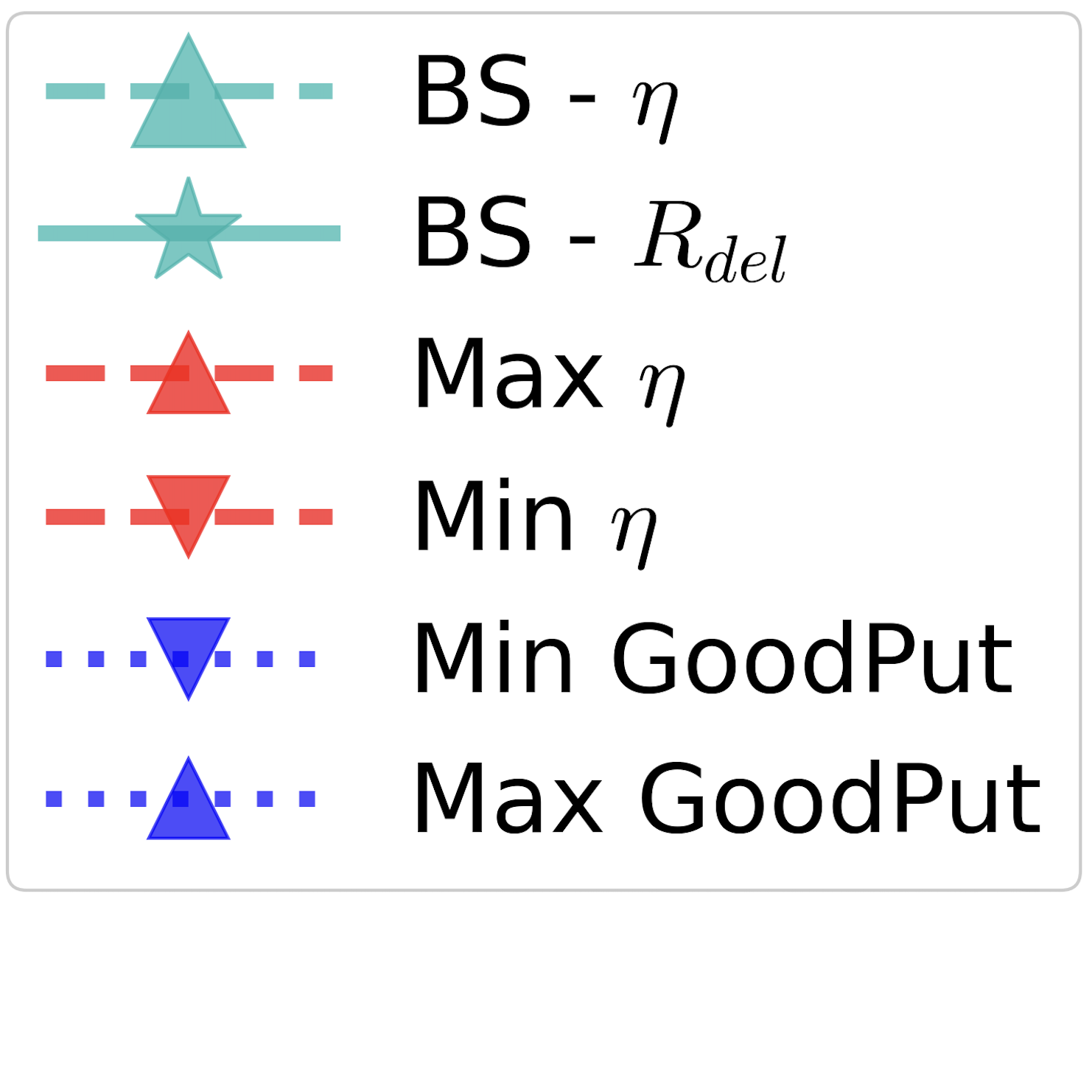}}
        \vspace{0.4cm}
        \caption{Legend}
        \label{fig:1f}
    \end{subfigure}
    \caption{\small Multicast scenario in a 6-node network, evaluated as a function of $\epsilon_2 \in [0.2, 0.6]$, with fixed erasure probabilities: $\epsilon_0 = \epsilon_4 = 0.1$, $\epsilon_1 = 0.4$, and $\epsilon_3 = 0.3$. The black dashed line indicates the end-to-end capacity.\off{ The turquoise dashed line shows the goodput, with its upper and lower bounds indicated by blue dashed lines. The turquoise solid line represents the throughput, and the red dashed lines denote its corresponding bounds.}}
    \vspace{-0.6cm}
\label{fig:composite}
\end{figure*}
\fi

\begin{corollary}
    The goodput at node $n$, denoted by $\Gamma_n$, in the multi-cast chain is bounded as
   $\Gamma_{\min} \leq \Gamma_n \leq \Gamma_{\max}$.
\end{corollary}
\begin{proof}
    The result follows directly from Theorem~\ref{theorem:goodputBounds}. By definition, the goodput at node $n$ is given by $\Gamma_n \triangleq \tau / (1 - \theta)$. Applying the per-link lower bound $\tau_{\min}$ and per-link upper bound $\theta_{\max}$ yields a lower bound on $\Gamma_n$, while the per-link upper bound $\tau_{\max}$ and per-link lower bound $\theta_{\min}$ yield an upper bound. Therefore,
    \ifpagelimit
    $ {\tau_{\min}}/{(1 - \theta_{\max})} \leq \Gamma_n \leq {\tau_{\max}}/{(1 - \theta_{\min})}, $
    \fi
    which corresponds to the definitions of $\Gamma_{\min}$ and $\Gamma_{\max}$ in this corollary.
\end{proof}

\begin{corollary}
    The throughput at node~$n$ in the multi-cast chain is bounded as follows, \ifpagelimit $\eta_{\min} \geq \min\{\eta_n^{\min}\}$ and $\eta_{\max} \leq \min\{\eta_n^{\max}\}$.\fi
\end{corollary}
\begin{proof}
    The bounds follow directly from Theorems~\ref{theorem:throughputUpper}--\ref{theorem:throughputLower}, and are consistent with the definition of multi-cast, where the throughput is defined as the rate at which all destination nodes can receive the transmitted raw data packets.
\end{proof}

The goodput and throughput bounds for the multi-cast network are illustrated in Fig.~\ref{fig:composite}. These results correspond to the evaluation scenario presented in Section~\ref{section:evaluation}.

\vspace{-0.1cm}
\section{Evaluation Results} \label{section:evaluation}
\label{sec:EvalRes}
In this section, we present simulation results demonstrating the effectiveness of the \ac{bs-acrlnc} protocol \footnote{The simulations code is available at \url{https://github.com/Adinawx/MH_NC}}.
We evaluate a 6-node network with BEC channels, where $\epsilon_0=\epsilon_4=0.1$, $\epsilon_1=0.4$, $\epsilon_3=0.3$, and middle link $\epsilon_2$ varies from $0.2$ to $0.6$ across simulations. The global bottleneck remains $\epsilon_{BN}=\epsilon_1=0.4$ until $\epsilon_2$ exceeds $0.4$ and becomes the new global bottleneck.
Considering the capacity limit, raw data packets arrive at the source according to a Bernoulli process with a rate slightly lower than the global bottleneck, $1-\epsilon_{BN} - 0.1$. We set ${\rm RTT}_n=20$ for each node $n$$^{\ref{note:rtt}}$, yielding a global $\rm RTT$ of $100$ slots over a time horizon of $T=5000$ slots. In addition, we set $\alpha = 2$\rg{,  $\kappa = 1$,} and use a window size of $w = 32$.

\begin{figure*}[tbp]
    \centering
    \begin{subfigure}[t]{0.258\textwidth}
        \includegraphics[width=\linewidth,height=4cm]{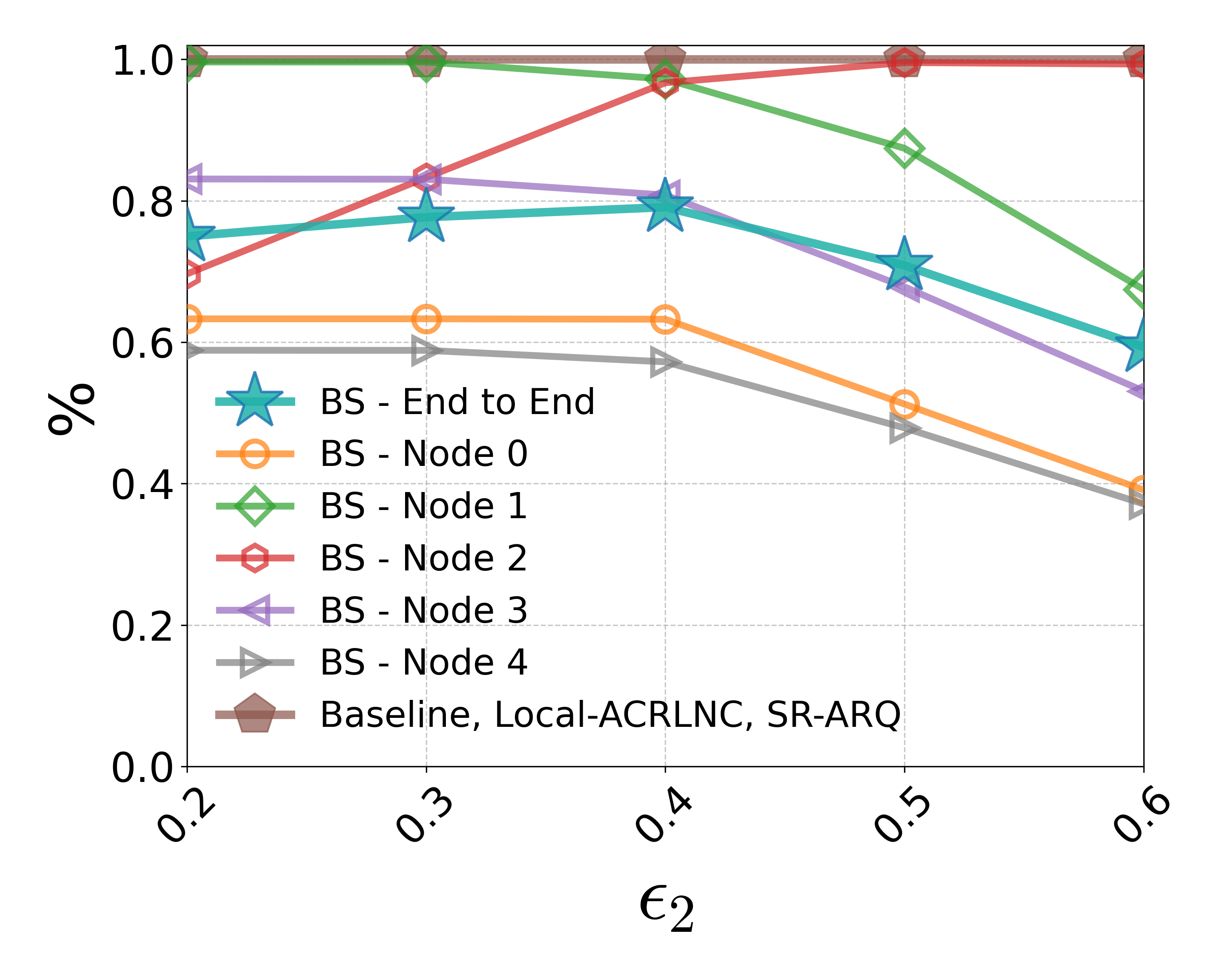}
        \caption{\footnotesize Channel Usage Rate - $U, U_n$}
        \label{fig:global_channel_use}
    \end{subfigure}
    \hspace{-0.45cm}
    \begin{subfigure}[t]{0.258\textwidth}
        \includegraphics[width=\linewidth,height=4cm]{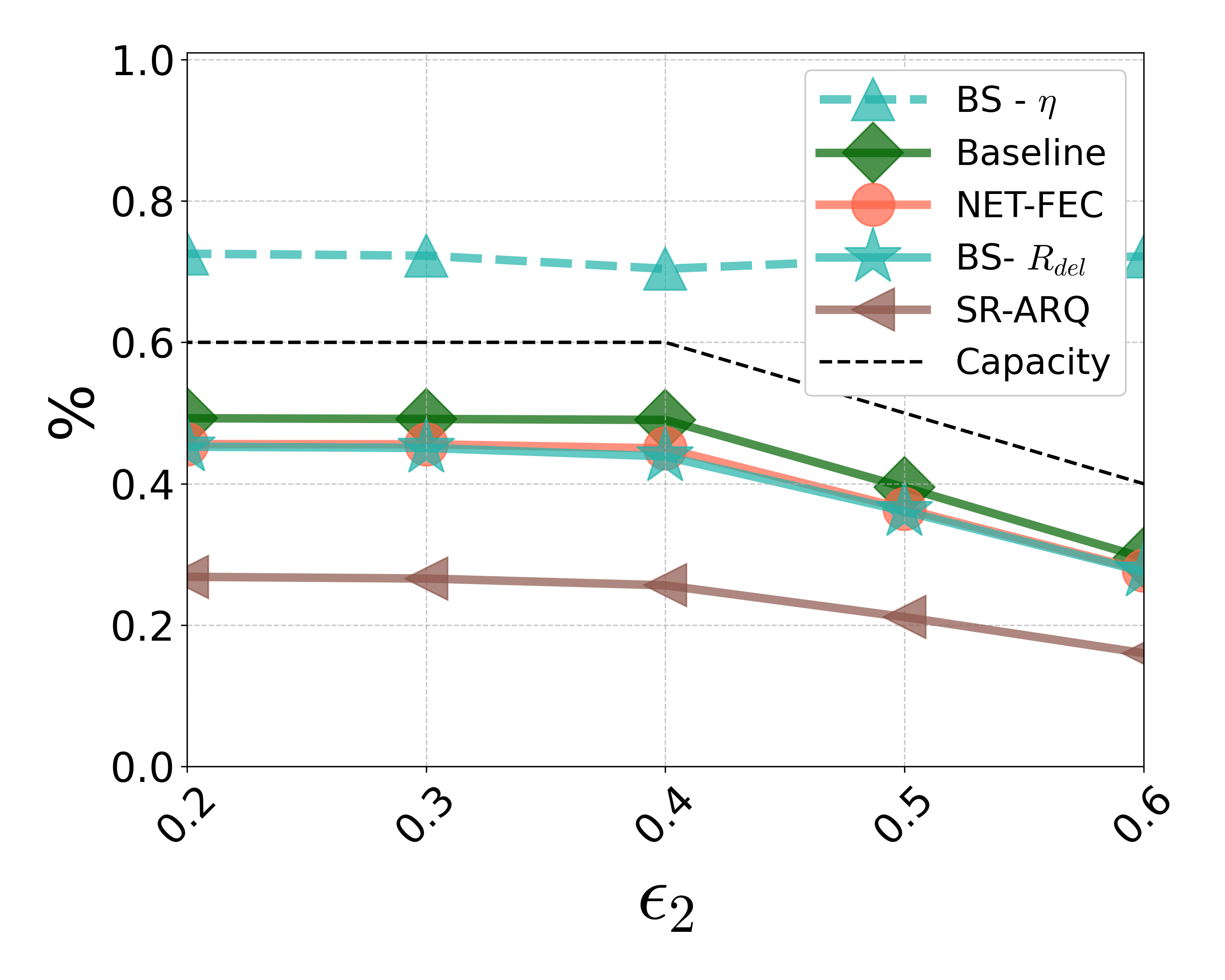}
        \caption{\footnotesize $R_{\rm del}$ and $\eta$}
        \label{fig:global_rate}
    \end{subfigure}
    \hspace{-0.45cm}
    \begin{subfigure}[t]{0.258\textwidth}
        \includegraphics[width=\linewidth,height=4cm]{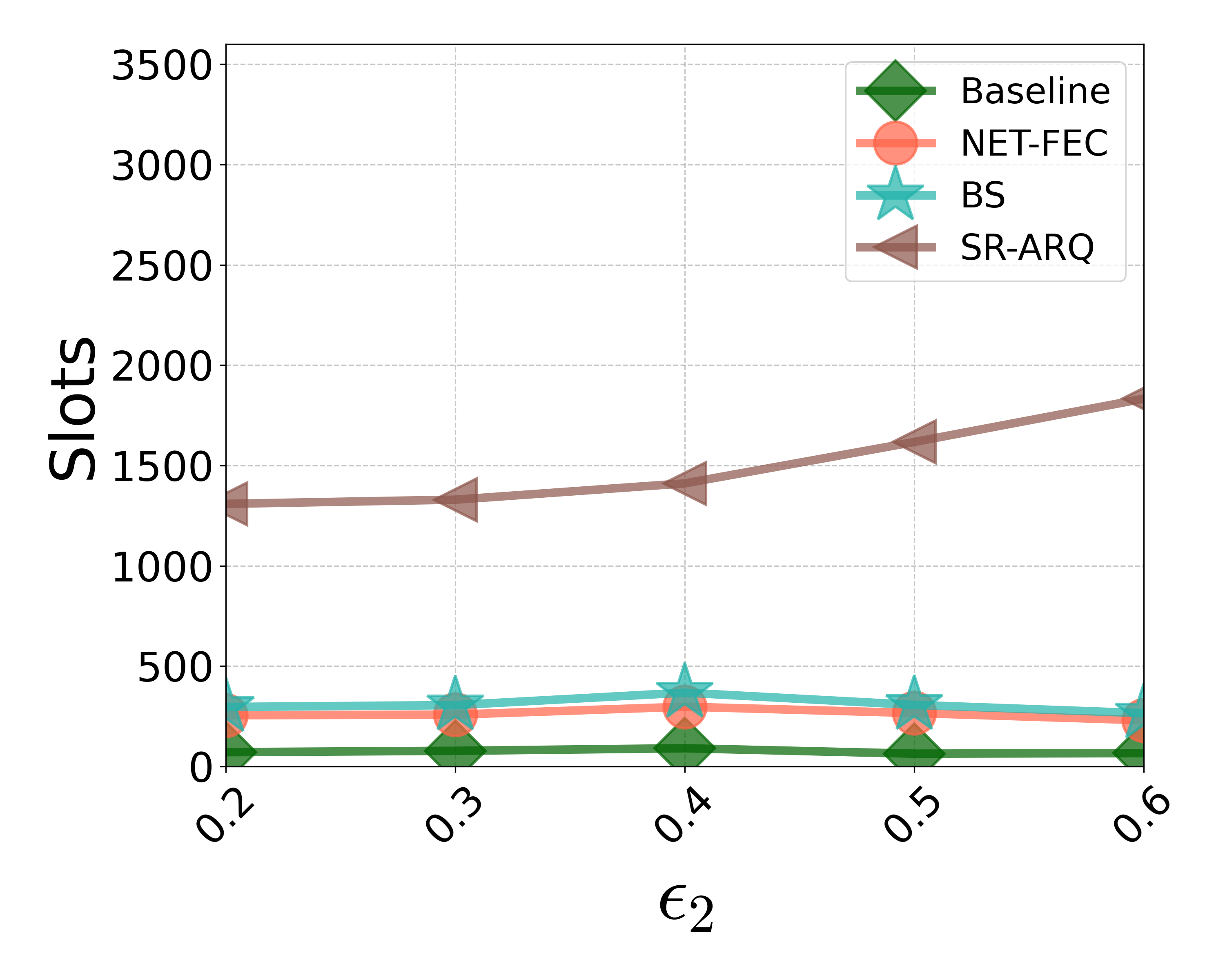}
        \caption{\footnotesize Mean Delay [Slots] - $D^{\rm mean}$}
        \label{fig:global_mean_d}
    \end{subfigure}
    \hspace{-0.45cm}
    \begin{subfigure}[t]{0.258\textwidth}
        \includegraphics[width=\linewidth,height=4cm]{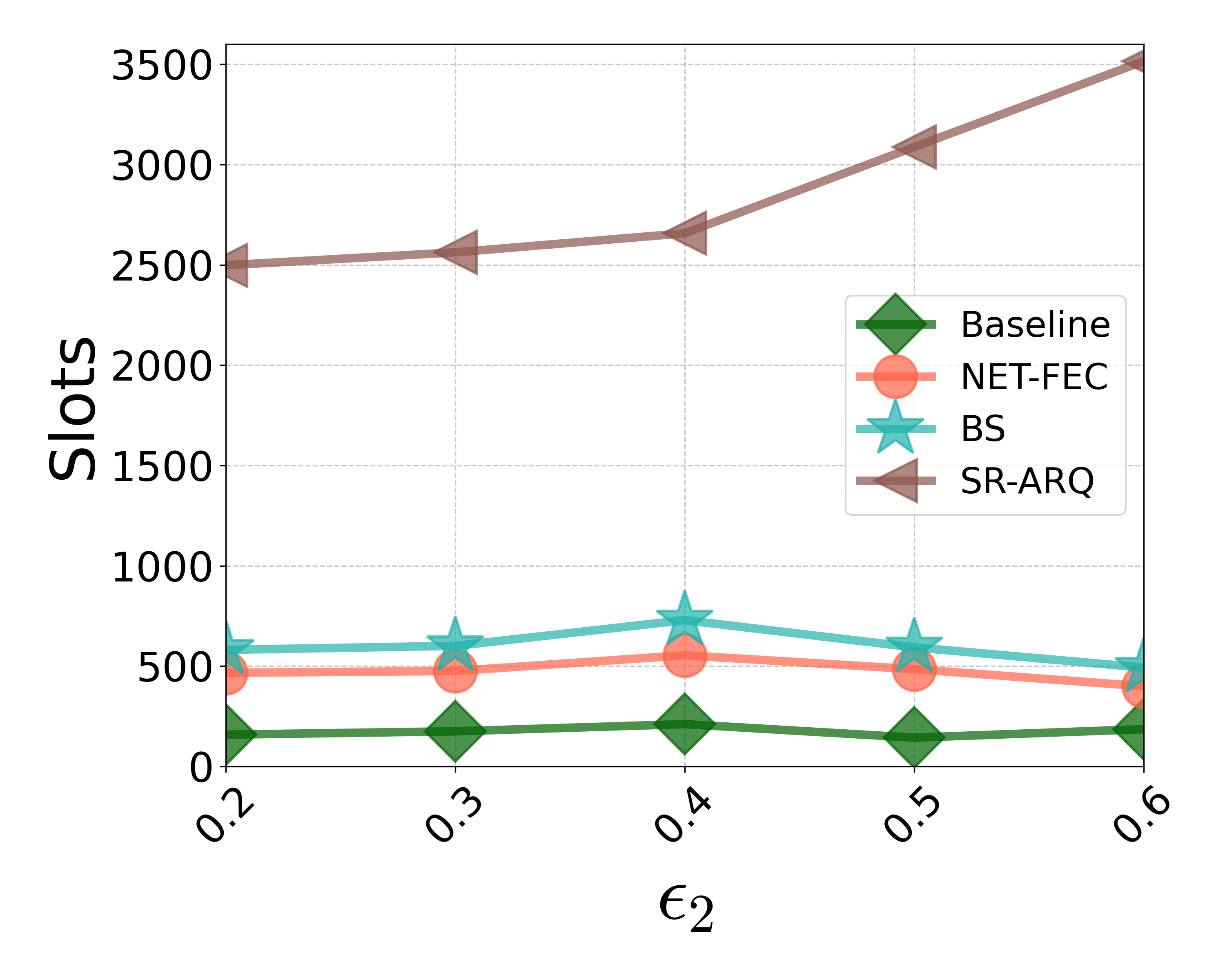}
        \caption{\footnotesize Max Delay [Slots] - $D^{\rm max}$}
        \label{fig:global_max_d}
    \end{subfigure}
    
    \caption{\small System performance of 6-nodes network with respect to $\epsilon_2 \in [0.2, 0.6]$ where $\epsilon_0=\epsilon_4=0.1$, $\epsilon_1=0.4$ and $\epsilon_3=0.3$. 
    }
    \label{fig:global_metrics}
    \ifdouble
    \vspace{-0.45cm}
    \else\fi
\end{figure*}

We compare four algorithms: 
1) The proposed \ac{bs-acrlnc}.
2) 'NET-FEC', a \ac{bs-acrlnc} version which performs \ac{net-acrlnc} without idle periods - the BS period is eliminated and No-New No-FEC pauses are replaced with FEC transmissions - This allows us to evaluate the effect of the idle periods distinctively.
3) 'Baseline' using MP-MH AC-RLNC, implementing AC-RLNC only at the source with basic re-encoding at intermediate nodes (see Sec.~\ref{subsec:MPMH}); 
and 4) common non-coded selective repeat ARQ ('SR-ARQ') \cite{weldon1982improved}.
%
In Fig.~\ref{fig:global_metrics}, we present the channel usage savings compared to the rate and delay metrics. 
In Fig.\ref{fig:global_channel_use}, we present the channel usage rate for each node separately and for the end-to-end network~\eqref{eqn:channel_use}. 
While NET-FEC, Baseline, and SR-ARQ operate at maximum channel utilization (rate of 1), \ac{bs-acrlnc} achieves significant reductions.

Examining each node demonstrates how \ac{bs-acrlnc} nodes adapt to network bottleneck constraints. We can see that usage generally decreases with increasing global bottleneck, except for the middle node (Node 2). This node increases activity with its erasure rate until reaching maximum utilization as it becomes the global bottleneck (for $\epsilon_2 > 0.4$). This causes nodes 1 and 2 usage to decrease via BS propagation effect, while nodes 3 and 4 decrease via No-New No-FEC, as fewer packets cross node 2.
The end-to-end performance turquoise stars) reveals that overall channel usage decreases with increasing erasure rate, highlighting \ac{bs-acrlnc}'s network-level benefits. We observe a $20\%$ reduction in channel usage when $\epsilon_1=0.4$ is the bottleneck, with even greater efficiency gains as the bottleneck shifts across the network.

Fig.~\ref{fig:global_rate} presents the goodput and delivery rates. For all algorithms except \ac{bs-acrlnc}, the two metrics are nearly identical, differing by no more than 0.01, and are therefore represented by a single line. \ac{bs-acrlnc} shows significantly higher goodput, indicated by the dashed line with upward triangles, due to its transmission pauses.
Notably, \ac{bs-acrlnc} reduces channel usage while maintaining a delivery rate comparable to the coding algorithms and outperforms the SR-ARQ. This demonstrates that the algorithm does not compromise its delivery performance for reducing channel usage.
%

The in-order delivery delay metrics of all the AC-RLNC solutions are upper-bounded in mean and max metrics by $400$ and $750$ time slots, respectively. The overall difference between the compared solutions remains within approximately two global $\rm RTT$. As demonstrated in \cite{dias2023sliding}, these delays meet the standard requirements for URLLC, and are one-order better than rateless RLNC \cite{bonello2011myths}, and two-orders better than non-coded solutions with UDP and selective repeat ARQ \cite{weldon1982improved}, which are typically used in TCP layers (discussed also in \cite{dias2023sliding,cohen2021bringing,cohen2020adaptiveMH}).
\ifdouble
\begin{figure}[ht]
   \centering
   \begin{subfigure}[b]{1\columnwidth}
       \includegraphics[width=1\columnwidth]{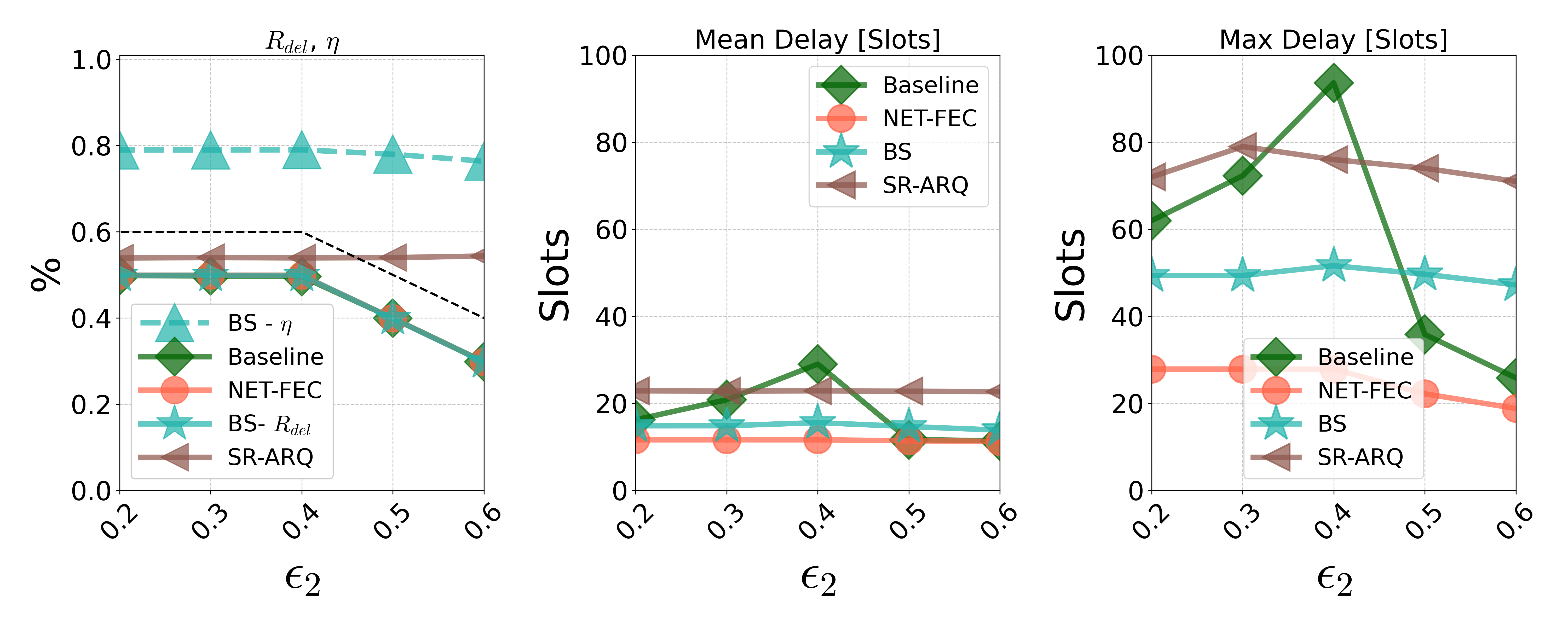}
       \caption{\small Node 1. The capacity for SR-ARQ is 0.9.}
       \label{fig:mc_1}
   \end{subfigure}
   \begin{subfigure}[b]{1\columnwidth}
       \includegraphics[width=\columnwidth]{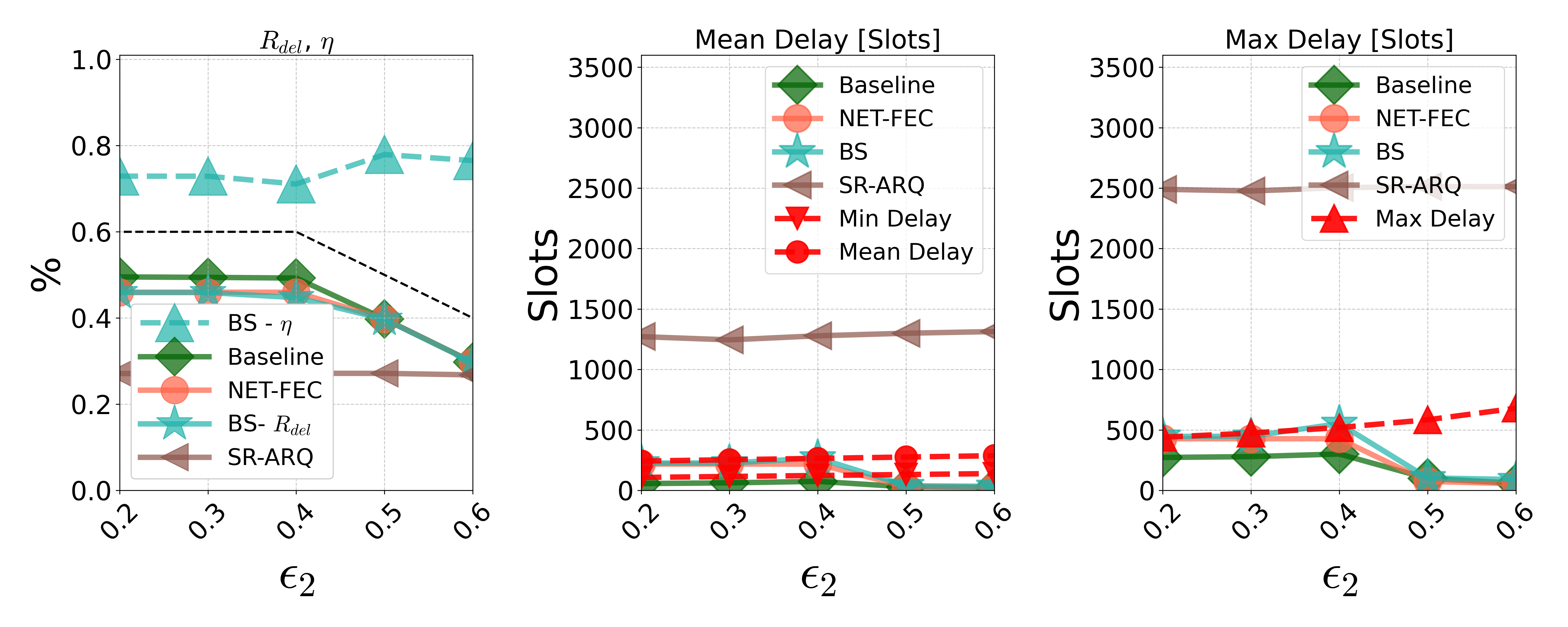}
       \caption{\small Node 2. The capacity for SR-ARQ is 0.6.}
       \label{fig:mc_2}
   \end{subfigure}
   \begin{subfigure}[b]{1\columnwidth}
       \includegraphics[width=\columnwidth]{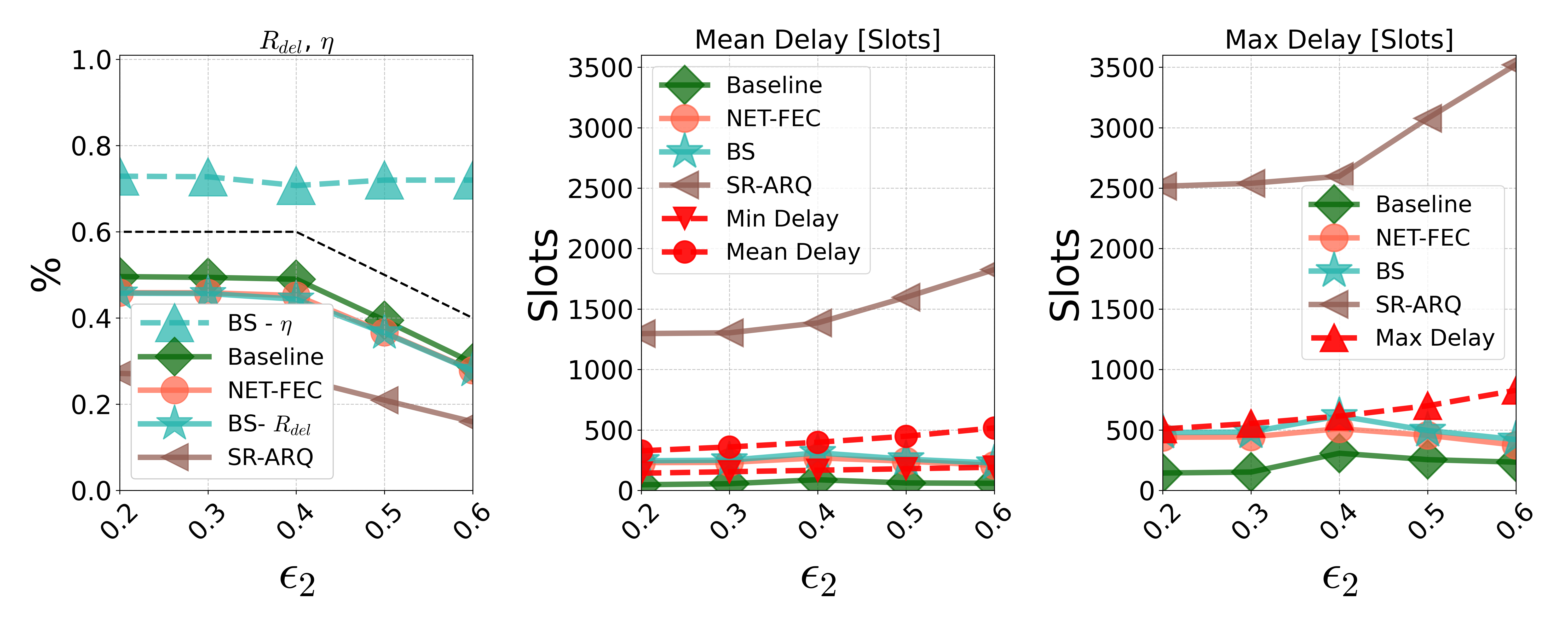}
       \caption{\small Node 3. The capacity for SR-ARQ is $\epsilon_2$.}
       \label{fig:mc_3}
   \end{subfigure}
   \\
   \begin{subfigure}[b]{1\columnwidth}
       \includegraphics[width=\columnwidth]{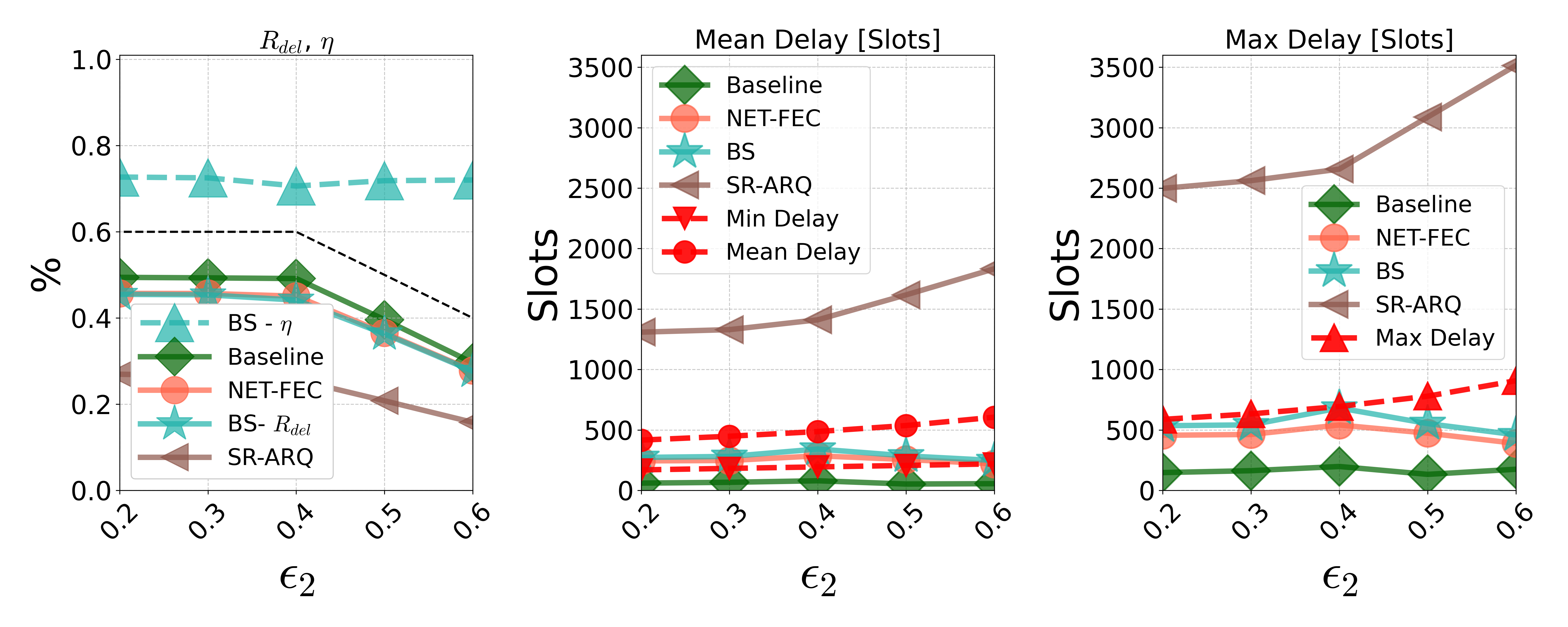}
       \caption{\small Node 4. The capacity for SR-ARQ is $\epsilon_2$.}
       \label{fig:mc_4}
   \end{subfigure}
   \begin{subfigure}[b]{1\columnwidth}
       \includegraphics[width=\columnwidth]{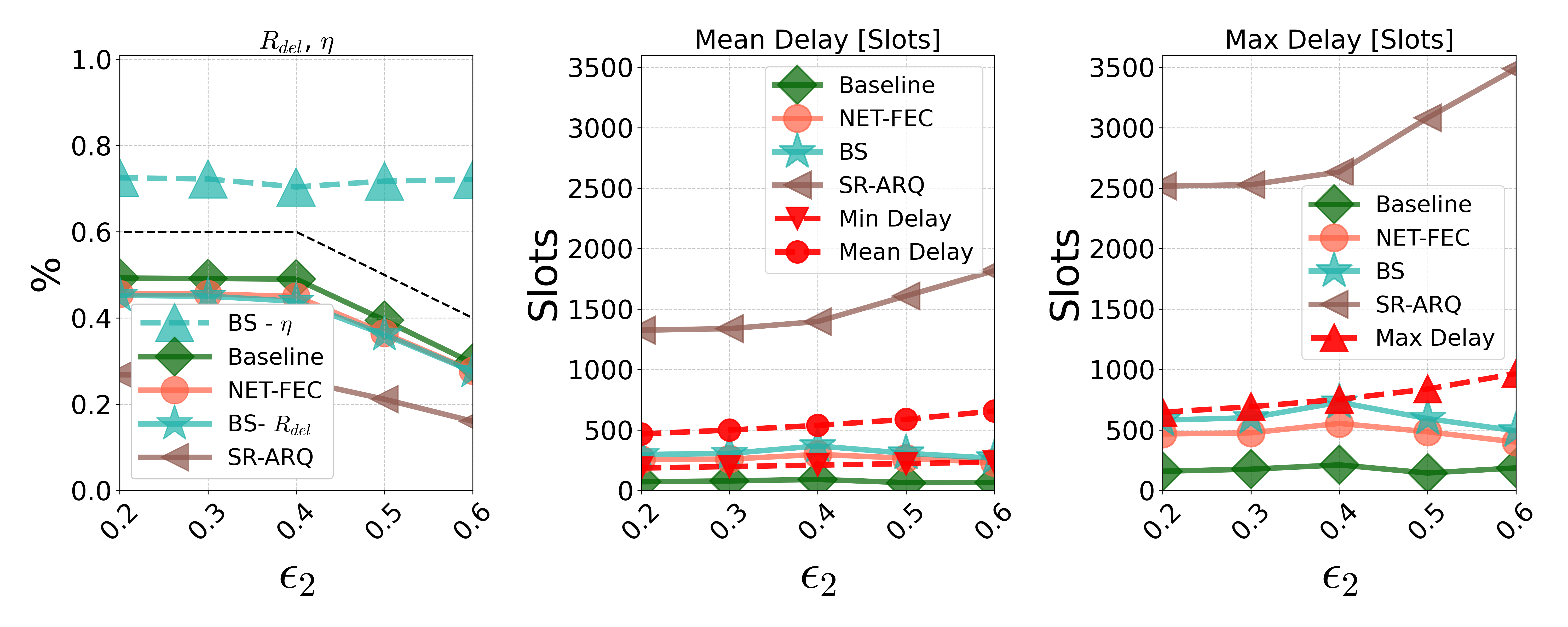}
       \caption{\small Node 5. The capacity for SR-ARQ is $\epsilon_2$.}
       \label{fig:mc_5}
   \end{subfigure}
   \caption{\small Multi-cast scenario when all nodes function as destinations.}
   \label{fig:multicast}
\end{figure}
\fi
%

In Fig.~\ref{fig:multicast}, we present a single-source multicast scenario where all intermediate nodes function as destinations, decoding the raw data packets. Note that this decoding does not affect the algorithm's operation, which continues to function in its original semi-decoding manner.
The results for both BS and NET-FEC demonstrate how delay varies with respect to the bottleneck channel position. The first node (Fig.~\ref{fig:mc_1}) experiences a minimal delay, consistent with the expected AC-RLNC algorithm performance. At the second node (Fig.~\ref{fig:mc_2}), we observe two distinct behaviors: high delay when $\epsilon\leq 0.4$ due to transmission over the bottleneck channel, and low delay when $\epsilon>0.4$ as the bottleneck shifts further downstream in the network. The remaining nodes  (Fig.~\ref{fig:mc_3}-Fig.~\ref{fig:mc_5}) exhibit higher delays as they are all affected by the bottleneck channel.

The goodput and delivery rate results align with our multi-hop findings (Fig.~\ref{fig:global_rate}). These metrics remain consistent across nodes since all algorithms consider the end-to-end network conditions rather than local information. While the delivery rate shows a slight decrease with distance from the source due to cumulative erasures, this could be improved through better erasure rate estimation.
The baseline solution demonstrates stable rate and delay performance due to its global approach, though at the cost of higher channel usage as shown in Fig.~\ref{fig:global_channel_use}. 
While SR-ARQ achieves competitive performance at the first node, its performance degrades significantly at subsequent nodes. Unlike the coding solutions, which consider the entire 6-node network, SR-ARQ operates locally at each node. This leads to high delivery rates at the initial nodes but declining performance when reaching bottleneck nodes. Since the multicast scenario requires uniform delivery rates across all nodes, this initial advantage becomes irrelevant.


These results demonstrate that \ac{bs-acrlnc} achieves significant improvements in channel usage and goodput while maintaining baseline delivery rate and low delay performance, validating the effectiveness of the proposed approach.
\section{Conclusions}\label{section:conclusions}

We proposed \acf{bs-acrlnc}, a novel adaptive and causal extension of AC-RLNC designed for multi-hop networks. The BS-AC-RLNC protocol introduces a new approach to mitigating the trade-off between throughput, delay, and efficiency by dynamically regulating traffic injection along the chain. Specifically, the proposed solution may prevent network flooding and buffer overflow conditions by allowing each node to independently estimate downstream bottlenecks.  Based on this local estimation, a node adaptively adjusts its packet transmission rate by inserting idle transmission slots, i.e., blank space period (BSP), thereby aligning its output rate with the forward path capacity. This decentralized regulation mechanism lays the foundation for broader scenarios, such as multi-access environments, where users must adaptively minimize their resource consumption to accommodate other concurrent applications.

To assess the impact of the proposed protocol on system performance, we derive theoretical bounds and validate them through experimental evaluation. We analytically establish bounds on in-order delivery delay, goodput, and throughput. The experimental results corroborate these theoretical findings, showing that BS-ACRLNC achieves delay and throughput performance on par with state-of-the-art schemes such as multi-hop multi-path adaptive causal RLNC (MP-MH AC-RLNC) and conventional selective repeat ARQ (SR-ARQ). Notably, these performance levels are attained with significantly reduced channel usage, highlighting the scheme's improved efficiency. Similar improvements are observed in the multi-cast setting, further demonstrating the robustness of the proposed approach across both unicast and multi-cast scenarios. Based on these promising results, we propose a further extension to multipath-multihop scenarios that would enable smarter transmission scheduling through idle paths, enhancing network efficiency.

\vspace{-0.5cm}
\bibliographystyle{IEEEtran}
\bibliography{refs_short}

\begin{thebibliography}{10}
\providecommand{\url}[1]{#1}
\csname url@samestyle\endcsname
\providecommand{\newblock}{\relax}
\providecommand{\bibinfo}[2]{#2}
\providecommand{\BIBentrySTDinterwordspacing}{\spaceskip=0pt\relax}
\providecommand{\BIBentryALTinterwordstretchfactor}{4}
\providecommand{\BIBentryALTinterwordspacing}{\spaceskip=\fontdimen2\font plus
\BIBentryALTinterwordstretchfactor\fontdimen3\font minus \fontdimen4\font\relax}
\providecommand{\BIBforeignlanguage}[2]{{%
\expandafter\ifx\csname l@#1\endcsname\relax
\typeout{** WARNING: IEEEtran.bst: No hyphenation pattern has been}%
\typeout{** loaded for the language `#1'. Using the pattern for}%
\typeout{** the default language instead.}%
\else
\language=\csname l@#1\endcsname
\fi
#2}}
\providecommand{\BIBdecl}{\relax}
\BIBdecl

\bibitem{yanmaz2018drone}
E.~Yanmaz, S.~Yahyanejad, B.~Rinner, H.~Hellwagner, and C.~Bettstetter, ``Drone networks: Communications, coordination, and sensing,'' \emph{Ad Hoc Networks}, vol.~68, pp. 1--15, 2018.

\bibitem{kumar2012overview}
M.~Kumar and R.~Mishra, ``An overview of {MANET}: History, challenges and applications,'' \emph{Ind. Jour. of Comp. Scie. and Eng. (IJCSE)}, vol.~3, no.~1, pp. 121--125, 2012.

\bibitem{radi2012multipath}
M.~Radi, B.~Dezfouli, K.~A. Bakar, and M.~Lee, ``Multipath routing in wireless sensor networks: survey and research challenges,'' \emph{sensors}, vol.~12, no.~1, pp. 650--685, 2012.

\bibitem{ahlswede2000network}
R.~Ahlswede, N.~Cai, S.-Y. Li, and R.~W. Yeung, ``Network information flow,'' \emph{IEEE Trans. on Inf. Theory}, vol.~46, no.~4, pp. 1204--1216, 2000.

\bibitem{li2003linear}
S.-Y. Li, R.~W. Yeung, and N.~Cai, ``Linear network coding,'' \emph{IEEE Trans. on Inf. Theory}, vol.~49, no.~2, pp. 371--381, 2003.

\bibitem{ho2006random}
T.~Ho, M.~M{\'e}dard, R.~Koetter, D.~R. Karger, M.~Effros, J.~Shi, and B.~Leong, ``A random linear network coding approach to multicast,'' \emph{IEEE Trans. on Inf. Theory}, vol.~52, no.~10, pp. 4413--4430, 2006.

\bibitem{dong2024throughput}
Y.~Dong, S.~Yang, J.~Wang, and F.~Cheng, ``Throughput and latency analysis for line networks with outage links,'' \emph{IEEE Journal on Selected Areas in Information Theory}, vol.~5, pp. 464--477, 2024.

\bibitem{4595268}
J.~K. Sundararajan, D.~Shah, and M.~Medard, ``{ARQ} for network coding,'' in \emph{2008 IEEE Int. Symp. on Inf. Theory}, 2008, pp. 1651--1655.

\bibitem{5061931}
J.~K. Sundararajan, D.~Shah, M.~Medard, M.~Mitzenmacher, and J.~Barros, ``{Network Coding Meets TCP},'' in \emph{IEEE INFOCOM}, 2009, pp. 280--288.

\bibitem{5688180}
J.~K. Sundararajan, D.~Shah, M.~Médard, S.~Jakubczak, M.~Mitzenmacher, and J.~Barros, ``{Network Coding Meets TCP: Theory and Implementation},'' \emph{Proc. of the IEEE}, vol.~99, no.~3, pp. 490--512, 2011.

\bibitem{luby2002lt}
M.~Luby, ``{LT} codes,'' in \emph{The 43rd Annual IEEE Symposium on Foundations of Computer Science, 2002. Proceedings.}\hskip 1em plus 0.5em minus 0.4em\relax IEEE Computer Society, 2002, pp. 271--271.

\bibitem{shokrollahi2006raptor}
A.~Shokrollahi, ``Raptor codes,'' \emph{IEEE Trans. on Inf. Theory}, vol.~52, no.~6, pp. 2551--2567, 2006.

\bibitem{joshi2012playback}
G.~Joshi, Y.~Kochman, and G.~W. Wornell, ``On playback delay in streaming communication,'' in \emph{2012 IEEE Int. Symp. on Inf. Theory}.\hskip 1em plus 0.5em minus 0.4em\relax IEEE, 2012, pp. 2856--2860.

\bibitem{cloud2015coded}
J.~Cloud, D.~Leith, and M.~M{\'e}dard, ``A coded generalization of selective repeat {ARQ},'' in \emph{2015 IEEE Conference on Computer Communications (INFOCOM)}.\hskip 1em plus 0.5em minus 0.4em\relax IEEE, 2015, pp. 2155--2163.

\bibitem{gabriel2018multipath}
F.~Gabriel, A.~K. Chorppath, I.~Tsokalo, and F.~H. Fitzek, ``Multipath communication with finite sliding window network coding for ultra-reliability and low latency,'' in \emph{2018 IEEE Int. Conf. on Comm. Workshops (ICC Workshops)}.\hskip 1em plus 0.5em minus 0.4em\relax IEEE, 2018, pp. 1--6.

\bibitem{9775949}
E.~Tasdemir, V.~Nguyen, G.~T. Nguyen, F.~H.~P. Fitzek, and M.~Reisslein, ``{FSW: Fulcrum Sliding Window Coding for Low-Latency Communication},'' \emph{IEEE Access}, vol.~10, pp. 54\,276--54\,290, 2022.

\bibitem{7249034}
J.~Du, N.~Sweeting, D.~C. Adams, and M.~Médard, ``Network reduction for coded multiple-hop networks,'' in \emph{2015 IEEE Int. Conf. on Comm. (ICC)}, 2015, pp. 4518--4523.

\bibitem{8638958}
D.~Malak, M.~Médard, and E.~M. Yeh, ``Tiny codes for guaranteeable delay,'' \emph{IEEE Jour. on Sel. Areas in Comm.}, vol.~37, no.~4, pp. 809--825, 2019.

\bibitem{cloud2013network}
J.~Cloud, D.~Leith, and M.~Medard, ``{Network coded TCP (CTCP) performance over satellite networks},'' \emph{arXiv preprint arXiv:1310.6635}, 2013.

\bibitem{6883489}
M.~Kim, J.~Cloud, A.~ParandehGheibi, L.~Urbina, K.~Fouli, D.~J. Leith, and M.~Médard, ``Congestion control for coded transport layers,'' in \emph{2014 IEEE Int. Conf. on Comm. (ICC)}, 2014, pp. 1228--1234.

\bibitem{8767270}
D.~Malak, A.~Schneuwly, M.~Médard, and E.~Yeh, ``Delay-aware coding in multi-hop line networks,'' in \emph{2019 IEEE 5th World Forum on Internet of Things (WF-IoT)}, 2019, pp. 650--655.

\bibitem{7117455}
Y.~Shi, Y.~E. Sagduyu, J.~Zhang, and J.~H. Li, ``Adaptive coding optimization in wireless networks: Design and implementation aspects,'' \emph{IEEE Trans. on Wireless Comm.}, vol.~14, no.~10, pp. 5672--5680, 2015.

\bibitem{9834704}
G.~Kasper~Facenda, E.~Domanovitz, M.~Nikhil~Krishnan, A.~Khisti, S.~L. Fong, W.-T. Tan, and J.~Apostolopoulos, ``On state-dependent streaming erasure codes over the three-node relay network,'' in \emph{2022 IEEE Int. Symp. on Inf. Theory (ISIT)}, 2022, pp. 1951--1956.

\bibitem{9174225}
E.~Domanovitz, A.~Khisti, W.-T. Tan, X.~Zhu, and J.~Apostolopoulos, ``Streaming erasure codes over multi-hop relay network,'' in \emph{2020 IEEE Int. Symp. on Inf. Theory (ISIT)}, 2020, pp. 497--502.

\bibitem{8835153}
S.~L. Fong, A.~Khisti, B.~Li, W.-T. Tan, X.~Zhu, and J.~Apostolopoulos, ``Optimal streaming erasure codes over the three-node relay network,'' \emph{IEEE Trans. on Inf. Theory}, vol.~66, no.~5, pp. 2696--2712, 2020.

\bibitem{10064107}
G.~K. Facenda, M.~N. Krishnan, E.~Domanovitz, S.~L. Fong, A.~Khisti, W.-T. Tan, and J.~Apostolopoulos, ``Adaptive relaying for streaming erasure codes in a three node relay network,'' \emph{IEEE Trans. on Inf. Theory}, vol.~69, no.~7, pp. 4345--4360, 2023.

\bibitem{cohen2020adaptive}
A.~Cohen, D.~Malak, V.~B. Bracha, and M.~M{\'e}dard, ``Adaptive causal network coding with feedback,'' \emph{IEEE Trans. on Comm.}, vol.~68, no.~7, pp. 4325--4341, 2020.

\bibitem{cohen2020adaptiveMH}
A.~Cohen, G.~Thiran, V.~B. Bracha, and M.~M{\'e}dard, ``Adaptive causal network coding with feedback for multipath multi-hop communications,'' \emph{IEEE Trans. on Comm.}, vol.~69, no.~2, pp. 766--785, 2020.

\bibitem{shannon2003zero}
C.~Shannon, ``The zero error capacity of a noisy channel,'' \emph{IRE Trans. on Inf. Theory}, vol.~2, no.~3, pp. 8--19, 2003.

\bibitem{gallager2003simple}
R.~Gallager, ``A simple derivation of the coding theorem and some applications,'' \emph{IEEE Trans. on Inf. Theory}, vol.~11, no.~1, pp. 3--18, 2003.

\bibitem{lovasz1979shannon}
L.~Lov{\'a}sz, ``On the shannon capacity of a graph,'' \emph{IEEE Trans. on Inf. Theory}, vol.~25, no.~1, pp. 1--7, 1979.

\bibitem{sahai2008block}
A.~Sahai, ``Why do block length and delay behave differently if feedback is present?'' \emph{IEEE Trans. on Inf. Theory}, vol.~54, no.~5, pp. 1860--1886, 2008.

\bibitem{polyanskiy2011feedback}
Y.~Polyanskiy, H.~V. Poor, and S.~Verdu, ``Feedback in the non-asymptotic regime,'' \emph{IEEE Trans. on Inf. Theory}, vol.~57, no.~8, pp. 4903--4925, 2011.

\bibitem{cohen2021bringing}
A.~Cohen, H.~Esfahanizadeh, B.~Sousa, J.~P. Vilela, M.~Luis, D.~Raposo, F.~Michel, S.~Sargento, and M.~Medard, ``Bringing network coding into {SDN}: Architectural study for meshed heterogeneous communications,'' \emph{IEEE Comm. Magazine}, vol.~59, no.~4, pp. 37--43, 2021.

\bibitem{cohen2022broadcast}
O.~Cohen, A.~Cohen, M.~Médard, and S.~S. Shitz, ``Broadcast approach meets network coding for ultra-reliable and low-latency data streaming,'' \emph{IEEE Transactions on Communications}, vol.~74, pp. 1301--1318, 2026.

\bibitem{dias2023sliding}
E.~Dias, D.~Raposo, H.~Esfahanizadeh, A.~Cohen, T.~Ferreira, M.~Lu{\'\i}s, S.~Sargento, and M.~M{\'e}dard, ``{Sliding window network coding enables NeXt generation URLLC millimeter-wave networks},'' \emph{IEEE Networking Letters}, vol.~5, no.~3, pp. 159--163, 2023.

\bibitem{ali2021urllc}
R.~Ali, Y.~B. Zikria, A.~K. Bashir, S.~Garg, and H.~S. Kim, ``{URLLC for 5G and beyond: Requirements, enabling incumbent technologies and network intelligence},'' \emph{IEEE Access}, vol.~9, pp. 67\,064--67\,095, 2021.

\bibitem{malak2019tiny}
D.~Malak, M.~M{\'e}dard, and E.~M. Yeh, ``Tiny codes for guaranteeable delay,'' \emph{IEEE J. Sel. Areas in Commun.}, vol.~37, no.~4, Apr. 2019.

\bibitem{karzand2015fec}
M.~Karzand, D.~J. Leith, J.~Cloud, and M.~M{\'e}dard, ``Fec for lower in-order delivery delay in packet networks,'' \emph{arXiv preprint arXiv:1509.00167}, 2015.

\bibitem{shi2013whether}
X.~Shi, M.~M{\'e}dard, and D.~E. Lucani, ``Whether and where to code in the wireless packet erasure relay channel,'' \emph{IEEE Jour. on Sel. Areas in Comm.}, vol.~31, no.~8, pp. 1379--1389, 2013.

\bibitem{PacketBEC}
M.~v.~d. Schaar and P.~A. Chou, \emph{Multimedia over IP and Wireless Networks: Compression, Networking, and Systems}.\hskip 1em plus 0.5em minus 0.4em\relax USA: Academic Press, Inc., 2007.

\bibitem{swrlnc}
S.~Wunderlich, F.~Gabriel, S.~Pandi, F.~H. Fitzek, and M.~Reisslein, ``{Caterpillar RLNC (CRLNC): A practical finite sliding window RLNC approach},'' \emph{IEEE Access}, vol.~5, pp. 20\,183--20\,197, 2017.

\bibitem{shrader2007queueing}
B.~Shrader and A.~Ephremides, ``A queueing model for random linear coding,'' in \emph{MILCOM 2007-IEEE Military Communications Conference}.\hskip 1em plus 0.5em minus 0.4em\relax IEEE, 2007, pp. 1--7.

\bibitem{domanovitz2022information}
E.~Domanovitz, T.~Philosof, and A.~Khina, ``The information velocity of packet-erasure links,'' in \emph{IEEE INFOCOM 2022-IEEE Conference on Computer Communications}.\hskip 1em plus 0.5em minus 0.4em\relax IEEE, 2022, pp. 190--199.

\bibitem{enenche2023network}
P.~Enenche, D.~H. Kim, and D.~You, ``{Network coding as enabler for achieving URLLC under TCP and UDP environments: A survey},'' \emph{IEEE Access}, vol.~11, pp. 76\,647--76\,674, 2023.

\bibitem{geil2008field}
O.~Geil, R.~Matsumoto, and C.~Thomsen, ``On field size and success probability in network coding,'' in \emph{International Workshop on the Arithmetic of Finite Fields}.\hskip 1em plus 0.5em minus 0.4em\relax Springer, 2008, pp. 157--173.

\bibitem{cohen2020bringing}
A.~Cohen, H.~Esfahanizadeh, B.~Sousa, J.~P. Vilela, M.~Lu{\'\i}s, D.~Raposo, F.~Michel, S.~Sargento, and M.~M{\'e}dard, ``Bringing network coding into sdn: a case-study for highly meshed heterogeneous communications,'' \emph{arXiv preprint arXiv:2010.00343}, 2020.

\bibitem{medard2025network}
M.~M{\'e}dard, V.~A. Vasudevan, M.~V. Pedersen, and K.~R. Duffy, \emph{Network Coding for Engineers}.\hskip 1em plus 0.5em minus 0.4em\relax John Wiley \& Sons, 2025.

\bibitem{patterson2014and}
S.~Patterson, ``How mit and caltech’s coding breakthrough could accelerate mobile network speeds’,'' \emph{Network World}, 2014.

\bibitem{katti2008xors}
S.~Katti, H.~Rahul, W.~Hu, D.~Katabi, M.~M{\'e}dard, and J.~Crowcroft, ``{XORs in the air: Practical wireless network coding},'' \emph{IEEE/ACM Transactions on networking}, vol.~16, no.~3, pp. 497--510, 2008.

\bibitem{chen2021exact}
W.~Chen, F.~Lu, and Y.~Dong, ``Exact decoding probability of sparse random linear network coding for reliable multicast,'' \emph{arXiv preprint arXiv:2108.11659}, 2021.

\bibitem{chen2020decoding}
------, ``The decoding success probability of sparse random linear network coding for multicast,'' \emph{arXiv preprint arXiv:2010.05555}, 2020.

\bibitem{michel2022flec}
F.~Michel, A.~Cohen, D.~Malak, Q.~De~Coninck, M.~M{\'e}dard, and O.~Bonaventure, ``{FlEC: Enhancing QUIC with application-tailored reliability mechanisms},'' \emph{IEEE/ACM Trans. on Net.}, vol.~31, no.~2, pp. 606--619, 2022.

\bibitem{9834410}
A.~Cohen, A.~Solomon, and N.~Shlezinger, ``{DeepNP: Deep Learning-Based Noise Prediction for Ultra-Reliable Low-Latency Communications},'' in \emph{2022 IEEE Int. Symp. on Inf. Theory (ISIT)}, 2022, pp. 2690--2695.

\bibitem{basu19972}
A.~Basu, I.~R. Harris, and S.~Basu, ``2 minimum distance estimation: The approach using density-based distances,'' \emph{Handbook of Statistics}, vol.~15, pp. 21--48, 1997.

\bibitem{yeung2008chapter18}
R.~W. Yeung, ``The max-flow bound,'' in \emph{Information Theory and Network Coding}.\hskip 1em plus 0.5em minus 0.4em\relax Springer Science \& Business Media, 2008, ch.~18, pp. 421--434.

\bibitem{polyanskiy2010channel}
Y.~Polyanskiy, \emph{Channel coding: Non-asymptotic fundamental limits}.\hskip 1em plus 0.5em minus 0.4em\relax Princeton University, 2010.

\bibitem{weldon1982improved}
E.~Weldon, ``An improved selective-repeat {ARQ} strategy,'' \emph{IEEE Trans. on Comm.}, vol.~30, no.~3, pp. 480--486, 1982.

\bibitem{bonello2011myths}
N.~Bonello, Y.~Yang, S.~Aissa, and L.~Hanzo, ``Myths and realities of rateless coding,'' \emph{IEEE Comm. Magazine}, vol.~49, no.~8, pp. 143--151, 2011.

\bibitem{CovThom06}
T.~M. Cover and J.~A. Thomas, \emph{Elements of Information Theory}, 2nd~ed.\hskip 1em plus 0.5em minus 0.4em\relax New-York: Wiley, 2006.

\bibitem{polyanskiy2009dispersion}
Y.~Polyanskiy, H.~V. Poor, and S.~Verd{\'u}, ``Dispersion of gaussian channels,'' in \emph{2009 IEEE International Symposium on Information Theory}.\hskip 1em plus 0.5em minus 0.4em\relax IEEE, 2009, pp. 2204--2208.

\end{thebibliography}

\ifpagelimit
\clearpage
\pagenumbering{arabic}
\setcounter{page}{1}
\begin{center}
  \huge Supplementary Materials
\end{center}
\else\fi

\appendices
\section{Example of the BS-AC-RLNC Algorithm}\label{appendix:example}
\begin{figure}[htb]
    \centering
    \ifdouble
    \includegraphics[width=0.8\linewidth]{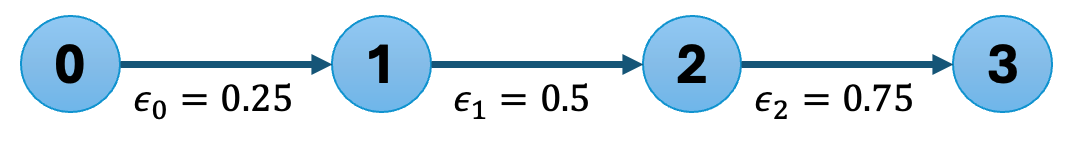}
    \fi
    \caption{\small An example of a multi-hop network with four nodes and three channels, where the erasure rates are $\epsilon_0 = 0.25$, $\epsilon_1 = 0.5$, and $\epsilon_2 = 0.75$.}
    \label{fig:example}
\end{figure}

\begin{figure}
    \centering
    \ifdouble
    \includegraphics[width=1\linewidth]{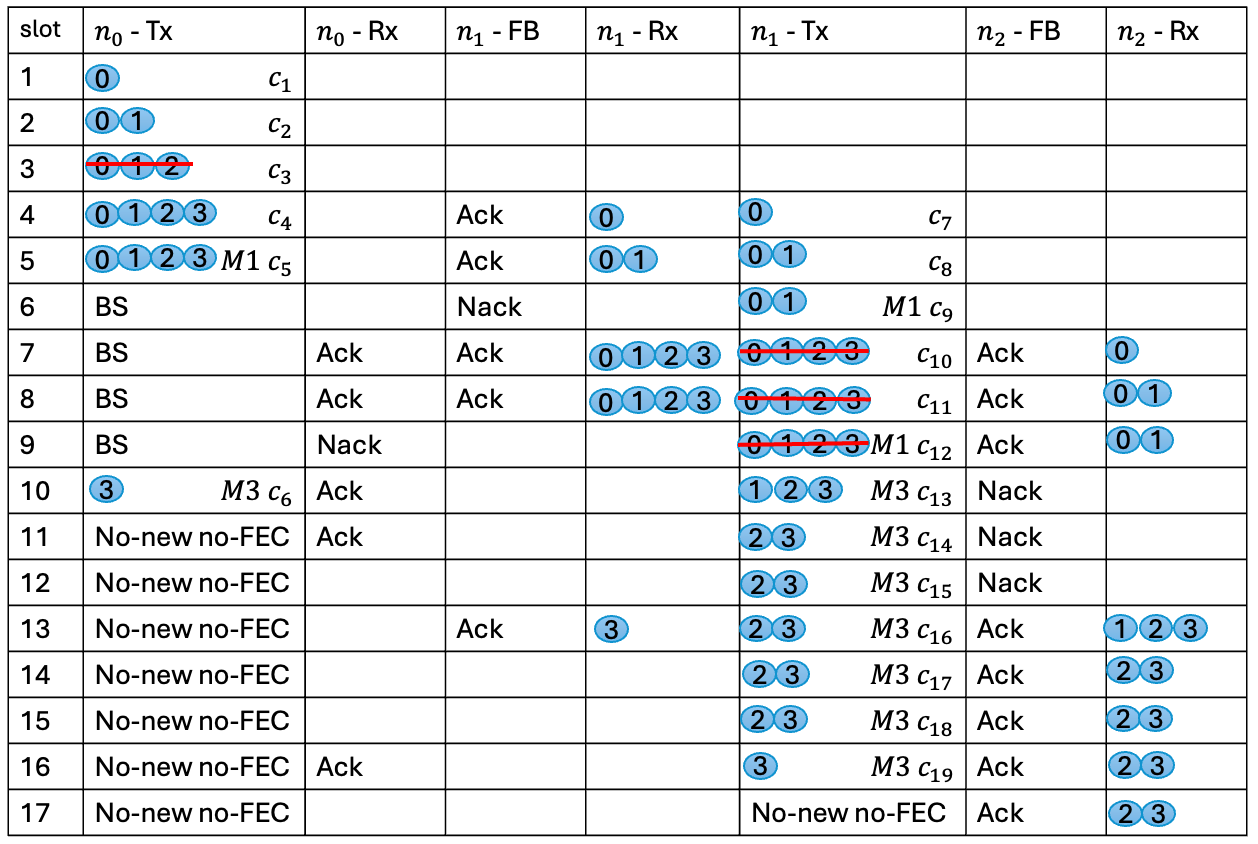}
    \fi
    \caption{\small Illustration of the \ac{bs-acrlnc} algorithm. The first column indicates the time slots. The second and third columns display the transmissions from, and the feedback received by, node $n_0$. The next three columns correspond to the sent feedback, received data, and transmissions of node $n_1$. The final two columns represent the sent feedback and received data at node $n_2$. The symbols $c_1$ through $c_{19}$ denote linear combinations of raw data packets, as referenced in the accompanying example. Blue dots represent the raw data packet $p_0$ through $p_3$, with each index shown on the corresponding dot. Erasures are indicated by red lines crossing the dots involved in each respective linear combination. Markers M1 and M3 correspond to the a-priori FEC and EOW conditions described in~\ref{itm:a-FEC} and~\ref{itm:EOW}, respectively.}
    \label{fig:example_algo}
\end{figure}

This appendix presents a detailed example demonstrating the operation of the \acf{bs-acrlnc} algorithm. The network configuration is shown in Fig.~\ref{fig:example}, and the execution of the algorithm is summarized in Fig.~\ref{fig:example_algo}. Throughout the example, we assume a round-trip time ($\rm RTT$) of 8 time slots for each channel, a window size of $w = 4$, BS parameters $\alpha = 1$ \eqref{eqn:BS_num} and $\kappa=1$ \eqref{eqn:delta_bs}, \rg{with intermediate nodes performing full decoding}.

\begin{itemize}[wide, labelwidth=0.3cm, labelindent=1pt]
    \item Consider a scenario in which node $n_0$ intends to transmit four raw data packets, denoted $p_0$, $p_1$, $p_2$, and $p_3$, to the destination node $n_3$. In time slot 1, $n_0$ sends a linear combination of $p_0$, given by $c_1 = \mu_0 p_0$, \rg{which corresponds to a new DoF (Definition~\ref{def:new})}. In time slot 2, it transmits a linear combination of $p_0$ and $p_1$, $c_2 = \mu_1 p_0 + \mu_2 p_1$, \rg{again yielding a new DoF}. This process continues, with each subsequent transmission incorporating one additional raw data packet, until time slot 4, in which a linear combination of all four raw data packets is transmitted.

    Given an erasure rate of $\epsilon_0 = 0.25$ and the transmission of packets $c_1-c_4$, the a-priori FEC mechanism~\ref{itm:a-FEC} requires one additional coded packet. Accordingly, node $n_0$ transmits an extra linear combination of all four raw data packets, denoted by $c_5$, \rg{corresponding to the same DoF (Definition~\ref{def:same})}.

    The forward bottleneck link for node $n_0$ has an erasure rate of $\epsilon_2 = 0.75$, resulting in a BSP of ten time slots. According to~\eqref{eqn:BS_num}, this is computed as \vspace{-1em} $$BS \leftarrow 1 \cdot \sum_{i=1}^{2} 8 \cdot \epsilon_i = 10.$$ The BSP termination condition, as defined in~\eqref{eqn:delta_bs} and~\eqref{BS_crit}, is not satisfied at time slot 9, since $$\Delta_{9}^{\mathrm{BS}} = \left[0.75 \cdot 7 + 2\cdot\ln(0.75 - 0.25)\right]^{-1} \approx 0.2151 < 0.25.$$ However, at time slot 10, the condition is met, $$\Delta_{10}^{\mathrm{BS}} = \left[0.75 \cdot 6 + 2\cdot\ln(0.75 - 0.25)\right]^{-1} \approx 0.2565 > 0.25.$$ Thus, the BSP terminates at time slot 10.
    
    \item Now suppose that coded packets $c_1$, $c_2$, $c_4$, and $c_5$ are successfully received by node $n_1$ during time slots 4–5 and 7–8, respectively, while $c_3$ is lost due to an erasure. In time slot 9, node $n_0$ receives a negative acknowledgment (NACK) for $c_3$. According to~\ref{itm:p-FEC}, a posteriori FEC is not required, since
    \[
    \textstyle  \Delta_9 = \frac{1 + 0.25 \cdot \rg{1}}{2 + 0.75 \cdot 1} - 1 < 0,
    \] 
    as defined in~\eqref{eqn:delta_new}. This expression reflects the current transmission state: one NACK and two ACKs have been received, \rg{one coded packet ($c_4$) is not yet acknowledged}, and one coded retransmission ($c_5$) has been sent. As $\Delta_9 < 0$, the condition for triggering a posteriori FEC is not met.

    During time slot 10, since node $n_0$ has reached the end of its transmission window and has received an acknowledgment for $c_1,c_2$ and $c_4$, it continues to transmit linear combinations of raw data packets $p_3$ as part of the end-of-window (EOW) phase~\ref{itm:EOW}. Finally, in time slot 11, an acknowledgment for $c_5$ confirms that node $n_1$ has acquired a sufficient number of DOFs to decode all raw data packets. As no new raw data packets are available at $n_0$, it follows the No-New No-FEC policy and terminates transmission.

    \item Node $n_1$ begins by transmitting a new linear combination of $p_0$, denoted by $c_{7}$, followed by $c_{8}$, a linear combination of $p_0$ and $p_1$. Since $c_3$ was lost due to an erasure and $n_1$ has no new raw data packets to transmit, it sends an a-priori FEC packet, $c_{9}$, which is a linear combination of $p_1$ and $p_2$. This additional coded transmission is required because two data packets have been sent, and the erasure probability is $\epsilon_1 = 0.5$.

    Since the forward bottleneck link for $n_1$ is characterized by $\epsilon_2 = 0.75$, the BSP duration is calculated as \ifpagelimit $BS \leftarrow 1 \cdot 8 \cdot 0.75 = 6$, \fi
    time slots. The BSP termination condition is evaluated at time slot 7, where \ifpagelimit   $\Delta_{7}^{\mathrm{BS}} = \left[0.5 \cdot 6 + \ln(0.75 - 0.5)\right]^{-1} \approx 0.4171 > 0.25$. \fi
    Since this condition is met, no BSP is initiated for node $n_1$.

    In time slots 7--8, $n_1$ receives packets $c_4$ and $c_5$, which it retransmits as $c_{10}$ and $c_{11}$, followed by an a-priori FEC packet, $c_{12}$. During time slots 10--12, $n_1$ receives acknowledgments from node $n_2$ for packets $c_7-c_9$, and accordingly updates its transmissions to send linear combinations of the remaining undecoded raw data packets, $p_2$ and $p_3$\footnote{\rg{Here we assume that intermediate nodes perform full decoding. If, instead, intermediate nodes only track the DoFs, node $n_1$ would transmit a new linear combination of a subset of the received coded packets (e.g., $c_4, c_5$, and $c_6$), which collectively involve the raw data packets $p_0$--$p_3$.}}. The EOW phase~\ref{itm:EOW} begins in time slot 10, after which $n_1$ continues transmitting until time slot 17, when it receives ACK indicating that $n_2$ has obtained a sufficient number of DOFs to complete the decoding.

    Note that since the EOW phase~\ref{itm:EOW} has been reached, we omit the analysis of the posterior FEC condition~\ref{itm:p-FEC}, as retransmissions continue in this case regardless.

    \item In a similar fashion, node $n_2$ transmits linear combinations of raw data packets $p_0-p_3$ to node $n_3$.
\end{itemize}

\ifpagelimit
\section{Tables}\label{appendix:tables}
The notation used throughout the paper is summarized in this section in Tables~\ref{fig:table1} and~\ref{fig:table_analytical_results}.

\ifpagelimit
\begin{table}[ht!]
    \footnotesize
    \centering
    \begin{tabular}{|l|l|l|}
        \hline
        {\bf Notation} & {\bf Definition} \\
        \hline
        \hline
        $t^-$& time slot ${\rm RTT}_n$ units before the current time\\
        $r_n(t^-)$& known erasure rate at node $n$\\
        $r_n(t)$& estimated rate at node $n$ in current time\\
        $V_n(t)$& rate variance\\
        $h$& number of hops between links $i$ and $j$\\
        $n_n^w$& total number of transmissions over channel $n$ \\&during a given transmission window\\
        $n_n^{\text{EW}}$& number of redundant transmission that occur \\&in channel $n$ at the end of the window\\
        $\text{erf}$& an error function\\
        $I_n$& total BSP for channel $n$\\
        $I_n^u$& unnecessary part of the BSP for channel $n$\\
        $D_{\min}$& end-to-end delay lower bound\\
        $D_{\text{mean}}$& expected end-to-end delay bound\\
        $D_{\max}$& end-to-end delay upper bound\\
        $\tau_{\min}$& lower bound of the achievable per-link \\&transmission rate\\
        $\tau_{\max}$& upper bound of the achievable per-link \\&transmission rate\\
        $\theta_{\min}$& per-link idle phase lower bound\\
        $\theta_{\max}$& per-link idle phase upper bound\\
        $\Gamma_{\min}$& normalized goodput lower bound\\
        $\Gamma_{\max}$& normalized goodput upper bound\\
        $\eta_n^{\min}$ & per-link lower throughput bound of channel $n$\\
        $\eta_n^{\max}$& per-link upper throughput bound of channel $n$\\
        $\eta_{\min}$& end-to-end throughput lower bound\\
        $\eta_{\max}$& end-to-end throughput upper bound\\
        \hline
    \end{tabular}
    \vspace{0.0cm}
    \caption{\small Notation referenced in the derivation and presentation of analytical results for the \ac{bs-acrlnc} algorithm.}
    \label{fig:table_analytical_results}
\end{table}
\else\fi

\begin{table}[ht!]\footnotesize
\centering
\footnotesize
\begin{tabular}{|l|l|l|}
\hline
{\bf Notation} & {\bf Definition} \\
    \hline
    \hline
    $N$ & nodes in the network, ranging from $0$ to $N-1$\\
    $e_n$ & noisy forward channel between nodes $(n,n+1)$\\
    $f_n$ & noiseless feedback channel\\
    $T > 0$ & time horizon\\ 
    $t',t < T$ & time slot\\
    $M$ &  \rg{number of last feedback messages received used}\\& \rg{to estimate erasure rate}\\
    $\epsilon_n$ & erasure probability of channel $n$\\
    $\hat{\epsilon}_n$ & estimated erasure rate of channel $n$\\
    $\hat{\epsilon}$& estimated erasure rate\\
    $\epsilon_{BN_n}$ & erasure probability of the forward bottleneck \\&channel\\
    $\hat{\epsilon}_{BN_n}$ & estimated erasure probability of the forward \\&bottleneck channel\\
    $r_n = 1 - \epsilon_n$ & channel rate\\
    $\hat{r}$& estimated channel rate\\
    $BN_n$ & forward channels bottleneck of node $n$\\
    $\text{BN}$ & global bottleneck\\ 
    ${\rm RTT}_n$ & round-trip time between nodes $(n,n+1)$\\
    $\rm RTT$ & global propagation delay\\
    $p_i$ & raw data packet, the ith arrived at the source\\
    $c_t^n$ & coded packet sent from node $n$ at time slot $t$\\
    $c_t$ & coded packet at time slot $t$\\
    $c_t^{\text{new}}$& number of new DoF in the current transmission \\&window\\
    $c_t^{\text{same}}$& number of retransmissions in the current \\&transmission window\\
    $l$& length of raw data packet\\
    $\mu_i$ & random coefficients\\
    $\mathbb{F}_q$ & finite field\\
    $\mathbb{F}_q^{\ell}$ & raw data packets of length $\ell$ over $\mathbb{F}_q$\\
    \hline
    \hline
    $T_1(p_i)$ & arrival time of $p_i$ at the source\\
    $T_d(p_i)$ & decoding time of $p_i$ at the destination\\
    $d(t)$ & cumulative number of raw data packets\\ & decoded at the destination in time slot $t$\\
    $O_n\rg{(t), O_n(T)}$ & number of idle slots for node $n$ \rg{at time $t$ and}
    \\& \rg{over the time horizon $T$, respectively}\\
    $\eta$ & normalized goodput\\
    $R_{del}$ & delivery rate\\
    $U$ & channel usage rate across the entire network\\
    $U_n$& channel usage rate per forward channel $n$\\
    $D$& in order delivery delay\\
    $D_i$& in order delivery delay per raw data packet $i$\\
    $D^{\text{mean}}$ & mean delay\\
    $D^{\max}$ & maximum delay\\
    \hline
    \hline
    $F_t^{n}$ & feedback from node $n$ for a packet sent at slot $t$\\
    $AF^n_t$& aggregate feedback from nodes $n+1$ to $N-1$\\ 
    $w$& \rg{maximum} size of the sliding window\\
    $\rg{\overline{w}}$& \rg{effective size of the sliding window, i.e. $\overline{w}\leq w$ }\\ 
    $w^{\min}_t,w_t^{\max}$& \rg{effective} sliding window bounds\\
    $md_t^{nack}$ &amount of NACK\\
    $ad_t^{ack}$& amount of retransmissions\\
    $th$& threshold\\
    $BS_n(t)$& blank space duration for node $n$ at time $t$\\
    $h$& distance in hop between the current channel\\& and the bottleneck\\
    $\Delta_t$& posterior FEC trigger\\
    $\Delta_t^{BS}$& blank space condition for termination\\
    \hline
\end{tabular}
\vspace{0.0cm}
\caption{\small Definitions of symbols. The table is organized into three sections: 1) system parameters, 2) performance metrics, and 3) solution-related variables.}
\label{fig:table1}
\end{table}
\else\fi

\clearpage

\rg{\section{Proof of Lemma~\ref{lemma:RedundantTransmissions}}\label{appendix:lemma11}

To upper bound the number of redundant transmissions at the end of the window, \(n_i^{\mathrm{EW}}\), consider the following scenario. At time \(t\), node \(i\) transmits the last coded packet, which is eventually acknowledged by node \(i+1\). Upon reception of the corresponding ACK, at time \(t+{\rm RTT}_i\), node \(i\) learns that node \(i+1\) has sufficient DoF to decode the raw data packets indexed from \(w_t^{\min}\) to \(w_t^{\max}\). Additionally, assume that node \(i\) has reached the maximum size of the sliding window \(w\). Suppose that \(i = BN_i\), and therefore no BSP is applied.

In this case, during the time-slot interval from \(t\) to \(t + {\rm RTT}_i\), node \(i\) reaches the EOW state (\ref{itm:EOW}) by following line~\ref{line:eow} in Algorithm~\ref{alg:full_alg}, and consequently retransmits the same DoF for \({\rm RTT}_i\) time slots. This implies that, overall, node \(i+1\) receives \((1-\epsilon_i){\rm RTT}_i\) redundant transmissions.

Regarding the term \(1 - \operatorname{erf}\!\left(\frac{1}{\sqrt{2}}\right)\), reaching the EOF state indicates that the coded packets transmitted during the FEC periods (\ref{itm:a-FEC} and~\ref{itm:p-FEC}) were insufficient for node \(i+1\) to decode.
That is, the deviation from the mean exceeds approximately one standard deviation, in accordance with the \(68\text{--}95\text{--}99.7\) rule. 
However, this rule holds only for sufficiently large \(t\). By applying the Berry--Esseen theorem~\cite[Theorem~13]{polyanskiy2010channel} to the independent erasure events, it can be shown that the error term of the asymptotic rule decreases as \(\frac{1}{\sqrt{t}}\), approaching zero for sufficiently large \(t\).}

\rg{\section{Proof of Lemma~\ref{lemma:rateConsecutiveNodes}}\label{appendix:lemma7}

The notation \(t^-\) refers to the time \(t - {\rm RTT}_n\) slots ago. Under the noiseless feedback channel assumption, all ACKs and NACKs for coded packets sent up to time slot \(t - {\rm RTT}_n\) have been received. The uncertainty arises only in the most recent interval, from \(t - {\rm RTT}_n\) to \(t\), since feedback for packets transmitted during this interval has not yet arrived.

The term \(r_n(t^-) = 1 - \hat{\epsilon}_n\) corresponds to the classical asymptotic regime, capturing the average behavior of the channel~\cite{shannon2003zero,CovThom06} for sufficiently large $M$. Meanwhile, the term \(\sqrt{V_n(t)}/{{\rm RTT}_n}\) accounts for the finite-blocklength regime, incorporating the second moment of the noise to characterize the channel variance~\cite{polyanskiy2010channel,polyanskiy2011feedback}.

Under the BEC assumption, the uncertainty is modeled as \({\rm RTT}_n\) independent trials, each with a success probability of \(1 - \hat{\epsilon}_n\). Consequently, the variance is given by \(V_n = {\rm RTT}_n \cdot \hat{\epsilon}_n (1 - \hat{\epsilon}_n)\), and \(\sqrt{V_n(t)}\) is divided by \({\rm RTT}_n\) to normalize with respect to the number of trials. Following Polyanskiy et al.~\cite{polyanskiy2009dispersion} the \(O(1)\) term in the finite-blocklength approximation of the BEC is neglected.}

\end{document}